\documentclass[%
 reprint,
nofootinbib,
amsmath,amssymb,
aps,
]{revtex4-2}

\usepackage{graphicx}
\usepackage{dcolumn}
\usepackage{bm}
\usepackage{appendix}
\usepackage{listings}
\usepackage{csquotes}
\usepackage[usenames,table,x11names]{xcolor}
\usepackage{cancel}
\usepackage{soul}
\usepackage{upgreek}
\usepackage{hyperref}
\usepackage{mathrsfs}
\usepackage{bbold}
\usepackage{enumerate}
\usepackage{orcidlink}

\newcommand{\beq}{\begin{equation}}
\newcommand{\eeq}{\end{equation}}
\newcommand{\bea}{\begin{eqnarray}}
\newcommand{\eea}{\end{eqnarray}}
\newcommand{\bit}{\begin{itemize}}
\newcommand{\eit}{\end{itemize}}
\newcommand{\ben}{\begin{enumerate}}
\newcommand{\een}{\end{enumerate}}
\newcommand{\nn}{\nonumber}

\newcommand{\Caltech}{Theoretical Astrophysics Group, California Institute of Technology, Pasadena, CA 91125, U.S.A.}
\newcommand{\Cardiff}{School of Physics and Astronomy, Cardiff University, Queens Buildings, Cardiff, CF24 3AA, United Kingdom}

\newcommand{\NBI}{Niels Bohr International Academy, Niels Bohr Institute, Blegdamsvej 17, 2100 Copenhagen, Denmark}
\newcommand{\soton}{Mathematical Sciences and STAG Research Centre, University of Southampton,Southampton SO17 1BJ, United Kingdom}

\begin{document}

\preprint{APS/123-QED}

\title{Hyperboloidal discontinuous time-symmetric numerical algorithm with higher order jumps for gravitational self-force computations in the time domain}
\author{Lidia J. Gomes Da Silva \orcidlink{0000-0001-9593-4217}}
\affiliation{School of Mathematical Sciences, Queen Mary University of London, E1 4NS, London, U.K.}
\email{Corresponding authors: lidiajoana@pm.me}
\author{Rodrigo Panosso Macedo \orcidlink{0000-0003-2942-5080}}
\affiliation{\NBI}
\affiliation{\soton}
\author{Jonathan E. Thompson \orcidlink{0000-0002-0419-5517} }
\affiliation{\Caltech}
\affiliation{\Cardiff}
\author{Juan A. Valiente Kroon \orcidlink{/0000-0001-9511-2520} }
\affiliation{School of Mathematical Sciences, Queen Mary University of London, E1 4NS, London, U.K.
}
\author{Leanne Durkan \orcidlink{0000-0001-8593-5793}}
\affiliation{University of Texas at Austin, Austin, TX, 78712, USA}
\author{Oliver Long \orcidlink{0000-0002-3897-9272} }
\affiliation{Max Planck Institute for Gravitational Physics (Albert Einstein Institute), Am M\"uhlenberg 1, Potsdam 14476, Germany}

\date{\today}

\begin{abstract}

Within the next decade the Laser Interferometer Space Antenna (LISA) is due to be launched, providing the opportunity to extract physics from stellar objects and systems, such as \textit{Extreme Mass Ratio Inspirals}, (EMRIs) otherwise undetectable to ground based interferometers and Pulsar Timing Arrays (PTA). Unlike previous sources detected by the currently available observational methods, these sources can \textit{only} be simulated using an accurate computation of the gravitational self-force. Whereas the field has seen outstanding progress in the frequency domain, metric reconstruction and self-force calculations are still an open challenge in the time domain. Such computations would not only further corroborate frequency domain calculations and models, but also allow for full self-consistent evolution of the orbit under the effect of the self-force. Given we have \textit{a priori} information about the local structure of the discontinuity at the particle, we will show how to construct discontinuous spatial and temporal discretisations  by operating on discontinuous Lagrange and Hermite interpolation formulae and hence recover higher order accuracy. In this work we demonstrate how this technique in conjunction with well-suited gauge choice (hyperboloidal slicing) and numerical (discontinuous collocation with time symmetric) methods can provide a relatively simple method of lines numerical algorithm to the problem. This is the first of a series of papers studying the behaviour of a point-particle prescribing circular geodesic motion in Schwarzschild in the \textit{time domain}. In this work we describe the numerical machinery necessary for these computations and show not only our work is capable of highly accurate flux radiation measurements but it also shows suitability for evaluation of the necessary field and it's derivatives at the particle limit. 
\end{abstract}
\maketitle
\onecolumngrid
\vspace{\columnsep}
  PACS: Classical Black Holes, Gravitational Wave Sources, Astronomical Black Holes 
\vspace{\columnsep}
\twocolumngrid

\section{Introduction} \label{introduction}

Detection of gravitational waves from extreme mass-ratio inspirals (EMRIs), in the low frequency band of $10^{-5}$ to $10^{-1}$ Hz, is one of the key targets for the Laser Interferometer Space Antenna, LISA  \cite{LISA, amaro2017laser, baker2019laser, barausse2020prospects, perkins2021probing}. EMRIs are composed of a supermassive black hole (BH) of mass $M$ in orbit with a compact object such as a neutron star (NS) or BH of mass $\mu$, where the mass-ratio $\epsilon = \mu/M \ll 1 $. Given the disparate scales in an EMRI, the \textit{radiation reaction} problem is best tackled through a perturbative treatment. Techniques that yield highly accurate models for comparable mass systems, like numerical relativity \cite{brugmann2018fundamentals} and post-Newtonian methods \cite{blanchet2014gravitational} will not work in their current state, for this case \cite{barackpoundREVIEW2018self}.

\subsection{Overview of current computational strategies for EMRI modelling} \label{introA}

Through the machinery of black hole perturbation theory, we approximate the smaller compact object as a \textit{point-particle} such that at zeroth order in $\mu$ it follows the geodesic of the background and at first order it deviates from this geodesic due to the interaction with its self-field. This deviation is viewed as a force acting on the smaller object, the self-force \cite{barackpoundREVIEW2018self, poundwardellREVIEW2022black, poissonpoundvegaREVIEW2011motion}. The metric is directly perturbed as:
\begin{equation}
    g_{\alpha\beta}(x) =  \bar{g}_{\alpha\beta}(x) + h_{\alpha\beta}(x) + \mathcal{O}(\epsilon^{2}), 
    \label{cha1_g_perturbed}
\end{equation}
where $\bar{g}_{\alpha\beta}$ is the background metric, taken to be a vacuum solution to the field equations, and $h_{\alpha \beta}$ is the perturbation of the metric to first order in the mass-ratio. In this work we shall ignore any higher-order terms and focus on first-order self-force computations. Incorporation of the local self-force is an essential step for a complete EMRI waveform model. Second-order calculations are imminent \cite{capra25monday, spiers2022second, durkan2022metricTHESIS, leather2022numerical}, with already simpler waveform models yielding promising results \cite{wardell2021gravitational, albertini2022comparing, albertini2022comparingii, van2023enhancing}. However, within the context of our numerical technique here, these will not be reviewed. 

In this paper we start by approximating the supermassive BH background spacetime to be spherically symmetric, described by the Schwarzschild metric. In reality, BHs are expected to be rotating \cite{teukolsky1973perturbations}, and more accurately approximated by a Kerr background, albeit at the price of adding extra degrees of complexity to what otherwise is a numerically friendly $1+1D$ model. Another decision that needs to be taken, strongly entangled with the choice of background, is the even harder choice of \textit{gauge}, which, ultimately will influence the final self-force result and the full EMRI's orbital evolution \cite{barackSFgaugeDep2001gravitational, barackpoundREVIEW2018self}. 

The foundations of the radiation reaction problem have traditionally been set in the \textit{Lorenz/harmonic/de Donder gauge} \cite{quinn1997axiomatic, mino1997gravitational, detweilerwhitting2003self, pound2012second, pound2017nonlinear}. In Schwarzschild, the Lorenz gauge is amenable to scalar and tensorial harmonic decomposition, facilitating gravitational self-force (GSF) computations that have been successful in multi-domain computational strategies. The regularisation of the metric perturbation in a Kerr background remains an open challenge \cite{whiting2005metric,van2015metric, dolan2022gravitational}, mostly due to the lack of separability into multi-pole modes. It was not until very recently that several promising strategies started emerging \cite{durkan2022slowderivatives, osburn2022newEllip, dolan2023metric}, paving the way to compute the full GSF from the Lorenz gauge, or, at least through a partial gauge approach \cite{toomani2021new, spiers2022second}.

Another common choice of gauge is the \textit{radiation gauge} (RG). Here the perturbations are most commonly calculated from curvature scalars in the Newman-Penrose framework where the Teukolsky equation is found to be separable \cite{teukolsky1973perturbations}. Radiation gauges can be subdivided into three main classes (for a detailed review we refer the reader to \cite{pound2014gravitational, poundwardellREVIEW2022black}).  The main difference among them is the resulting \enquote{string-like} singularity emerging from the point-particle. Most of the progress in Kerr in Table \ref{ch1_table_previouswork_graviational}, has come from the implementation of the \textit{no-string} RG \cite{van2018gravitational, lynch2022eccentric}. 

The \textit{Regge-Wheeler gauge} \cite{regge1957stability} choice yields a well-posed numerical set-up in Schwarzschild mostly thanks to the fact that a full tensorial harmonic decomposition is possible and one is left to solve a set of $1+1D$ \textit{wave-like} equations  
which are distributionally sourced and have a potential. These are known as the Regge-Wheeler-Zerilli, RWZ, \cite{regge1957stability,zerilli1970gravitational} equations. A full regularised framework has been implemented in \cite{thompson1611, thompson1811} for circular geodesic motion at first order. 
Unfortunately, the extension of the RW gauge to Kerr has remained unsolved to date \cite{Chandrasekhar:1985kt}, with a recent progress in the slow rotating regime \cite{franchini2023slow}. In this work we choose to work in the RW gauge and solve for the RWZ master functions. 

There are currently \textit{four main different computational strategies} to tackle self-force computations. In what follows we give a quick overview and refer the reader to \cite{barackREVIEW2009gravitational, wardellREVIEW2015self} for a more comprehensive review of these techniques.  
\begin{itemize}
    \item \textit{GSF from flux balance laws} - One can calculate the gravitational self-force dissipative components in the time $t$ and angular $\phi$ direction, $\mathcal{F}_{t}$ and 
     $\mathcal{F}_{\phi}$ by computing the total radiated fluxes at future null infinity and the horizon one can compute the dissipative components of the gravitational self-force, \textit{e.g.},
     \cite{Galtsov:1982hwm, Sago_2006,Mino:1996nk}. For the full self-force computation one needs to include any of the other three computational strategies discussed below.
    \item \textit{GSF from mode sum} - The $4D$ gravitational field is first decomposed into spherical harmonics modes allowing for the $1 + 1D$ numerical treatment of the problem \cite{barack2000mode}. This wave equation is then solved mode by mode. With adequate regularisation techniques the scalar, \textit{e.g.,} \cite{barack2000self,Detweiler:2002gi}, and the gravitational, \textit{e.g.,} \cite{heffernan2012high,van2016gravitational}, self-force can then be computed by summing over all modes. 
    \item \textit{GSF from effective source/puncture methods} - Regularisation is done before solving the wave equation. The final numerical solution to the wave equation accounts for the regularised field \cite{vega2008regularization, barack2007scalar, barack2007m,jaranowski2007analytic, lousto2008new, wardell2012generic,Wardell:2015ada}. 
    \item \textit{GSF from Green functions} -  The regularised retarded field is computed as the integral of the matched retarded Green function along the particle’s worldine \cite{anderson2005matched,casals2009self, o2021characteristic}. 
\end{itemize} 

All these computational strategies, background metric and gauge choices result in a final differential equation which, like the Schr\"odinger's problem can be tackled as a time-dependent or time-independent problem: either we choose to solve it as an ordinary differential equation (ODE) for the radial component in the \textit{frequency domain}, (FD), or as a partial-differential equation (PDE), where we must solve for two-dimensions in time and space (\textit{time domain}, TD)\footnote{One should note, depending on the gauge and computational strategy choices, the problem can be tackled in $1+1D$, $2+1D$ or $3+1D$ given that simplifications of the angular dimensions are not always possible.}. In Table \ref{ch1_table_previouswork_graviational}, we highlight the current state of \textit{first-order} GSF calculations.

\subsection{Motivation for \textit{time-domain} algorithm development}  \label{introB}

Most of the progress has happened in the frequency domain. This is mainly due to the fact that the equations to solve become fully separable into frequency harmonic modes, reducing the numerical problem to solving a set of ODEs. However, FD methods still show some disadvantages. Due to the number of higher modes required in this domain, highly eccentric orbital models remain an open challenge (the more eccentric, the more modes are required for convergence), especially in the Lorenz gauge \cite{akcay2011fast, akcay2013frequency, van2018gravitational}. In the RG some progress has recently been achieved at first-order in the FD \cite{lynch2022eccentric}, though there is still room for improvement as the current results do not reach the required sub-radian accuracies from EMRI data and inclusion of second-order effects is expected to improve current results. 

Another disadvantage of FD methods is that they are largely restricted to bound orbits due to the need for a finite range of discrete orbital frequencies. Recent work has relaxed this requirement allowing for calculations of self-force quantities for scattering orbits \cite{whittall2023frequency}. However, the most accurate calculations have been performed in the TD, where the methods can naturally handle the full range of frequencies \cite{long2021time, barack2022self, barack2023PM}. 

Local self-force computations are just the first step towards the production of a full EMRI waveform model, on top of more efficient self-force computations for highly eccentric and unbound orbits, time-domain methods are also expected to provide two alternative ways for waveform generation \cite{poundwardellREVIEW2022black, wardellREVIEW2015self}.  One is by computing all the required dynamical quantities for an EMRI model through a self-consistent evolution \cite{diener2012self, samuelcuppC25} and another is as part of a two-timescale expansion, where, a balancing act between frequency- and time-domain methods may be possible \cite{poundwardellREVIEW2022black}. Furthermore, accurate TD codes can help improve current 0PA adiabatic-level inspirals attained through FD methods and provide a more flexible route to evolve from the adiabatic inspiral to the transition to plunge \cite{taracchini2014small, apte2019exciting, lim2019exciting, rifat2020surrogate, islam2022surrogate}.

\subsection{Difficulties associated with \textit{time-domain} numerical methods} \label{introC}

With all these motivations in mind, we consider some of the key difficulties associated with the implementation of a time-domain numerical PDE solver:
\begin{itemize}
    \item \textit{Difficulty 1} - representation of the Dirac-$\delta$ distribution emerging from the point-particle model of the compact object;
    \item \textit{Difficulty 2} - the numerical domain of integration encompasses an infinite domain $(-\infty\ , \infty)$, whereas boundary conditions are imposed at a finite artificial hypersurface; 
    \item  \textit{Difficulty 3} - Time-domain methods must ensure that sufficient accuracy is maintained for signals potentially lasting the entire $3 \ \pm $ year LISA mission duration.
    \item \textit{Difficulty 4} - Owing to the distributional nature of our source, one needs care when choosing initial data and effective metrics to determine it's effects in the full radiative process of EMRI systems. Smooth initial data choices are ill-posed and must be carefully considered.
\end{itemize}

\subsubsection{Difficulty~1 - Dirac $\delta$ distribution representation} \label{introC-D1}

There have been many computational implementations solving the RWZ master functions in the TD, the main difference in these algorithms being the treatment of the Dirac-$\delta$ distribution, \textit{difficulty 1}. The first attempt at solving the RWZ equations in the TD was implemented by \textit{Lousto \& Price} \cite{lousto1997head,lousto1997understanding, lousto1998improved} using a finite-difference representation of the $\delta$ distribution and integrating in the time-domain using a Runge-Kutta scheme. Later, \textit{Martel \& Poisson} calculated the TD energy and angular momentum fluxes by correcting the finite-difference scheme to second-order convergence \cite{martel2002one, martel2004gravitational}. Their work became the standard accuracy test for all the TD algorithms that followed (including this work). The method was then fine-tuned by \textit{Lousto} \cite{lousto2005time} to fourth-order. Following their attempts came a new algorithm from \textit{Sopuerta \& Laguna} \cite{sopuerta2006finite}, who used an adaptive finite-element approach to represent the distributional source terms. Another novel approach by \textit{Sopuerta~et~al.} was to use a multi-domain spectral collocation method representing the Dirac-$\delta$ by finite elements \cite{canizares2009simulations,jaramilloSopCan2011time} following the promising implementations of \cite{jung2007spectral, jung2007spectralNote}. Later \textit{Field~et~al.} \cite{field2009discontinuous, field2010persistent} put forward a novel discontinuous Galerkin algorithm registering significant improvements to previous physical flux results. Another attempt at numerically representing the Dirac-$\delta$ distribution was as a Gaussian function by \textit{Nagar \& Bernuzzi~et~al.} \cite{nagar2007binary, bernuzzi2010binary}. 

\subsubsection{Difficulty~2 - How to solve the outer radiation problem}
\label{introC-D2}
The answer to \textit{difficulty~2} by most of these numerical algorithms \cite{lousto2005time, martel2004gravitational, sopuerta2006finite, canizares2009simulations, nagar2007binary, bernuzzi2010binary, field2009discontinuous} was to use radiation boundary conditions put forward by \cite{lau2004rapid, lau2005analytic}. The main issue with this is known as the outer radiation problem: where it is not entirely clear whether extraction of information by extrapolating data to the horizon/infinity is contaminated with new information stemming from the implementation of these conditions. An alternative that gets rid of this issue altogether is the use of \textit{hyperboloidal slicing}, a mathematical technique arising naturally in the study of asymptotics of the gravitational field by means of conformal methods \cite{penrose1963asymptotic,friedrich1983cauchy}. With the spacetime parameterised by a compact radial coordinate defined on a hyperboloidal time hypersurface, one is able to access information directly at the black hole horizon and future null infinity. The first successful application of hyperboloidal slicing in black hole perturbation theory (BHPT) came from \textit{Zenginoglu}, when solving the Bardeen-Press equation \cite{zenginouglu2009gravitational}. Later, along with \textit{Bernuzzi \& Nagar}, \textit{Zenginoglu} extended his work from solving the homogeneous RWZ equations \cite{zenginouglu2010asymptotics} to the inhomogeneous case using the aforementioned Gaussian-$\delta$ numerical treatment \cite{zenginouglu2010asymptotics, bernuzzi2011binary}. Altogether, their innovative implementation showed significant improvement relative to their previous TD work \cite{nagar2007binary, bernuzzi2010binary}, motivating the adaptation of hyperboloidal methods within the radiation-reaction community.

\subsubsection{Difficulty~3 - Time integration numerical evolution schemes}\label{introC-D3}

Finally, we discuss \textit{difficulty~3} with regards to time evolution methods. There are two main classes of integration methods: explicit and implicit methods. Explicit methods depend on the Courant-Friedrichs-Lewy condition, i.e., a limitation on the time discretisation with respect to the spatial grid which ensures the numerical algorithm's velocity does not surpass the physical velocity. Besides, these methods tend to violate energy conservation by not preserving an underlying symplectic structure. Examples of such widely used methods are explicit Runge-Kutta methods, with the most common version, RK4,  ensuring fourth-order convergence. Implicit methods such as geometric integrators do not violate energy conservation nor symplectic structure nor are constrained by any time steps. Historically, all of the aforementioned numerical strategies have performed time evolution by using a RK4 time integration scheme. Ref.~\cite{sopuerta2006finite} is the only exception as it employs a  Bulirsch-Stoer extrapolation evolution method. Ideally, any suitable time-integrator for EMRI modelling should allow for long, stable and accurate time evolution.

\subsubsection{Difficulty~4 - Initial data choices}

We consider these to be the three main difficulties associated with designing a core numerical evolution algorithm in the time domain. However, even if the algorithms address all of these difficulties we also expect there to be two further potential complications stemming from the choice of initial data. The initial value problem (IVP) has been attempted before at the light of solving the RWZ equations. The first attempt at understanding the problem was made by \textit{Lousto \& Price} \cite{lousto1997head,lousto1997understanding} who derived exact ID from an initially conformally flat three-metric for the case where a particle is plunging into the SMBH. Later, these works were generalised by \textit{Martel \& Poisson} as a one-parameter family of time-symmetric ID, where the parameter measures the GW content of the initial surface. With this model, they were able to show the choice of ID strongly influenced the waveforms produced and the total energy radiated \cite{martel2002one}. These initial data was then used in \cite{barack2002computing}, for the first, and only to date, TD GSF computation in the RW gauge. A key observation concerning ID by their computation was the presence of a spurious effect at the onset of the plunge which was resolved by waiting a sufficiently long time allowing for the extraction of the GSF to good analytical agreement. Recently, \textit{O'Toole \& Wardell} \cite{o2021characteristic, o2022green} have derived a set of ID to calculate the Green function of the RWZ equations by taking a characteristic initial value approach. Relatively, to their previous work \cite{wardell2014self}, where a smooth narrow Gaussian was used as ID, the number of modes required to compute the self-force \textit{halved}, without significant accuracy loss further motivating future ID investigations.

In spite of these attempts most TD work for a point-particle on a circular orbit \cite{martel2004gravitational, sopuerta2006finite, field2009discontinuous, bernuzzi2010binary} has been treated as an ill-posed IVP where smooth trivial initial data is chosen, i.e the field and it's time derivative are initially \textit{zero}, given rise to spurious, \textit{junk} radiation, which dominates at earlier times. Two main concerns arise \cite{field2009discontinuous}: 
\begin{itemize}
    \item \textit{Concern I} -  TD simulations must ensure that simulations have been evolved for a sufficiently long time such that the \textit{junk} radiation dominating the simulation initially is negligible. TD codes must thus have reasonable metrics to gauge when the simulating has reached \textit{steady-state}. 
    
    \item \textit{Concern II} - Potential presence and/or effects of \textit{persistent junk} radiation - Given the distributional nature of our source, choosing smooth initial data, such as \textit{zero}, results in an ill-posed problem. Understanding how/if this affects the numerical physical results extracted at longer times is paramount to ensure TD performs accurately.  
\end{itemize}
In previous work, \cite{field2009discontinuous, bernuzzi2010binary} found necessary, due to \textit{concern's I and II}, to slowly turn-on the source through the means of an error function acting on the distributional source terms. \textit{Field~et~al} \cite{field2010persistent} observed that a \textit{Jost} solution would appear i.e a \textit{persistent junk solution}, with it's contamination extent depending greatly on how the distributional forcing term is treated and not the numerical method. Shortly after, \textit{Jaramillo~et~al} \cite{jaramilloSopCan2011time} argued that no \textit{Jost} solutions would be observed provided the numerical method is capable of correctly implementing the $\delta$ distributional source terms at late times. They justify the presence of \textit{Jost} solutions in \cite{field2010persistent} as a consequence of them modifying the distributional source terms to make them compatible with trivial ID choices, i.e by means of an error function that slowly turns them on.\\

\subsection{This work}\label{introD}

With all these difficulties in mind, we introduce a new method to address all concerns. To benchmark the algorithm we follow refs.~\cite{martel2004gravitational, lousto2005time, sopuerta2006finite, field2009discontinuous, bernuzzi2011binary} and considered the problem of a particle on a circular orbit on a Schwarzschild background, with the perturbation equations represented in the Regge-Wheeler gauge. We crosscheck our time domain results against those obtained in the frequency-domain at \textit{first-order} by ref.~\cite{thompson1811}.

In our novel algorithm we incorporate three main numerical methods to handle the aforementioned difficulties. To handle \textit{difficulty 2} we will use an alternative to both radiation boundary conditions and hyperboloidal layers and instead use a hyperboloidal chart known as the \textit{minimal gauge}, introduced by \cite{schinkel2014initial, ansorg_rpm_1604.02261, jaramillo_rpm_2004} which automatically ensures outflow behaviour at the boundaries and direct access to relevant physical quantities at both the horizon $\mathcal{H}$ and future null infinity $\mathscr{I}^+$. 
To address \textit{difficulty 3} and ensure we can compute long-term evolutions, we will apply an implicit geometric integrator in an explicit form as demonstrated in \cite{o2022conservative}. Unlike predicted in the literature \cite{teukolsky2000stability}, we can use implicit-turned-explicit schemes to ensure not just energy conservation to higher order accuracy, but also recover symplectic structure \cite{lidiaICN, o2022conservative}. \textit{Difficulty 1}, will be handled by using discontinuous collocation methods to represent the Dirac-$\delta$ distribution. These were initially introduced by \cite{2014arXiv1406.4865M} and further refined to the hyperboloidal wave equation and generic cases by \cite{24thCapraTalk, 25thCapraTalk, reviews-lidia, phdthesis-lidia}. As we know the local structure of the discontinuity at the point-particle \textit{a priori}, we can adapt the Lagrange interpolation scheme by adding these known amplitudes, such that it holds for the case where the problem is smooth everywhere except at the point-particle's location. 

Finally we address \textit{difficulty 4}. Most TD RWZ circular orbits results have been obtained with algorithms using trivial initial data. In this work, in line with the results of \textit{Jaramillo~et~al} \cite{jaramilloSopCan2011time}, we too opt for trivial ID when addressing \textit{difficulty~4}. By design our algorithm adequately incorporates the jump conditions at all times in the evolution system, ensuring the correct initial value is passed on. We expect by addressing \textit{concern I}, to solve for \textit{difficulty 4} altogether, expecting the effects of spurious radiation to be minimal/if present at all at late times. Furthermore, in other gauges, namely in the \textit{Lorenz gauge} the GSF has been computed in the TD for a particle on a circular geodesic in Schwarzschild by \textit{Barack \& Sago} \cite{barack2007gravitational} through the use of trivial ID and monitoring of \textit{concern I} by studying the time evolution of the metric perturbations evaluated at the particle location at different times.  We further note by opting for trivial ID, we also ensure a more faithful comparison with \cite{barack2007gravitational}, who seemingly seem to further validate \cite{jaramilloSopCan2011time} findings. We found though important to highlight the problem of initial data should not be overlooked. A good example is the work carried out by \textit{Dolan \& Barack}, \cite{dolan2013self}, who were able to compute the GSF in Schwarzschild in the TD through an effective source approach in the \textit{Lorenz gauge}. For radiative modes, they too found trivial ID to suffice, however, for the $m=\{0,1\}$ modes a linear instability growing in time was observed. To remedy this, for $m=0$, they were able to derive analytically exact initial data, effectively taming the instability. However, for $m=1$ this was not possible, so they mitigated the effects of the growing instability by applying a frequency filter. For  $m=\{0,1\}$ imposing ID was \textit{indispensable}. Remarkably, in their recent work, \cite{dolan2023metric}, their approach was validated through comparison against their recent novel FD framework computing metric perturbations in Kerr (see their Fig.~12) allowing for the error to be quantified.

This is the first of a series of upcoming papers studying circular geodesic motion of a point-particle. Here we will show competitive accuracy to frequency domain methods with a simple yet elegant proof-of-concept implementation in Mathematica, and we will then show a preliminary evaluation of the fields and it's derivatives. In an upcoming paper \cite{paper2} we will show how we can compute the scalar self-force surpassing all the aforementioned difficulties with not one but two numerical methods in the time-domain, the second method following the recipe of \cite{rpm_scalar_fd2202}. Lastly, in future work \cite{paper3}, we will aim to finalise the computation for the fully regularised GSF corroborating the results of \cite{thompson1611, thompson1811}.

This paper is organised as follows. In Section \ref{sectionii} we revisit the necessary theoretical background to understand the problem of modelling gravitational perturbations induced by the point-particle on a Schwarzschild background prescribing circular geodesic motion. This section is further complemented by Appendix \ref{appA}. In Section \ref{sectioniii} we introduce our numerical framework to solve this problem, complemented by Appendix \ref{AppB_BHPT_RWG} and \ref{AppC}. In Section \ref{sectioniv} we introduce our results substantiating them with Appendix \ref{AppD_complement}. Finally in Section \ref{conclusion} we summarise our results and motivate our upcoming work.

\begin{table*}
\begin{ruledtabular}
\begin{tabular}{l l l l l}
\textrm{Case}&
\textrm{Author(s)}&
\textrm{Domain }&
\textrm{Gauge}&
\textrm{Strategy}\\
\colrule
Newtonian Potential:& Pfenning \& Poisson \cite{pfenning2002scalar} &  &  & direct, analytic\\
\colrule
Schwarzschild:  & Barack \& Lousto \cite{barack2002computing} & Time  & Regge-Wheeler & mode sum, numerical,\\
radial geodesics  &  &  &    & ($1+1D$ evolution)\\
\colrule
Schwarzschild:  & Keidl \textit{et al} \cite{keidl2007finding} &   & Radiation & analytic,\\
static geodesics  &  &  &    & regularisation\\
\colrule
Schwarzschild:  & Barack \& Sago \cite{barack2007gravitational} & Time & Lorenz  & mode-sum, ($1+1D$ evolution)\\
circular geodesics  & Dolan \& Barack \cite{dolan2013self} &   &   &  effective-source, ($2+1D$ evolution) \\
\rowcolor{lightgray} & This work &  & Regge-Wheeler & mode sum \& flux balance, \\
\rowcolor{lightgray}& & & &($1+1D$ evolution) \\ \cline{2-5} 
& Ak\c{c}ay \cite{akcay2011fast} &  Frequency  & Lorenz & mode sum, numerical \\
& Merlin \& Shah \cite{merlin2015self}&    &   &     \\ 
& Berndtson \cite{berndtson2007harmonic} & & & \\
& Durkan \cite{durkan2022metricTHESIS} & & & \\
& Keidl \textit{et al} \cite{keidl2010gravitational} &  & Radiation & \\
& Shah\textit{et al} \cite{shah2011conservative} &  &  & \\
& Detweiler \cite{detweiler2008consequence} &   & Regge-Wheeler &  \\
& Berndtson \cite{berndtson2007harmonic} & & &\\
& Thompson \textit{et al} \cite{thompson1611, thompson1811} &  &  & \\ 
& Durkan \cite{durkan2022metricTHESIS} & & & \\
with spinning secondary & Mathews \textit{et al} \cite{mathews2022self} &  &  & \\
& &  &  & \\
\colrule
Schwarzschild:  & Barack \& Sago \cite{barack2009gravitationalEccentric, barack2010gravitational} & Time & Lorenz  & mode-sum,\\
eccentric geodesics  &  &   &   &  ($1+1D$ evolution) \\\cline{2-5} 
& Ak\c{c}ay \textit{et al} \cite{akcay2013frequency} & Frequency  & Lorenz & mode-sum, numerical \\
& Osburn \textit{et al} \cite{osburn2014lorenz} & & & \\
osculating & Warburton \textit{et al} \cite{warburton2012evolution}  &   &   & \\
 & Osburn \textit{et al} \cite{osburn2016highly}  &   &   & \\
\colrule
Kerr:  & Shah \textit{et al} \cite{shah2012extreme} & Frequency & Radiation & mode-sum, numerical\\
circular geodesics  &  &   &   &   \\
osculating geodesics  & Lynch \textit{et al}  \cite{lynch2023self} &   &   &   \\
& Isoyama \textit{et al} \cite{isoyama2014gravitational} & Frequency, & Radiation, & effective, numerical \\ 
 &  &  Time  & Lorenz &  
\\\cline{2-5} 
\colrule
Kerr:  & van de Meent \& Shah \cite{van2015metric} & Frequency &  Radiation  & mode-sum, numerical\\
eccentric geodesics & & & \\
  &  &   &   &  \\
equatorial  & van de Meent \cite{van2015metric, van2016gravitational}  &   &    &   \\
generic & van de Meent \cite{van2018gravitational} &   &  &  \\ 
NIT\footnote{near-identity (averaging) transformations} \& osculating & Lynch \textit{et al} \cite{lynch2022eccentric} &   &   & \\
\end{tabular}
\caption{Summary table of the previous work on computing the \textit{first-order} gravitational self-force by the radiation reaction community. This tables builds up from Table 3 and Table 1 of, \cite{barackREVIEW2009gravitational, wardellREVIEW2015self}, respectively, though strictly restricted to works where an explicit GSF result is numerically given.}
\label{ch1_table_previouswork_graviational}
\end{ruledtabular}
\end{table*}

The conventions that we follow throughout this work are: Greek indices denote $4D$ spacetime indices; lower-case Latin letters are used for indices in the 2 dimensional \((t,r)\) Lorentzian manifold $\mathcal{M}^{2}$; and capital Latin letters are used to refer to the $2D$ spherical space $\mathcal{S}^{2}$. We use physical units in which $G = c = 1$ and metric signature $(-,+,+,+)$.

\section{Gravitational Perturbations on Schwarzschild in the Time Domain}\label{sectionii}
In this section we briefly review the perturbation formalism we adapt to model the point-particle behaviour on a circular geodesic around the non-rotating supermassive BH. We show how this model yields a simple $1+1D$ \textit{wave-like} inhomogeneous PDE problem which admits a \textit{weak-form}.  

\subsection{Point-particle on circular geodesic motion around a non-rotating supermassive black hole}\label{sectioniia}
We can express the Schwarzschild spacetime as a 4-dimensional manifold $\mathcal{M}$ with coordinates $(t,r,\theta,\phi)$ and metric: 
\begin{eqnarray}
    ds^{2} &=& \bar{g}_{ab}dx^{a}dx^{b} +  r^{2}\bar{\omega}_{AB}d\theta^{A}d\theta^{B}. \nn \\
   & =& -f(r) dt^2 + \dfrac{dr^2}{f(r)} + r^2 \bigg( d\theta^2 + \sin\theta^2 d\phi^2\bigg),
    \label{cha2_2d_2d_split}
\end{eqnarray}
with $f(r) = 1- 2M/r$.

The standard approach in numerical BHPT is to split this spacetime into two sub manifolds, $\mathcal{M}= \mathcal{M}^{2}\times \mathcal{S}^{2}$: a 2D Lorentzian manifold $\mathcal{M}^{2}$ with coordinates $x^{a} = \{t,r\}$, covariant under two-dimensional coordinate transformations $x^{a} \rightarrow x'^{a}$, and a 2D spherical space, $\mathcal{S}^{2}$, with coordinates $\theta^{A} =\{\theta,\phi\}$. We further note the $2D$ tensor $\bar{g}_{ab}$ and the scalar coordinate $r$  are functions of the coordinates $x^{a}$ and $\bar{\omega}_{AB} = diag(1,\sin^{2}\theta)$.

Following our ansatz solution in Eq.~\eqref{cha1_g_perturbed}, we expand the Einstein tensor $G_{\alpha\beta}$ to linear order in $h_{\alpha\beta}$, such that $G_{\alpha\beta} = \bar{G}_{\alpha\beta} - \delta G_{\alpha\beta}/2 + O(\epsilon^{2})$.
We know that the Schwarzschild metric is a solution to the Einstein vacuum equations, $\bar{G}_{\alpha\beta}= 0$, thereby to \textit{first-order} in the mass-ratio $\epsilon$ it suffices to write
\begin{equation}
    \delta G_{\alpha\beta}= -16\pi T_{\alpha\beta}.
    \label{cha2_einstein_tensor}
\end{equation}
Considering the case where we have some external matter represented by the stress-energy tensor $T_{\alpha\beta}$, written in terms of Dirac-$\delta$ distributions which reflect the point-particle model of our compact object of mass $\mu$ following a geodesic given as $z^{\mu}(\uptau) = \{t_{p}(\uptau), r_{p}(\uptau), \theta_{p}(\uptau), \phi_{p}(\uptau)\}$, we have
\begin{eqnarray}
         T_{\alpha\beta} = \mu \int^{\infty}_{\infty} \frac{u_{\alpha}u{\beta} }{\sqrt{-g}}\delta^{4}(x^{\mu} - z^{\mu}(\uptau)) \ d\uptau  \ \ \ \ \ \ \ \ \ \ \ \ \ \ \ \ \ \ \  \ \   \nonumber \\ 
  = \mu \frac{u_{\alpha}u_{\beta}}{u^{t}  r_{p}(t)^{2} } \delta(r - r_{p}(t))\delta(\theta - \theta_{p}(t))\delta(\phi - \phi_{p}(t)), \ \ 
    \label{cha2_sem_pointparticle}
\end{eqnarray}
where $\uptau$ is the particle's proper time.
We benchmark our numerical framework with the motion of the particle along a circular equatorial geodesic in the Schwarzschild spacetime, where $z^{\mu}(\uptau) = \{t_{p}(\uptau), r_{p}(\uptau)=r_{p}, \theta_{p}(\uptau) = \frac{\pi}{2}, \phi_{p}(\uptau)\}$. One can parameterise the particle's four-trajectory $z^{\mu}(\uptau)$ in terms of its specific energy and angular momentum $ \{\mathcal{E},\mathcal{L}\}$ given as 
\begin{equation}
    \mathcal{E} =\frac{f}{\sqrt{1 - 3M/r}} , \ \ \ \mathcal{L} = \frac{\sqrt{r M}}{1 - 3Mr}.
    \label{cha2_physicalenergymomentum}
\end{equation}
The four-velocity $u^{\mu} = dz^{\mu}/ ds$ has then components 
\begin{equation}
    u^{t}= \frac{\mathcal{E}}{f}, \ (u^{r})^{2} = \mathcal{E}^{2} - U^{2}, \ \ u^{\theta} = 0 , \ \ u^{\phi} = \frac{\mathcal{L}}{r^{2}},
    \label{cha2_fourvelocity_CO}
\end{equation}
where $U^{2} = f(r)(1 + \mathcal{L}^{2}/r^{2})$. 
For circular orbits we can further show the orbit's \textit{physical} frequency satisfies Kepler's law given by
\begin{equation}
    \bigg(\frac{d \phi}{ dt} \bigg)^{2}= \Omega^{2} = \frac{M}{r_p^{3}}.
    \label{cha2_physical_orbitalfrequency}
\end{equation}

\subsection{The Regge-Wheeler-Zerilli Formalism}\label{sectioniib}
We first test our algorithm by solving for two \textit{scalar} master functions obtained by Regge-Wheeler \cite{regge1957stability} and Zerilli \cite{zerilli1970gravitational}, RWZ, in terms of the metric perturbations for the polar and axial cases. We follow the gauge invariant approach of \cite{thompson1611, thompson1811} with the axial or odd-parity, ``$a$'', perturbative treatment given by the Regge-Wheeler master function \cite{regge1957stability} and the polar or even-parity, ``$p$'' given by the Zerilli-Moncrief master functions \cite{moncrief1974gravitational}. Each of these master functions satisfies a \(1+1D\) hyperbolic wave equation
\begin{equation}
\left[ -\partial_{t}^{2}  + \partial_{r^{*}}^{2}  - V_{l}^{a/p}(r) \right] \Psi(t,r)^{a/p}_{lm} = S_{lm}^{a/p}(t,r),
   \label{cha2_generalwave}
\end{equation}
with $r^{*} = r + 2M\log(r/2M - 1)$ the tortoise coordinate. 
Furthermore the potential $V_{l}^{a/p}(r)$ refers to the axial and polar potentials respectively, given as, 
\begin{eqnarray}
       V_{l}^{a}(r)& = &\frac{f}{r^{2}}\bigg[ (\lambda +2) - \frac{6M}{r} \bigg],  \\
       \label{cha2_cpm_potential}
        V_{l}^{p}(r) &= &\frac{2 f}{r^{2}} \bigg[
        \frac{\lambda^{2}(\lambda +2)r^{3} + 6M (\kappa \lambda r + 12M^{2}) }{r\kappa^{2}}
        \bigg], 
\end{eqnarray}
where $\lambda = (l+2)(l-1)/2$ and $\kappa = 6 M + \lambda r$.\\ 

The source $S_{lm}^{a/p}(t,r)$ is usually given as, 
\begin{eqnarray}
   \nonumber S_{lm}^{a/p}(t,r) &=& G^{a/p}_{lm}(t,r)\delta\left(r-r_{p}(t)\right) \\
    &&+ F^{a/p}_{lm}(t,r)\delta'(r-r_{p}(t)). 
    \label{cha2_sourceterm}
\end{eqnarray} 
We can further expand this with respect to the particle worldline by using selection property Eq.~\eqref{appendix_a_dirac_selectionII}, 

\begin{eqnarray}
       S_{lm}^{a/p}(t,r_{p}(t)) =  F^{a/p}_{lm}(t,r_{p}(t))\delta'(r-r_{p}(t)) \nonumber\\
      + \bigg[G^{a/p}_{lm}(t,r) - \partial_{r} F_{lm}^{a/p}(t,r))\bigg]\bigg|_{r=r_{p}} \delta(r-r_{p}(t)). 
      \label{cha2_sourceterm_particle}
\end{eqnarray}
For simplicity, and following \cite{hopper2010gravitational} we re-define the source term functions given in Eq.~\eqref{cha2_sourceterm} as,

\begin{eqnarray}
       \bar{F}_{lm}^{a/p}(t) &=& F^{a/p}_{lm}(t,r_{p}(t)), \\
       \bar{G}_{lm}^{a/p}(t) &=& \bigg[ G^{a/p}_{lm}(t,r) - \partial_{r} F^{a/p}_{lm}(t,r) \bigg]\bigg|_{r=r_{p}}. 
       \label{cha2_sourceterm_wrt_partworldline}
\end{eqnarray}

\subsection{The \textit{weak-form solution} to the inhomogenous RWZ master functions}\label{sectioniic}
Given the presence of distributional term $\delta(r-r_{p})$ and its radial derivative we need to extend the homogeneous solutions for the master functions $\Psi^{a/p}$ to a \textit{weak-form solution} \cite{hopper2010gravitational} \footnote{We further clarify here, by \textit{weak-form solution} we refer to a general weak solution to the PDE, which is smooth but may have a set of measure zero non-differentiable or limited differentiable points.} as they approach the particle position, $r_{p}$, from the left $(-)$ and right $(+)$ limits,   
\begin{equation} 
    \Psi_{lm}(t,r)=  \Psi^{+}_{lm}(t,r) \Theta[r-r_{p}(t)] + \Psi^{-}_{lm}(t,r)\Theta[r_{p}(t)-r], 
    \label{ch3_weakformsolutionRWZ}
\end{equation}
where $\Theta(z)$ is the Heaviside function defined via 

\begin{equation}\Theta(z)= \left\{ 
\begin{array}{ccc}
1 &\text{for}& z>0, \\
\frac{1}{2}&\text{for}& z=0, \\
0 &\text{for}& z<0.  
\end{array}
\right. 
\label{ch3_HeavisideStepFunction}
\end{equation}

The functions $\Psi_{lm}^{\pm}(t,r)$ satisfy the homogeneous RWZ equations and the jump of the retarded field (as well as its derivatives) at the particle's trajectory is defined via, 
\begin{eqnarray}
\label{ch3_j0} 
   [[\Psi]](t) &=& \lim_{\epsilon \rightarrow 0^{+}}\big[\Psi(t, r_{p}(t) + \epsilon)  - \Psi(t, r_{p}(t) - \epsilon)\big], \nonumber \\ 
  &=& \frac{\mathcal{E}^{2}}{f^{2}_{p} U^{2}_{p}} \bar{F}^{a,p}_{lm}(t),\\  
  {}[[\Psi_{r}]](t) &=&  \frac{E^{2}}{f_{p}^{2} U^{2}_{p}} \bigg[ \bar{G}^{a,p}_{lm}(t) \nonumber \\
&&+ \frac{1}{U^{2}_{p} r^{2}_{p}} \bigg( 3M - \frac{\mathcal{L}^{2}}{r_{p}}+ \frac{5 M \mathcal{L}^{2}}{r^{2}_{p}}  \bigg) \bar{F}^{a,p}_{lm}(t) \nonumber \\
&&- 2 \dot{r}_{p} \frac{d}{dt} \big( [[\Psi]](t) \big) \bigg];\label{ch3_j1}\\
  {}\label{ch3_jt}[[\Psi_{t}]](t) &=& \partial_{t}[[\Psi]](t) - \dot{r}_{p}[[\Psi_{r}]](t);\\
     {}\label{ch3_jx}[[ \Psi_{r^{*}}]](t) &=& f_{p} [[ \Psi_{r}]](t). 
\end{eqnarray}

Here, we follow the jump formalism of ref.~\cite{hopper2010gravitational} to facilitate future work and comparisons and indicate a derivative with a subscript. We further include in Appendices \ref{appA} and \ref{AppB_BHPT_RWG} a full derivation of the jump conditions and the explicit form of terms $\bar{F}^{a/p}_{lm}(t)$ and $\bar{G}^{a/p}_{lm}(t)$, respectively.

\section{Hyperboloidal Time-Symmetric Discontinuous Collocation Numerical Algorithm with Higher Order Jumps }\label{sectioniii}
Here we describe our novel numerical algorithm, addressing all three main difficulties reviewed in Section \ref{introduction}. Numerically, we tackle this problem through the method of lines (MoL) framework, which usually prescribes four essential stages: picking boundary and initial conditions, followed by reducing the system to first-order with spatial discretisation and then integrating in time. 
Our $1 +1D$ wave-like inhomogeneous equation is a hyperbolic PDE of the type 
\begin{equation}
    \partial_{t} \textbf{U} = L \ \textbf{U} + \mathcal{S}, \ \ \ \ \textbf{U} = \begin{pmatrix}\Psi\\\Pi, 
    \end{pmatrix}
    \label{ch3_general_pdes}
\end{equation}
where $L$ is a spatial differential operator and $\Psi^{a/p}_{lm}(t,r)$, $\Pi^{a/p}_{lm}(t,r)$ are the RWZ master functions and their partial derivatives with respect to time, respectively, given explicitly in Eq.~\eqref{cha2_generalwave}. 

We then show how we build hyperboloidal discontinuous spatial and temporal discretisations by building upon well-established Lagrange interpolation methods. Our work takes full advantage of the fact that we know all the information associated with the distributional part of the problem represented by the coefficients on the RHS of Eq.~\eqref{cha2_generalwave}, here represented by $\mathcal{S}$. 

\subsection{Boundary conditions  - a hyperboloidal slicing approach}\label{sectioniiiA}

As reviewed in Section \ref{introduction}, \textit{difficulty 2}, associated with the outer radiation problem, can be avoided altogether with the use of hyperboloidal methods. We choose to fully map our problem into a new set of coordinate transformations through applying a hyperboloidal slice, such that $t \rightarrow t(\tau,\sigma), \ \sigma \rightarrow r(\sigma)$ \cite{zenginouglu2009gravitational}. We employ the so-called ``scri-fixing technique"~\cite{Zenginoglu:2007jw} to construct the hyperboloidal slices. Specifically, the coordinate transformations are given by, 
\begin{equation}
    t = \lambda\left(\tau - H(\sigma)\right), \ \ r = \frac{2M}{\sigma}, 
    \label{cha3_hyper_transf}
\end{equation}
where $\lambda$ is a length scale of the spacetime, conveniently fixed here to $\lambda=4M$. We follow the the strategy in \cite{ansorg_rpm_1604.02261,PanossoMacedo:2018hab, PanossoMacedo:2019npm, jaramillo_rpm_2004}, which fixes the height function to,  
\begin{equation}
    H(\sigma) = \frac{1}{2}\bigg[ \ln(1-\sigma) - \frac{1}{\sigma} + \ln(\sigma) \bigg].
    \label{cha3_heightfunction}
\end{equation}
In these coordinates, future null infinity is located at $\sigma=0$, whereas the black hole horizon is at $\sigma = 1$.

To obtain the inhomogeneous wave equation in hyperboloidal coordinates, we begin by abusing the notation and identifying $\Psi^{a/p}_{lm}(\tau, \sigma)$ directly with $\Psi^{a/p}_{lm}(t(\tau,\sigma), r(\sigma))$. Then, the coordinate transformation in Eq.~\eqref{cha3_hyper_transf} implies
\beq
\label{eq:hyp_chainrules}
\partial_t = \lambda^{-1} \partial_\tau, \quad 
\partial_r =- \dfrac{\sigma^2}{2M \, } \bigg( \partial_\sigma + H'(\sigma)\partial_\tau \bigg).
\eeq
Applying these chain rules to the wave equation \eqref{cha2_generalwave} yields
\begin{eqnarray}
   \big[ \square - V_{l}(\sigma) \big]\Psi^{a/p}_{lm}(\tau, \sigma) &=& {\cal S}^{a/p}_{\ell m}(\tau, \sigma)
   \label{ch3_rwz_fieldvariables_hyper}
\end{eqnarray}
with 
\begin{eqnarray}
\square \Psi =
\bigg( -\Gamma(\sigma) \partial_{\tau}^{2}\Psi + \upvarepsilon(\sigma)\partial_{\sigma}  \partial_{\tau}\Psi + \upvarrho(\sigma) \partial_{\tau}\Psi\nonumber \\
+ \upchi(\sigma)\partial^{2}_{\sigma}\Psi + \iota(\sigma)\partial_{\sigma}\Psi \bigg). \ \ \ 
\label{cha3_hyper_transformed_operator}
\end{eqnarray}
The coefficients on the above expression read
\begin{eqnarray}
\Gamma(\sigma) &=& \frac{\sigma^{2}}{4M^{2}}(1+\sigma),\\
\label{ch3_sb2_diffOperators_gamma}
\upvarepsilon(\sigma) &=& \frac{\sigma^{2}}{4M^{2}} (1-2\sigma^{2}), \\
\label{ch3_sb2_diffOperators_alpha}
 \upvarrho(\sigma) &=& \frac{\sigma^{2}}{4M^{2}} (2\sigma),  \\
\label{ch3_sb2_diffOperators_beta}
\upchi(\sigma) &=& \frac{\sigma^{2}}{4M^{2}} \sigma^{2}(1-\sigma),\\
\label{ch3_sb2_diffOperators_chi}
\iota(\sigma) &=& \frac{\sigma^{2}}{4M^{2}} \sigma(2-3\sigma),
\label{ch3_sb2_diffOperators_iota}
\end{eqnarray}
where we observe two features: (i) an overall factor $\sigma^2/(4M^2)$ in all functions; and (ii) the vanishing of the function $\upchi(\sigma)$ at the boundaries $\sigma=0$ and $\sigma=1$. The latter behaviour is a practical way of identifying the desired outflow behaviour at future null infinity and the horizon. Indeed, boundary conditions in this setup arise directly from the regularity conditions that follow by imposing the wave equation directly at $\sigma=0$ and $\sigma=1$.

We also write $\tilde{\mathcal{S}}^{a/p}(\tau)$ as a time-dependent only function to conserve the convention established in Eq.~\eqref{cha2_sourceterm_wrt_partworldline} which ensures our numerical algorithm is with respect to the particle worldline. For the moment this is placeholder notation and we will show in the next section how the discontinuous nature of our problem will be correctly incorporated into the algorithm.  
Furthermore we reduce our problem to a set of coupled ODEs as given in Eq.~\eqref{ch3_general_pdes}, simplifying Eq.~\eqref{cha3_hyper_transformed_operator} to the following system of equations, 
\begin{equation}
    \partial_{\tau} \textbf{U} = L \ \textbf{U} + \mathcal{S}, 
    \label{ch3_generalhyperPDE}
\end{equation}
where $\textbf{U}$ is now defined in hyperboloidal coordinates $(\tau, \sigma)$ and $L$ is given as, 
\begin{eqnarray}
    L = \begin{pmatrix} 0 & {\mathbb 1} \\ \textbf{L}_{1} & \textbf{L}_{2} \end{pmatrix},
    \label{hyper_l_op}
\end{eqnarray}
with the operators
\beq
\textbf{L}_{1} = \dfrac{1}{\Gamma(\sigma)}\bigg({\upchi}(\sigma)\partial^{2}_{\sigma} + {\iota} (\sigma) \partial_{\sigma} -  {V}_{l}(\sigma) \bigg)
\label{l1_hyper}
\eeq
and
\beq
\textbf{L}_{2} = \dfrac{1}{\Gamma(\sigma)}\bigg({\upvarepsilon}(\sigma) \partial_{\sigma} - {\upvarrho}(\sigma) \bigg).
\label{l2_hyper}
\eeq
The source vector $\mathcal{S}$ in Eq.~\eqref{ch3_general_pdes} reads
\beq
{\cal S} = \begin{pmatrix} 0 \\{\cal S}^{a/p}_{\ell m}(\tau)/\Gamma(\sigma) \end{pmatrix}. \label{prelim_source}
\eeq
For convenience we introduce the tilde notation to denote division of the coefficients in the operator by $\Gamma(\sigma)$, \textit{e.g.}, $ \tilde{\upvarepsilon}(\sigma) = \upvarepsilon(\sigma)/ \Gamma(\sigma)$. In what follows from the rest of the numerical method, we will work entirely on the Lorentzian manifold, $\mathcal{M}^{2}$, parameterised by the hyperboloidal coordinates $x^{a} = (\tau, \sigma)$, with  $\sigma_{p}=2M/r_p$ referring to the particle position. To extract the wanted physical quantities inherent to the $(t,r,\theta,\phi)$ coordinates, we revert back to the original system by careful application of high-order chain rules. 

\subsection{Discontinuous Spatial Discretisation }\label{sectioniiiB}
In this section we address \textit{difficulty 1} on how we will represent the Dirac-$\delta$ distribution. Before we introduce this machinery,  we need to address how we will discretise in space the differential operators present in Eq.~\eqref{hyper_l_op}. We will then correct these operations such that the distributional nature of the problem is fully accounted for. 

\subsubsection{Spatial discretisation through Lagrangian interpolation}

We can discretise our field, given generically in Eq.~\eqref{ch3_general_pdes}, and further specified in hyperboloidal coordinates in Eqs.~\eqref{ch3_generalhyperPDE}-\eqref{hyper_l_op}, $\textbf{U}(\tau,\sigma)$, in space such that $\textbf{U}(\tau,\sigma) \rightarrow \textbf{U}(\tau, \sigma_{i}) := \textbf{U}_{i}(\tau)  = \textbf{U}(\tau)$ where $0 \leq \sigma_{i} \leq 1  $ with the collocation nodes ranging from $0<i<N$. Essentially, we discretise in space by building the collocation polynomial of degree $N$, 
\begin{equation}
    p(\sigma) = \sum^{N}_{j=0} c_{j} \sigma^{j}, 
    \label{smooth_interpol_gen}
\end{equation}
determined by solving the linear algebraic system of conditions specifically given as, 
\begin{equation}
    p(\sigma_{i}) = U_{i}, 
    \label{smooth_interpol_gen_conditions}
\end{equation}
for the coefficients $c_{j}$. 
Rewriting it in Lagrangian formulae we have the Lagrange interpolating polynomial (LIP), 
\begin{equation}
    p(\sigma) = \sum^{N}_{j=0} U_{j} \pi_{j}(\sigma)
    \label{smooth_interpol_LIP}
\end{equation}
where $\pi_{j}(\sigma)$ is the Lagrange basis polynomial (LBP) given as, 
\begin{equation}
    \pi_{j}(\sigma) = \prod^{N}_{\substack{k=0,\\ k\neq j}} \frac{\sigma - \sigma_{k}}{\sigma_{j} - \sigma_{k}}.
     \label{ch3_sb2_LBP}
\end{equation}
We can then, by acting on the LIP as given in Eq.~\eqref{smooth_interpol_LIP}, differentiate (or integrate) our field $U$ any \textit{n-th} times, explicitly, 

\begin{equation}
    U^{(n)}(\sigma_{i}) \approx p^{(n)}(\sigma_{i}) = \sum^{N}_{j=0} D^{(n)}_{ij} U_{j},  
    \label{spatial_disc}
\end{equation}
where, 
\begin{equation}
    D^{(n)}_{ij} = \pi^{(n)}_{j} (\sigma_{i}). 
    \label{diffmatrix_gen}
\end{equation}
We will now explicitly give the form of spatial differential operators given in Eq.~(\eqref{ch3_generalhyperPDE},\eqref{hyper_l_op}) and later discretised as Eq.~\eqref{spatial_disc} by using both spectral and finite-difference collocation methods \cite{o2022conservative}. 

For the spectral method, the Chebyshev-Gauss-Lobatto collocation nodes are given by, 
\begin{eqnarray}
    \sigma_{i} = \frac{a + b}{2} + \frac{b-a}{2}z_{i}, \nonumber \\ 
     z_{i} = -\cos{\theta_{i}}, \ \ \theta_{i} = \frac{i\pi}{N}, \ i = 0,1,...,N
    \label{chebyshev_lobatto_nodes} 
\end{eqnarray}
yielding, 
\begin{equation}
    {D}^{(1)}_{i j} = 
\frac{2}{b-a}   \begin{cases}
        \frac{c_i (-1)^{i+j}}{c_j (z_i-z_j)} & i \neq j\\
        -\frac{z_j}{2(1-z_j^2)} & i = j \neq 0,N\\
        -\frac{2 N^2 + 1}{6} & i = j = 0\\
        \frac{2 N^2 + 1}{6} & i = j = N
    \end{cases}
    \label{cha3_firstorder_diffmatrix}
\end{equation}
for the first-derivative matrix, and
\begin{small}
\begin{equation}
D_{ij}^{(2)} = {\left( {\frac{2}{{b - a}}} \right)^2}\left\{ {\begin{array}{*{20}{l}}
{\frac{{{{( - 1)}^{i + j}}}}{{{c_j}}}\frac{{z_i^2 + {z_i}{z_j} - 2}}{{(1 - z_i^2){{({z_i} - {z_j})}^2}}}}&{i \ne j,{\; }i \ne 0,N }\\
{\frac{2}{3}\frac{{{{( - 1)}^j}}}{{{c_j}}}\frac{{(2{N^2} + 1)(1 + {z_j}) - 6}}{{{{(1 + {z_j})}^2}}}}&{i \ne j,{\;}i = 0}\\
{\frac{2}{3}\frac{{{{( - 1)}^{j + N}}}}{{{c_j}}}\frac{{(2{N^2} + 1)(1 - {z_j}) - 6}}{{{{(1 - {z_j})}^2}}}}&{i \ne j,{\;}i = N}\\
{ - \frac{{({N^2} - 1)(1 - z_j^2) + 3}}{{3{{(1 - z_j^2)}^2}}}}&{i = j,{\; }i \ne 0,N}\\
{\frac{{{N^4} - 1}}{{15}}}&{i = j = 0{\; \rm{ or }\; }N}
\end{array}} \right.
\label{cha3_secondorder_diffmatrix}
\end{equation}
\end{small}
for the second-derivative matrix. 
We also further note, alternatively we could compute the dot product from Eq.~\eqref{cha3_firstorder_diffmatrix}, such that $D^{(2)}_{ij} = D^{(1)}_{ij} \cdot D^{(1)}_{ij}$. 

For finite-difference methods we use equidistant nodes, 
\begin{equation}
    \sigma_{i} = a + i\frac{b-a}{N}, \ \ \ \ i=0,1,..,N
    \label{ch3_sb2_EquiNodes}
\end{equation}
and resort to the fast library of the Wolfram Language which uses Fornberg's algorithm to compute higher-order derivatives, $D^{(n)}_{ij}$, obtained through the command, 
\begin{verbatim}
Dn = 
NDSolve‘FiniteDifferenceDerivative
[Derivative[n],X,
"DifferenceOrder" -> "4"]
@ "DifferentiationMatrix"//Normal//
Developer`ToPackedArray//SparsedArray.
\end{verbatim}
It is important to note in the minimal gauge the interval $[a,b]$ is defined as $[0,1]$. 

\subsubsection{Discontinuous generalisation of the Lagrange interpolation method by incorporating higher order jumps}
We then construct the discontinuous generalisation to the Lagrange interpolation briefly reviewed above. This was put forward generically in \cite{2014arXiv1406.4865M} and later improved by \cite{24thCapraTalk,25thCapraTalk, phdthesis-lidia}. This method uses higher order jumps 
as input, where here by jump we mean exactly that we can take advantage of the fact we know the location of the particle, $\sigma_{p}$, \textit{a priori}, such that it holds to write the master field RWZ variables a combination of jumps in their fields and derivatives. Hence, 
\begin{equation}
    \Psi^{(m)}(\sigma_{p}^{+}) -  \Psi^{(m)}(\sigma_{p}^{-}) = J_{m}(\tau), \ \ \  m=0,1, ..., \infty
    \label{ch3_b2_higherorderjumps}
\end{equation}
where explicitly, albeit in $(t,r)$ coordinates this have been given for the first $m={0,1}$ orders in Eq.~(\eqref{ch3_j0}- \eqref{ch3_j1}) and will be later specified in hyperboloidal coordinates after it has been adequately incorporated within the Lagrangian method described above. 
Essentially we take our weak-form of the solution to the master RWZ functions as given in Eq.~\eqref{ch3_weakformsolutionRWZ} and rewrite it as a \textit{generic} collocation polynomial,
\begin{equation}
    p(\sigma) =  p_{+}(\sigma) \Theta(\sigma-\sigma_{p}) + p_{-}(\sigma)\Theta(\sigma_{p}-\sigma),  
    \label{ch3_sb2_gen_p_poly}
\end{equation}
where the right/left interpolating polynomials are given respectively as
\begin{equation}
    p_{+}(\sigma) = \sum^{N}_{j=0} c^{+}_{j}(\sigma_{p})^{j}, \ \ \ 
    p_{-}(\sigma) = \sum^{N}_{j=0} c^{-}_{j}(\sigma_{p})^{j}.
    \label{ch3_sb2_gen_IPsLR}
\end{equation}

By solving Eqs.~\eqref{ch3_sb2_gen_IPsLR} as a system of algebraic equations with the collocation conditions given by 
\begin{equation}  \Psi_i=\left\{ 
\begin{array}{ccc}
p_{+}(\sigma_{i}), & \qquad\sigma_{i} > \sigma_{p}, \\
p_{-}(\sigma_{i}), & \qquad\sigma_{i} < \sigma_{p},
\label{ch3_sb2_collcond_gen}
\end{array}
\right. 
\end{equation}
we determine half of the $(2N+2)$ polynomial coefficients, $c^{\pm}_{j}$.
The remaining coefficients are then determined by imposing the jump conditions in Eq.~\eqref{ch3_sb2_gen_p_poly} as,
\begin{equation} 
p_{+}^{(m)}(\sigma_{p}) - p_{-}^{(m)}(\sigma_{p}) = \left\{ 
\begin{array}{ccc}
J_{m}, & m =0,1, ..., M \\
0,  & m = M+1, ..., N 
\end{array}
\right. .
\label{ch3_sb2_coeffFR_JC_gen}
\end{equation}
where here $M$ ranges from $[-1,...,N]$ and the $M=N+1$ jumps are left unspecified until when studying the number of jumps optimal to the algorithm's implementation for the particular physical model here studied. 
We then rewrite everything in the LBP, $\pi_{j}(\sigma)$, as given in Eq.~\eqref{ch3_sb2_LBP} which varies 
depending on whether we choose to work in a spectral Eq.~\eqref{chebyshev_lobatto_nodes} or a finite-difference Eq.~\eqref{ch3_sb2_EquiNodes} collocation methods. It then suffices to solve algebraically the \textit{interpolating piecewise polynomial}, 
\begin{equation}
    p^{\pm}(\sigma) = \sum^{N}_{j=0} C^{\pm}_{j}(\sigma_{p}) \pi_{j}(\sigma).
    \label{ch3_sb2_IPP}
\end{equation}
Specifically we have the algebraic conditions,
\begin{eqnarray}
    C^{+}_{j}(\sigma_{p}) = \Psi_{j} + \Theta(\sigma_{p} - \sigma_{j}) g(\sigma_{j} - \sigma_{p}), \\
    \label{ch3_sb2_ipp_collcright}
    C^{-}_{j}(\sigma_{p}) = \Psi_{j} - \Theta(\sigma_{j} - \sigma_{p}) g(\sigma_{j} - \sigma_{p}), 
    \label{ch3_sb2_ipp_collcleft}
\end{eqnarray}
where, 
\begin{equation}
    g(\sigma_{j} - \sigma) = \sum^{M}_{m=0} \frac{J_{m}}{k!}(\sigma_{j} - \sigma_{p})^{m}
    \label{ch3_sb2_gVector_expSum}
\end{equation}
are the weights computed from the jump conditions derived at the discontinuity $\sigma_{p}$. Finally, substituting Eqs.~(\eqref{ch3_sb2_IPP}-\eqref{ch3_sb2_ipp_collcleft}) into Eq.~\eqref{ch3_sb2_gen_p_poly} we get the \textit{generic} interpolating piecewise polynomial, 
\begin{equation}
    p(\sigma) =  \sum^{N}_{j=0} \bigg[ \Psi_{j} + \Delta_{\Psi}(\sigma_{j} - \sigma_{p}; \sigma - \sigma_{p}) \bigg] \pi_{j}(\sigma)
\end{equation}
where the $\Delta_{\Psi}$ function is given by
\begin{small}
\begin{eqnarray}
    \Delta_{\Psi}(\sigma_{j} - \sigma_{p}; \sigma -\sigma_{p}) = \nonumber \\
    =\bigg[ \Theta(\sigma-\sigma_{p})\Theta(\sigma_{p} -\sigma_{j})\nonumber \\
     -  \Theta(\sigma_{p}-\sigma)\Theta(\sigma_{j} - \sigma_{p})
    \bigg]  g(\sigma_{j} - \sigma_{p}) \nonumber \ \ \ \ \ \ \  \\
    = \bigg[\Theta(\sigma_{i} - \sigma_{p}) - \Theta(\sigma_{j} - \sigma_{p})\bigg]g(\sigma_{j} - \sigma_{p}).  \ \ \ \ \ \ 
    \label{ch3_sb2_delta}    
\end{eqnarray}
\end{small}
\noindent In the end we approximate our master RWZ field variables as given in Eq.~(\ref{ch3_rwz_fieldvariables_hyper}) by
\begin{equation}
    \Psi(\tau,\sigma) \approx \sum^{N}_{j=0} \bigg[ \Psi_{j}(\tau) + \Delta_{\Psi}(\sigma_{j} - \sigma_{p}(\tau); \sigma - \sigma_{p}(\tau))  \bigg] \pi_{j}(\sigma).
    \label{ch3_sb2_WFsolution}
\end{equation}
To be precise, all the differential operators in Eq.~\ref{ch3_rwz_fieldvariables_hyper} and further specified in Eqs.~(\eqref{ch3_sb2_diffOperators_gamma}-\eqref{ch3_sb2_diffOperators_iota}) will be computed through \textit{discontinuous differentiation} as, 
\begin{eqnarray}
  \partial_{\sigma}^{n}(\tau,\sigma)\Psi|_{\sigma = \sigma_{i}}  &=&  p^{(n)}(\sigma) \nonumber  \\
   &=& \sum^{N}_{j=0}  D^{(n)}_{ij}  \Psi_{j} + s_{i}^{(n)}(\tau), \ \ \ \ 
\label{cha2_spatial_disc_generic}
\end{eqnarray}
where $  s^{(n)}_{i}(\tau)$ is given as 
\begin{equation}
    s^{(n)}_{i}(\tau) = \sum^{N}_{j=0}D^{(n)}_{ij} \Delta_{\Psi}\big(\sigma_{j} - \sigma_{p}(\tau); \sigma_{i} - \sigma_{p}(\tau)\big)  
    \label{ch3_sb2_SpaceSpource}
\end{equation}
and the user-specifiable high-order jumps in Eq.~\eqref{ch3_sb2_gVector_expSum} obtained through the computation of the higher order recurrence relation given as,

\begin{widetext}
\begin{eqnarray}
    J_{m+2}(\tau) &=&  -\gamma^{2} \sum^{\infty}_{m=0}  \bigg[ \sum^{m}_{k=0} {m \choose k} \bigg(\upvarepsilon^{(k)}(\sigma_{p}) \dot{J}_{n+1-k} 
+ \upvarrho^{(k)} (\sigma_{p}) (\dot{J}_{m-k} 
- \dot{\sigma}_{p} J_{m+1-k}) \nonumber \\ 
&&+ \iota^{(k)}(\sigma_{p}) J_{m+1-k} - V^{(k)}(\sigma_{p}) J_{m-k}   
+ \ \Gamma^{(k)}(\ddot{J}_{m-k} - 2 \dot{J}_{m+1-k}\dot{\sigma}_{p} - J_{m+1-k}\ddot{\sigma_{p}}  ) \bigg)   \nonumber \\
&&+  \sum^{m}_{k=1} {m \choose k} J_{m+2-k}  \big(\upvarepsilon^{(k)}(\sigma_{p})\dot{\sigma}_{p} + \upchi^{(k)}(\sigma_{p}) \big) \bigg].
\label{cha3_rec_relation_njumps_RWZ}
\end{eqnarray}
\end{widetext}
where here for simplicity the time dependence on the RHS of the jumps has been suppressed and we have, $\gamma^{-2} = \big(  \dot{\sigma}_{p}^{2} \Gamma(\sigma_{p}) -  \dot{\sigma}_{p} \upvarepsilon(\sigma_{p}) - \upchi(\sigma_{p})\big)$. 
Furthermore we note the initialising jumps $J_{0}(\tau), J_{1}(\tau)$, given explicitly in Eqs.~(\eqref{hyper_j0}, \eqref{hyperx_j1}), are obtained through careful implementation of the chain rule and Dirac-$\delta$ distribution composition rules as given in Eqs.~(\eqref{appendix_a_dirac_compI}-\eqref{appendix_a_dirac_compositionII}) acting on the $J_{0}(t), J_{1}(t)$ jumps as given in Eqs.~(\eqref{ch3_j0}-\eqref{ch3_jt}) following the derivation in Appendices \ref{appA_frobs} and \ref{AppC}.  We also note for the derivation of the jumps, as described thoroughly in Eqs.~(\eqref{hyper_partialtau_rho}- \eqref{hyper_partialrhopartialtau_epsilon}) we work with the source term without dividing by the $\Gamma(\sigma)$, as one may be misled by Eq.~\eqref{prelim_source}. This is then accounted for by incorporation of the source terms through Eq.~\eqref{cha2_spatial_disc_generic} and explicitly given in Eqs.~(\eqref{s1psi_iota} - \eqref{spi_epsilon}).

\subsection{Discontinuous Time Integration}\label{sectioniiiC}
Finally we address \textit{difficulty 3} by applying a new class of geometric integrators as put forward in our previous work \cite{o2022conservative}. That work introduced a numerical evolution scheme based on Hermite integration which has time-reversal symmetry and is unconditionally stable, ensuring symplectic structure and energy are conserved throughout long-time evolutions. 
Mathematically, we  apply the fundamental theorem of calculus and discretise in time, as:
\begin{equation}
    \textbf{U}(\tau_{n+1}) = \textbf{U}_{n} + \int^{\tau_{n+1}}_{\tau_{n}} \textbf{L} \cdot \textbf{U}(\tau) \ d\tau, 
    \label{cha3_fundamentaltheo_disctime}
\end{equation}

By building a 2 point-Hermite interpolant, we then get the \textit{fourth} order Hermite rule 
\begin{eqnarray}
  \textbf{U}^{n+1} &=& 
   \textbf{U}^{n}  + 
  \frac{\Delta \tau}{2}\textbf{L}\cdot(\textbf{U}_{n} + \textbf{U}_{n+1}) \nonumber \\
  &&+\frac{\Delta \tau^{2}}{12}\textbf{L}\cdot(\dot{\textbf{U}}_{n} - \dot{\textbf{U}}_{n+1}) + \mathcal{O}(\Delta \tau^{5}). 
 \label{cha3_h4_exampl_homogeneous1}
\end{eqnarray} 
Furthermore we can write this in an \textit{explicit form} as,
\begin{eqnarray}
  \textbf{U}^{n+1} &=& 
  \bigg(\textbf{I} - \frac{\Delta \tau}{2} \textbf{L} + \frac{\Delta \tau^{2}}{12} \textbf{L}^{2} + \cdots  \bigg)^{-1} \cdot \nonumber \\
  &&\bigg( \textbf{I} + \frac{\Delta \tau}{2} \textbf{L} + \frac{\Delta \tau^{2}}{12} \textbf{L}^{2} + \cdots \bigg)\cdot \textbf{U}^{n}, 
  \label{cha3_h4_exampl_homogeneous2}
\end{eqnarray}
where we have replaced the time derivatives as $\dot{\textbf{U}} = \textbf{L} \cdot \textbf{U}$. This effectively amounts to a fourth-order numerical evolution scheme as showed by our preliminary results in~\cite{o2022conservative}.

We note for second-order implementation of Eq.~\eqref{cha3_fundamentaltheo_disctime}, it is essentially the \textit{trapezium rule}, and Eq.~\eqref{cha3_h4_exampl_homogeneous1} yields the Crank-Nicholson scheme, which can be written in a similar fashion, \textit{i.e.,} in an \textit{implicit-turned-explicit form} and solved through self-consistent iteration (Iterated Crank-Nicholson, ICN). In the past, when attempting to use \textit{implicit-turned-explicit} evolution schemes hoping to retain the benefits of a geometric integrator schemes, such as ICN \cite{teukolsky2000stability}, it was expected that iterating more than twice would not lead to improvements, however, as we show in \cite{phdthesis-lidia, 25thCapraTalk} we recover time-symmetry and conserve energy with increasing number of iterations. 

To reduce the number of stored matrix operations and minimise \textit{round-off} error, we further rewrite Eq.~\eqref{cha3_h4_exampl_homogeneous2} as
\begin{equation}
    \textbf{U}^{n+1} = \textbf{U}^{n} + \bigg[ \textbf{I} - \frac{\Delta t}{2} \textbf{L}\cdot \bigg( \textbf{I} - \frac{\Delta t}{6} \textbf{L} \bigg) \bigg]\cdot \textbf{U}^{n}.
    \label{reduced_form_evolution_homoge_integration}
\end{equation}
To account for the discontinuous nature of the numerical problem we will use a similar scheme attained through similar logic as described above in Section \ref{sectioniiiB} This has been outlined in detail in \cite{2014arXiv1406.4865M} and later adapted to the hyperboloidal case by \cite{24thCapraTalk, 25thCapraTalk, phdthesis-lidia}. Specifically, 
Eq.~(\ref{cha3_fundamentaltheo_disctime}) has the additional integrals, 
\begin{eqnarray}
\textbf{U}^{n+1} = \textbf{U}^{n} + \textbf{L}.\int^{\tau_{n+1}}_{\tau_{n}} \textbf{U}^{n} \ d\tau 
+ \int^{\tau_{n+1}}_{\tau_{n}} \tilde{\textbf{s}}(\tau) \ d\tau +  [[\Upsilon]]_{i}, \nonumber \\ \label{cha3_final_integral}
\end{eqnarray}
where $\tilde{\textbf{s}}(\tau)$ accounts for the source terms introduced by the discontinuous discretisation highlighted in Section \ref{sectioniiiB} by Eq.~\eqref{ch3_sb2_SpaceSpource} describing all the differential operators associated with Eq.~\eqref{ch3_rwz_fieldvariables_hyper}, given specifically as 
\begin{equation}
    \tilde{\textbf{s}}(\tau) = \begin{pmatrix}
        0 \\
        \tilde{s}_{\Psi}^{(1)} + \tilde{s}_{\Psi}^{(2)} + \tilde{s}_{\Pi}^{(1)}
    \end{pmatrix}, 
    \label{s_spatial_disc_vector}
\end{equation}
and explicitly given in Eq.~(\eqref{s1psi_iota}-\eqref{spi_epsilon}). Additionally we have in Eq.~\eqref{cha3_final_integral}, 
\begin{equation}
    [[\Upsilon]]_{i} = \int^{\tau_{n+1}}_{\tau_{n}}  \tilde{\mathcal{S}}(\tau) \delta(\sigma_{i} - \sigma_{p}) d\tau 
    \label{source_time_integration}
\end{equation}
which as shown in \cite{2014arXiv1406.4865M}, only \textit{switches on} when the particle worldline crosses the numerical grid  $\sigma_{i}$ at a time $ t_{i} = [\tau_{n}, \tau_{n+1}]$. The $\mathcal{\tilde{S}}(\tau)$ vector follows directly from the RHS of Eq.~\eqref{ch3_rwz_fieldvariables_hyper} and is given as, 
\begin{equation}
    \tilde{\mathcal{S}}(\tau) = \begin{pmatrix}
\bar{F}(\tau)  \\
\bar{G}(\tau) 
\end{pmatrix}.
    \label{}
\end{equation}
where here we loosely use the $\bar{F}(\tau),\;\bar{G}(\tau) $ notation to refer to the coefficients associated with the distributional sources. 

Finally, we solve the PDE problem in Eq.~\eqref{ch3_general_pdes} through the MoL recipe. We obtain the following generic fourth-order Hermite evolution scheme, 

\begin{eqnarray}
       \textbf{U}^{n+1} &=& \textbf{U}^{n}  + \nonumber \\
       &&
      \Delta \tau \bigg[ \bigg(I - \frac{\Delta \tau}{2} \cdot \bigg(I - \frac{\Delta \tau}{6} \textbf{L} \bigg) \bigg)^{-1}\cdot \nonumber \\
      && \bigg(   \textbf{L} \cdot \bigg( \textbf{U}^{n} +  \frac{\Delta \tau}{12} (s^{n} - s^{n+1}) \bigg) \bigg)  \nonumber \\ 
      && + \ \textbf{J}(\Delta\tau, \Delta \tau_{i})   \nonumber \\ 
      && + \ \frac{1}{2} (s^{n} + s^{n+1})  +  \frac{\Delta \tau}{12} (\dot{s}^{n} - \dot{s}^{n+1}) \nonumber \\ 
      && + \  [[\Upsilon]]_{i} \bigg], \ \ \ \ \ 
       \label{cha2_h2_finalevolution}
\end{eqnarray}
where ${s,\dot{s}}$ are given by Eq.~\ref{s_spatial_disc_vector}; $\Delta \tau_{i}$ is the interval from $\tau_{n}$ to the crossing time $\tau_{i}$, satisfying $\sigma_{p}(\tau_{i}) = \sigma(\tau)_{i}$ for some $i$ and  $\textbf{J}(\Delta \tau, \Delta \tau_{i})$ is a vector including the jumps in $\textbf{U}$ resulting from integrating in time with the discontinuous collocation algorithm, again where the Lagrange formula has been corrected to account for the temporal jumps. Explicit form of this can be found in Appendix \ref{AppC_DiscTime}. We note that as we are studying circular geodesic motion, this jump vanishes, but when we apply our algorithm to eccentric/radial in-fall motion this will be necessary due to the inherent time dependencies.  

\begin{figure*}
\includegraphics[width=85mm]{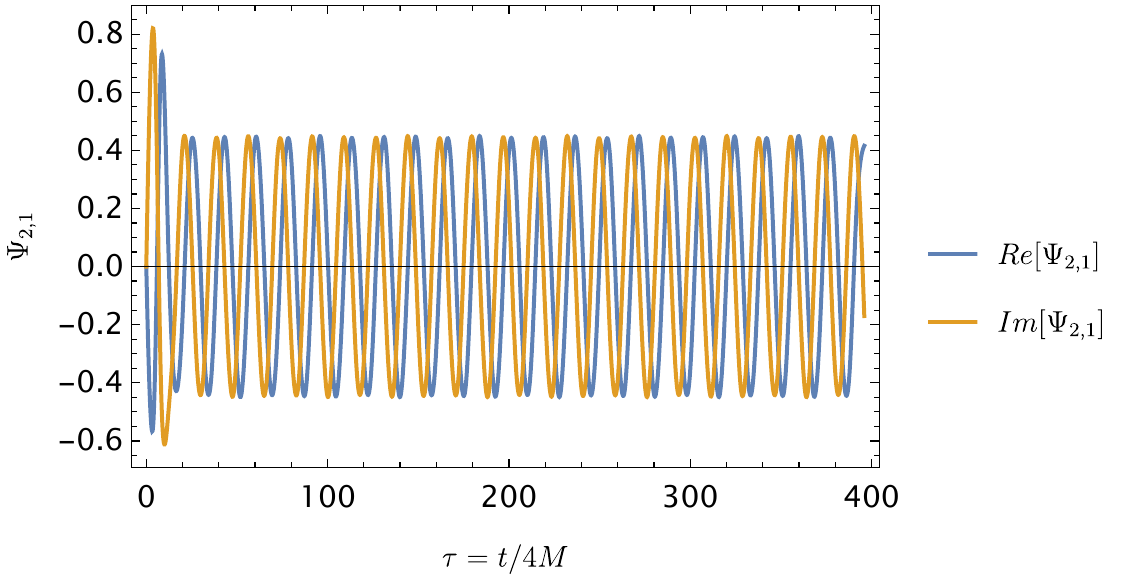}
\quad
\includegraphics[width=85mm]{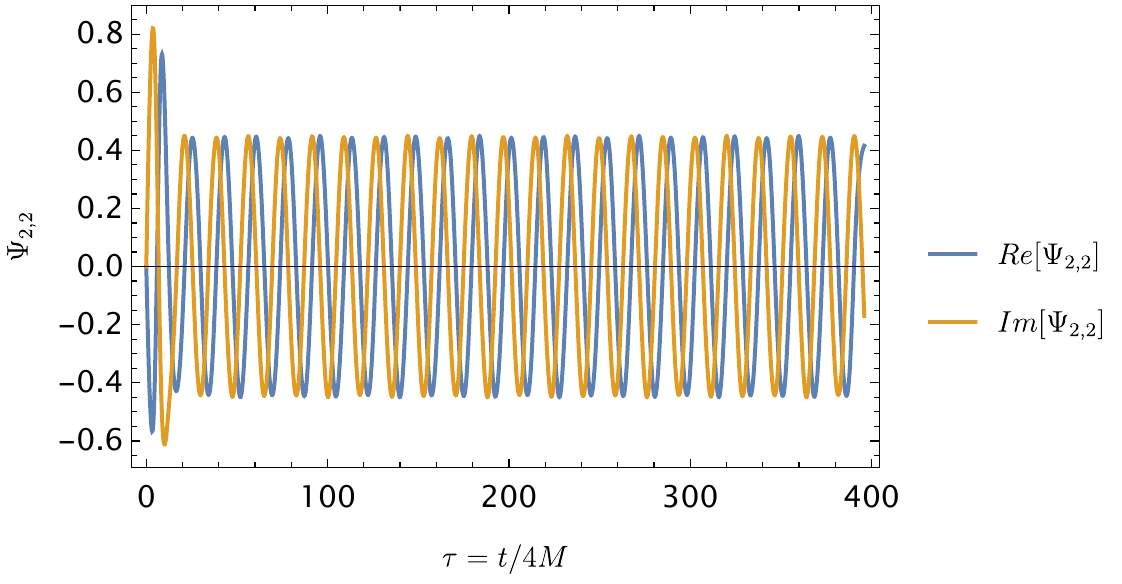}
\caption{\label{ch3_l2m2Fieldsolution} Gravitational waveforms for a point-particle on a circular geodesic at $\sigma_{p} =(2M/7.9456M)$ on a Schwarzschild background. \textbf{Top:} Waveform for the axial component, $(l,m) = (2,1)$, of the Regge-Wheeler master function. \textbf{Bottom:} Waveform for the polar component, $(l,m) = (2,2)$, of the Zerilli master function.}
\end{figure*}

\begin{figure*}
\includegraphics[width=80mm]{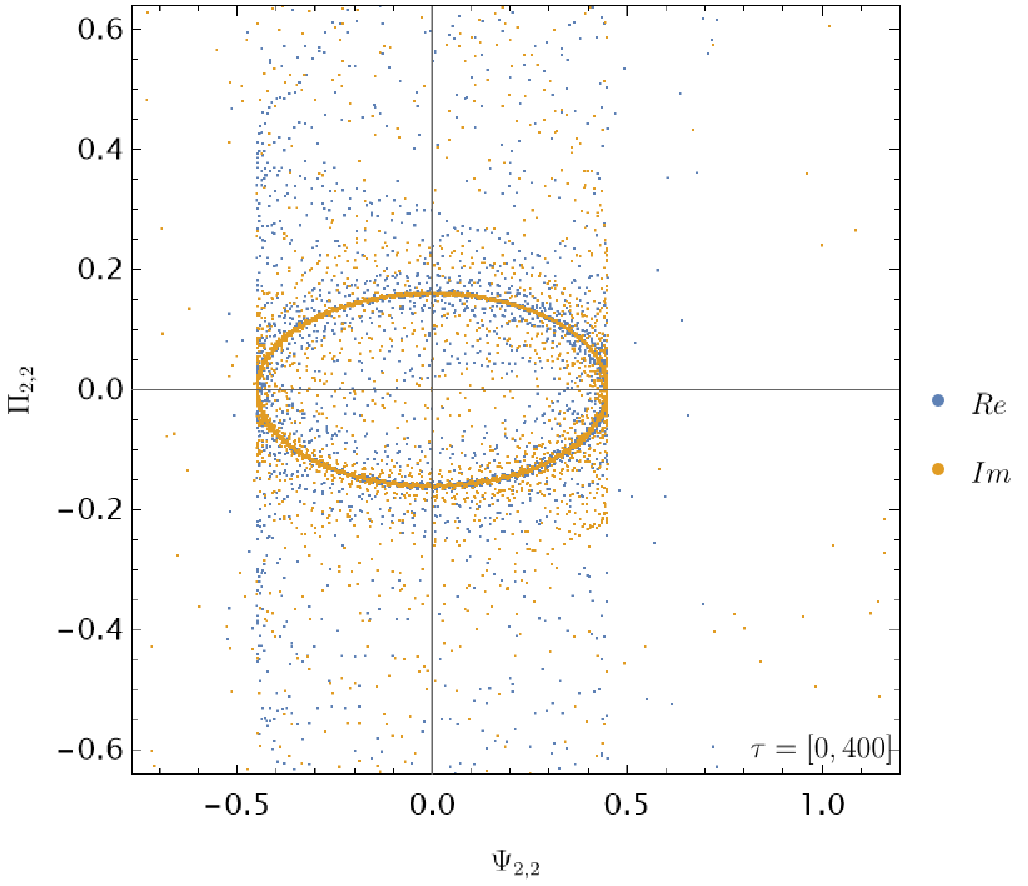}
\quad
\includegraphics[width=80mm]{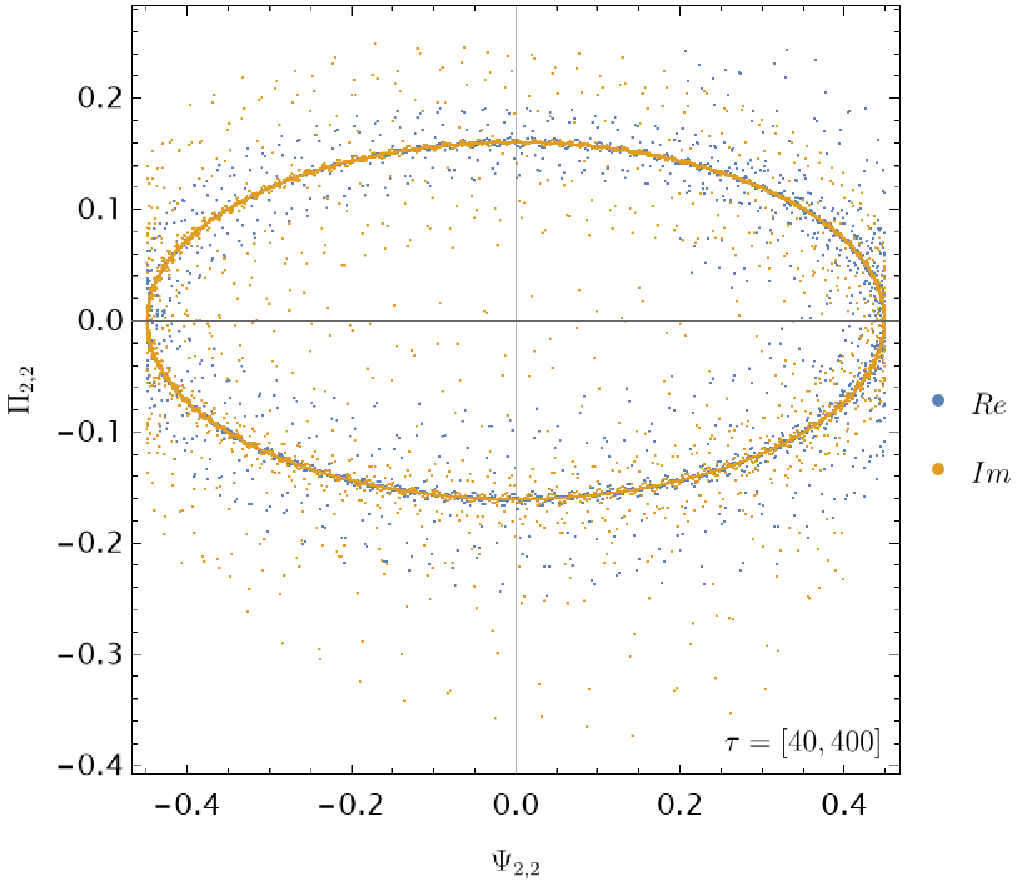}
\quad
\includegraphics[width=80mm]{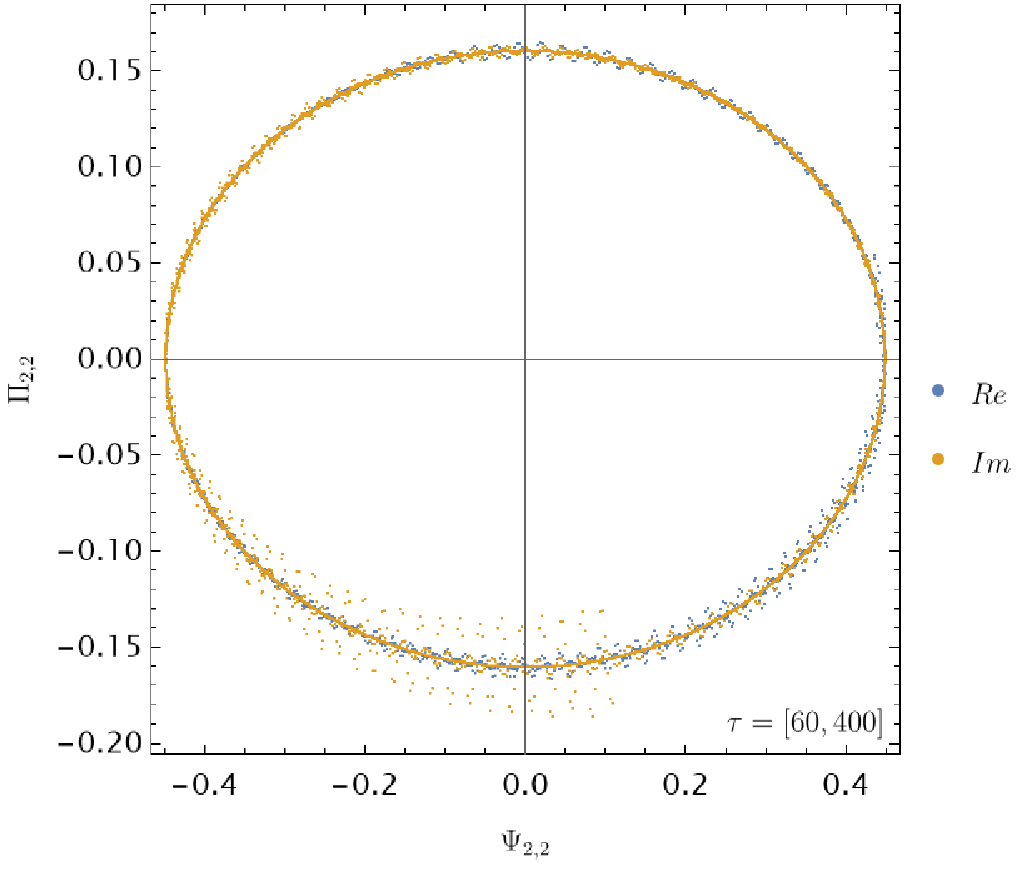}
\quad
\includegraphics[width=80mm]{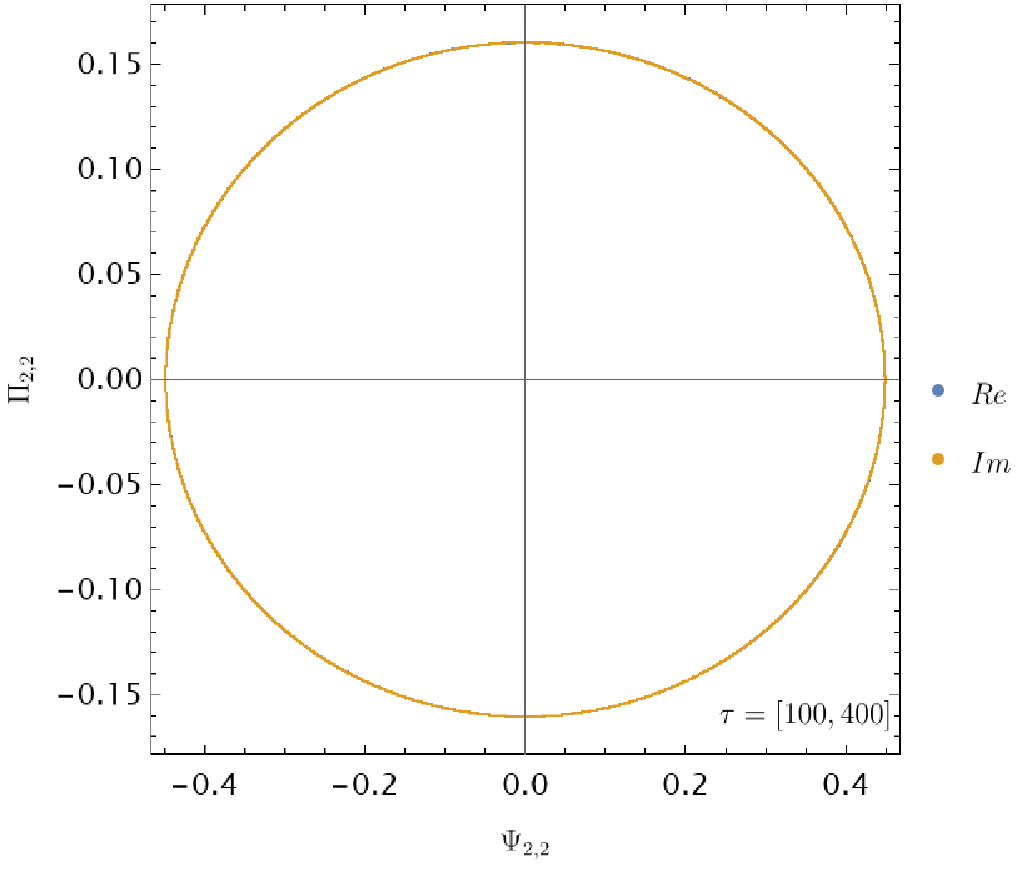}
\caption{\label{ch3_l2m2solutionPhasePs}Here we show the phase portrait of our numerical evolution for the field $\Psi_{2,2}^{p}(\tau,\sigma)$ at $\mathscr{I}^{+}$. From left to right and then top to bottom, we include the evolution from an initial time of $\tau = \{0, 40, 60, 100\}$. We determine a reasonable time to assume all junk radiation to dissipate away and our simulation to be in \textit{steady-state} to be \textit{at least} $\tau > 100$, corresponding to ignoring the first 2.8 orbits as per Fig.~\ref{ch3_l2m2Fieldsolution}. We will then show in the proceeding section that this can be more accurately measured when all numerical optimisation factors are taken into consideration.}  
\end{figure*}

\subsection{Initial Data }\label{sectioniiiD}
Our main goal is to compare our work to the reference frequency domain work of \cite{thompson1811, durkan2022slowderivatives} and the time domain codes of \cite{martel2004gravitational, sopuerta2006finite, bernuzzi2010binary, field2009discontinuous}. With the exception of \cite{field2009discontinuous} all of the aforementioned TD computations used trivial initial data,  where $\Psi(0,\sigma) = \Pi(0,\sigma) = 0 $ and, here, we too choose trivial initial data.  This non-physical initial data produces spurious ``junk" radiation which decays as $t^{-2l-3}$ \cite{Barack:1999st}. We discard data from the first few orbits to ensure the system has reach a \textit{steady-state}. Studying the potential contributions by Jost solutions \cite{field2010persistent, jaramilloSopCan2011time} into our hyperboloidal implementation for a full self-force computation goes beyond the scope of this paper and it may be subject of further works. We will briefly revisit the IVP in Section \ref{sectionivA} and \ref{conclusion}.

\section{Computing the numerical solution to the RWZ master functions}\label{sectioniv}

In this section we present our results emphasising the improvement our numerical framework brings to time-domain methods when solving a distributionally sourced equation of the type Eq.~\ref{ch3_rwz_fieldvariables_hyper}, allowing us to compute relevant physical quantities to high accuracy. We focus on the case for a point-particle on a circular geodesic with orbital radius of $r_{p} = 7.9456 M$. \footnote{We note this specific value choice of $r_{p} = 7.9456 M$ was originally made by Martel in \cite{martel2004gravitational} and later adapted by the community, though the choice has no specific physical significance we are aware of \cite{poissonPC} and can be interpreted as arbitrary.} The main numerical test from previous time-domain works in the Regge-Wheeler gauge, see refs.~\cite{martel2004gravitational, sopuerta2006finite, field2009discontinuous, bernuzzi2011binary}, has focused on calculating the radiation fluxes at both infinity and the horizon. Here we too compute these quantities and compare to their results. However, most self-force computational strategies require the evaluation of the fields and their derivatives at the particle limit from the right (which we call exterior solutions) and left (interior solutions), thus we include this test as a sensible measure of whether or not our algorithm can handle a full self-force computation.
Appendix \ref{AppD_PriceLaw} includes the implementation of our algorithm to the simpler problem, where the vacuum equations of the scalar perturbation problem are solved by studying the late-time behaviour obeying Price's Law \cite{price1972nonspherical_i}. These results complement our numerical work, validate our time-integration scheme and allow for a self-contained understanding of the numerical strategy described in the previous section.

\subsection{Numerical Solution: symplectic structure and discontinuities}\label{sectionivA}
We begin our numerical investigations by assessing if the orbital behaviour of our system is as expected of a point-particle on a circular geodesic. 
In Fig.~\ref{ch3_l2m2Fieldsolution} we show the gravitational waveforms of axial and polar parity for the modes $(l,m) = \{(2,1), (2,2)$\}, respectively. As we can see at early times the field oscillates irregularly due to the junk radiation resulting from trivial initial conditions. Once we have waited sufficient time, the effect dissipates away and the radiation emitted dominates, reaching a \textit{steady-state}. We observe, as expected for circular motion, a periodic pattern where the field oscillates with an angular frequency of $m \Omega$. 

The conservation of symplectic structure is of particular importance for our \textit{implicit-turned-explicit} time integrator. To assess this property, one must study the $2D$ phase space trajectory.  Fig.~\ref{ch3_l2m2solutionPhasePs} displays the field $\Psi_{\ell m}$ plotted versus its time derivative $\Pi_{\ell m}$ at $\mathscr{I}^{+}$. The conservation of the symplectic structure implies that the phase space trajectory is closed. The particular closed trajectory must 
be a circle due to the constant energy associated with the circular geodesic motion.

Here, we can also see from the first three plots, at early times of $\tau = \{0, 40, 60\}$ symplectic structure is not preserved, but only when \textit{steady-state} is reached we observe this feature. Both Figs.({\ref{ch3_l2m2Fieldsolution}, \ref{ch3_l2m2solutionPhasePs}) allow us to determine with more confidence that to effectively extract relevant physical quantities, one should allow the evolution to run for times longer than $\tau > 100 $, i.e ignoring the first 2.8 full orbits as per Fig.~\ref{ch3_l2m2Fieldsolution}. We find this observation to be of high value given from Fig.~\ref{ch3_l2m2Fieldsolution} alone, one could be misled into stopping at earlier times due to the periodic pattern observed. 
\begin{figure}
\includegraphics[width=90mm]{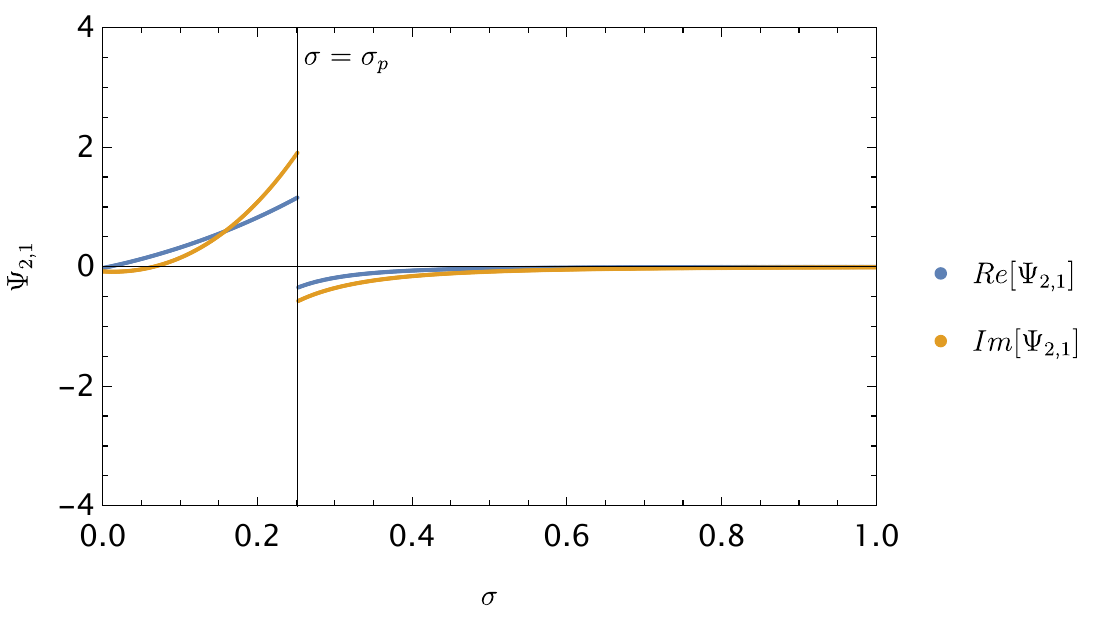}
\quad
\includegraphics[width=90mm]{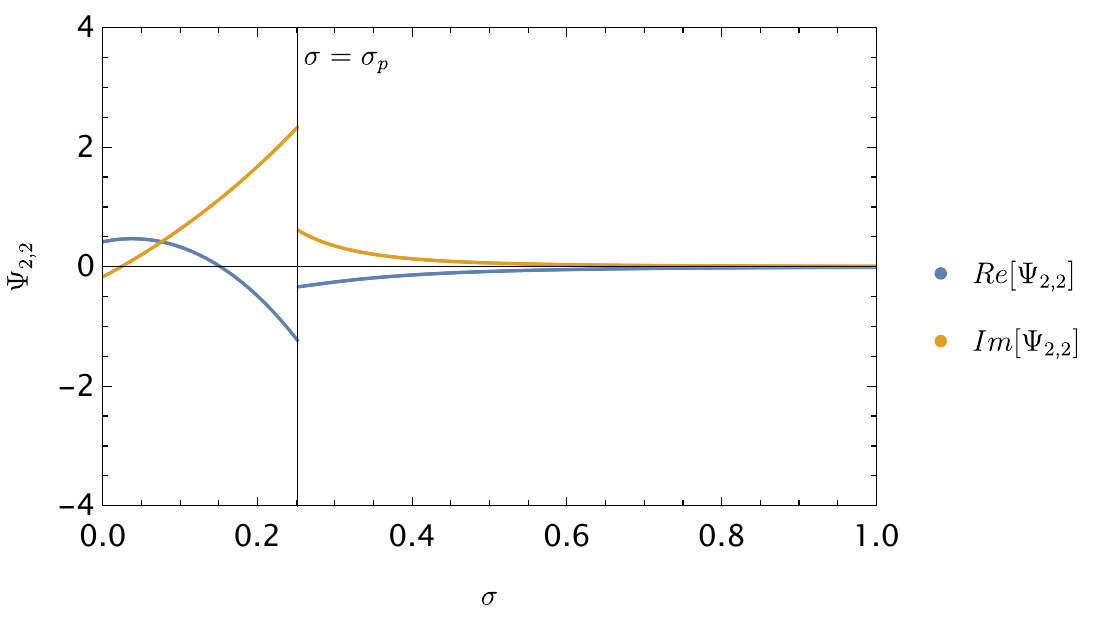}
\caption{\label{ch3_l2m2FieldAtParticle} Retarded hyperboloidal field, $\Psi(\tau,\sigma)$ for a point-particle on a circular geodesic at $\sigma_{p}$ as calculated from Eq.~\eqref{cha3_hyper_transf} where $r_{p} = 7.9456 M$  and $\tau = 100$ on a Schwarzschild background extending from the future-null infinity, $\mathscr{I}^{+}$ at $\sigma = 0 $ to the future event horizon $\mathcal{H}$ at $\sigma = 1$. \textbf{Top:} Regge-Wheeler field's axial component, $(l,m) = (2,1)$ \textbf{Bottom:} Zerilli's polar component, $(l,m) = (2,2)$.} 
\end{figure}
In Appendix \ref{AppD_complementSecIVA} we further corroborate this observation with Fig.~\ref{ch3_l2m2solutionPhasePsh} showing the same minimal time requirement for the radiation at the $\mathcal{H}$. In Fig.~\ref{ch3_l2m2FieldAtParticle} we can clearly observe that the discontinuous collocation method is capable of resolving the problem, where we see the discontinuities clearly around the particle position $\sigma_{p} = 2M/r_{p}$ for either parity of the master functions. 

\subsection{Radiation measurements at infinity and the event horizon}\label{sectionivB}
\begin{table*}[t!]
\begin{ruledtabular}
\begin{tabular}{l|| c  || c ||c || c }
\textrm{$(l,m)$}& 
\textrm{$\dot{E}^{\infty}_{lm}$}&
\textrm{$\dot{E}^{\infty}_{lm,{\rm MP}}$ \cite{martel2004gravitational}}&
\textrm{$\dot{E}^{\infty}_{lm,{\rm SL}}$ \cite{sopuerta2006finite}}&
\textrm{$\dot{E}^{\infty}_{lm,{\rm BNA}}$ \cite{bernuzzi2011binary}}\\
\colrule
(2,2) & $1.70621954 \times10^{-4} $  & $1.7051 \times10^{-4}$    & $1.7064 \times10^{-4}$ & $1.7065\times10^{-4}$  \\
(2,1) &$8.16304023 \times 10^{-7}$ &$8.1623 \times10^{-7}$  & $8.1662 \times10^{-7}$   & $8.1632\times10^{-7}$\\\hline
(3,3) &$2.54706161\times10^{-5}$  &$2.5432 \times10^{-5}$ & $2.5475 \times10^{-5}$  & $2.5481\times10^{-5}$ \\
(3,2) &$2.5198449 \times10^{-7}$   &$2.5164 \times10^{-7}$ & $2.5204 \times10^{-7}$ &  $2.5203\times10^{-7}$  \\
(3,1) &$2.1730303\times10^{-9}$&$2.1741 \times10^{-7}$    & $2.1732 \times10^{-9}$   &  $2.1740\times10^{-9}$  \\\hline
(4,4) &$4.7253849 \times10^{-6}$   &$4.7080 \times10^{-6}$   & $4.7270 \times10^{-6}$  &  $4.7289\times10^{-6}$ \\
(4,3) & $5.774899 \times10^{-8}$ &$5.7464 \times10^{-8}$  & $5.7765 \times10^{-8}$   &  $5.7777\times10^{-8}$  \\
(4,2) &$2.508984 \times10^{-9}$  & $2.4986 \times10^{-9}$    & $2.5099 \times10^{-9}$  &   $2.5112\times10^{-9}$ \\
(4,1) &$8.39527 \times10^{-13}$   & $8.3507 \times10^{-13}$  & $8.4055 \times10^{-13}$& $8.4001\times10^{-13}$ \\ \hline
(5,5) &$9.455964 \times10^{-7}$   &$9.3835 \times10^{-7}$  & $9.4616 \times 10^{-7}$   &  $9.4660\times10^{-7}$  \\
(5,4) &$1.232377 \times10^{-8}$  &$1.2193 \times10^{-8}$   & $1.2329\times10 ^{-8}$  &  $1.2334\times10^{-8}$  \\
(5,3) &$1.093226 \times10^{-9}$   & $1.0830 \times10^{-9}$  & $1.0936\times10^{-9}$ &  $1.0948\times10^{-9}$  \\
(5,2) &$2.78952\times10^{-12}$    &$2.7587 \times10^{-12}$  & $2.7909\times10^{-12}$  & $2.7925\times10^{-12}$  \\
(5,1) &$1.2593\times10^{-15}$  &$1.2544 \times10^{-15}$  &  $1.2607\times10^{-15}$  & $1.2612\times10^{-15}$  \\ \hline 
Total & $2.02907692\times10^{-4}$  & $2.0273 \times10^{-4}$  & $2.0293\times10^{-4}$ & $2.0292\times10^{-4}$ \\ \hline
$\eta$ \footnote{We note that here, for the other time domain codes, the error is as stated in their publications}& $1.1 \times10^{-9}$  & $0.2\%$ & $0.005\%$ &  $0.02\%$ \\
\end{tabular}
\caption{Comparison of energy fluxes at  $\mathscr{I}^{+}$, in units of \((M/\mu)^2\), against reference FD values and other time domain codes. Column 2 displays the tabulated flux values by \textit{Martel $\&$ Poisson}; Columns 3 and 4 show the fluxes computed by \textit{Sopuerta et al.} and \textit{Bernuzzi et al.} teams, \cite{sopuerta2006finite, bernuzzi2011binary}, respectively.   We note we do not include comparison with Table I of \cite{field2009discontinuous} though one can visibly see a discrepancy against the reference values and the value attained with their method.  } 
\label{ch3g3_table_energyInfinity_timedomain}
\end{ruledtabular} 
\end{table*}
\begin{table*}[t!]
\begin{ruledtabular}
\begin{tabular}{l||c||c||c||c   }
\textrm{$(l,m)$}& 
\textrm{$\dot{L}^{\infty}_{lm}$}&
\textrm{$\dot{L}^{\infty}_{lm,{\rm MP}}$ \cite{martel2004gravitational}}&
\textrm{$\dot{L}^{\infty}_{lm,{\rm SL}}$\cite{sopuerta2006finite}}&
\textrm{$\dot{L}^{\infty}_{lm,{\rm BNA}}$ \cite{bernuzzi2011binary}}\\
\colrule
(2,2)  &$3.82142165\times 10^{-3}$  &$3.8164\times 10^{-3}$    &  $3.8219\times 10^{-3}$& $3.8220\times 10^{-3}$ \\
(2,1)  &$1.8282769\times 10^{-5}$ & $1.8270\times 10^{-5}$  & $1.8283\times 10^{-5}$   &$1.8283\times 10^{-5}$ \\\hline
(3,3)  &$5.7046564 \times 10^{-4} $ &$5.6878\times 10^{-4}$   & $5.7057\times 10^{-4}$   & $5.7070\times 10^{-4}$ \\
(3,2) &$5.6436993\times 10^{-6}$  &$5.6262\times 10^{-6}$ &  $5.6450\times 10^{-6}$   & $5.6448\times 10^{-6}$  \\
(3,1) &$4.8669382\times 10^{-8}$  &$4.8684\times 10^{-8}$ & $4.8675\times 10^{-8}$  & $4.8691\times 10^{-8}$  \\\hline
(4,4) &$1.058344951\times 10^{-4}$  &$1.0518\times 10^{-4}$  & $1.0586\times 10^{-4}$& $1.0591\times 10^{-4}$  \\
(4,3)&$1.2934048\times 10^{-6}$ &$1.2933\times 10^{-6}$&  $1.2937\times 10^{-6}$   & $1.2940\times 10^{-6}$ \\
(4,2)  &$5.619375\times 10^{-8}$   &$5.5926\times 10^{-8}$    &  $5.6215\times 10^{-8}$  & $5.6243\times 10^{-8}$ \\
(4,1)   &$1.880290\times 10^{-11}$  &$1.8692\times 10^{-11}$  &  $1.8825\times 10^{-11}$& $1.8814\times 10^{-11}$  \\ \hline
(5,5)  &$2.1178533 \times 10^{-5} $  &$2.0933\times 10^{-5}$   & $2.1190\times 10^{-5}$ & $2.1201\times 10^{-5}$ \\
(5,4) &$2.760156\times 10^{-7}$   & $2.7114\times 10^{-7}$ &  $2.7613\times 10^{-7}$  & $2.7625\times 10^{-7}$ \\
(5,3) &$2.4485\times 10^{-8}$ &$2.4227\times 10^{-8}$   & $2.4494\times 10^{-8}$  & $2.4520\times 10^{-8}$ \\
(5,2) &$6.2477\times 10^{-11}$ &$6.1679\times 10^{-11}$   & $6.2509\times 10^{-11}$ & $6.2543\times 10^{-11}$ \\
(5,1)  &$2.8205\times 10^{-14}$  &$2.8090\times 10^{-14}$  & $1.8237\times 10^{-14}$  & $1.8814\times 10^{-14}$ \\ \hline 
Total & $4.54452565 \times 10^{-3} $&  $4.5399\times 10^{-3}$ & $4.5446\times 10^{-3}$ & $4.5454\times 10^{-3}$ \\ \hline
$\eta$  \footnote{We note that here, for the other time domain codes, the error is as stated in their publications}  & $2.8 \times 10^{-10} $ &  $0.1\%$&  $0.02\%$ & $0.02\%$ \\
\end{tabular}
\caption{Comparison of the angular momentum fluxes at $\mathscr{I}^{+}$, in units of \((M/\mu^2)\),  against reference FD values and other time domain codes. Column 2 displays the first tabulated flux values by \textit{Martel $\&$ Poisson}; Columns 3 and 4 show the fluxes computed by \textit{Sopuerta et al.} and \textit{Bernuzzi et al.} teams, \cite{sopuerta2006finite, bernuzzi2011binary}, respectively.   We note we do not include comparison with Table I of \cite{field2009discontinuous} though one can visibly see a discrepancy against the reference values and the value attained with their method. }
\label{ch3g3_table_L_Infinity_timedomain}
\end{ruledtabular} 
\end{table*}

The accuracy of time domain methods is traditionally checked by computing the total power radiated through the energy and angular momentum fluxes at infinity $\mathscr{I}^{+}$ and the horizon $\mathcal{H}$ \cite{martel2004gravitational, sopuerta2006finite, bernuzzi2011binary, field2009discontinuous}. In this work we follow the gauge invariant formalism of \cite{phdthesis-jonathan, brizuela2009complete, gleiser2000gravitational} and compute the energy fluxes radiated down to the $\mathcal{H}$ and out to $\mathscr{I}^{+}$ the black hole as, 
\begin{small}
\begin{eqnarray}
         \dot{E}^{a}(r)\bigg|_{\mathscr{I}^{+}/\mathcal{H}} = \frac{1}{16 \pi} \sum_{l,m} \frac{l(l+1)}{(l-1)(l+2)} \bigg[ \frac{\partial \Psi^{a}_{lm}(t,r)}{\partial t}\bigg]^{2}\bigg|_{\mathscr{I}^{+}/\mathcal{H}}, \ \ \ \ \ \ \\
         \label{energyflux_odd}
       \dot{E}^{p}(r)\bigg|_{\mathscr{I}^{+}/\mathcal{H}} = \frac{1}{4 \pi} \sum_{l,m} \frac{(l+2)(l-1)}{l(l+1)} \bigg[ \frac{\partial \Psi^{p}_{lm}(t,r)}{\partial t}\bigg]^{2}\bigg|_{\mathscr{I}^{+}/\mathcal{H}}. \ \ \ \ \ \ 
       \label{energyflux_even}
\end{eqnarray}
   \end{small}
For the angular momentum fluxes we use, 
\begin{small}
\begin{eqnarray}
         \dot{L}^{a}(r)\bigg|_{\mathscr{I}^{+}/\mathcal{H}} =  \nonumber \\ 
         = \frac{1}{16\pi}  \sum_{l,m} \frac{l(l+1) i m }{(l-1)(l+2)} \bigg[ \frac{\partial \Psi^{a}_{lm}(t,r)}{\partial t}  \Psi^{*, a}(t,r)\bigg]
\bigg|_{\mathscr{I}^{+}/\mathcal{H}}, \ \ \ \ \ \label{angflux_odds} \\
\dot{L}^{p}(r)\bigg|_{\mathscr{I}^{+}/\mathcal{H}} =\nonumber \\
= \frac{1}{4\pi}  \sum_{l,m} \frac{(l-1)(l+2) i m }{l(l+1)} \bigg[ \frac{\partial \Psi^{p}_{lm}(t,r)}{\partial t}  \Psi^{*, p}(t,r)\bigg]
\bigg|_{\mathscr{I}^{+}/\mathcal{H}} . \ \ \ \ \ 
\label{angflux_evens}
\end{eqnarray}
\end{small}}

We note that these flux formulas differ from the convention adapted in \cite{martel2004gravitational, sopuerta2006finite, bernuzzi2011binary, field2009discontinuous, hopper2010gravitational}. Regardless, one could choose to work in their conventions. For example in \cite{hopper2010gravitational}, where the Cunningham-Price-Moncrief and Zerilli-Moncrief versions of the RWZ master functions are used, one can correct for the difference by multiplying the jump terms given in Eq.~(\eqref{ch3_j0} - \eqref{ch3_jx}) by a factor of 2 and use their flux formulas as given by their Eq.~(4.1).\footnote{We also note the Black Hole Perturbation Toolkit \cite{BHPToolkit} uses the same convention as us in their Regge-Wheeler package.}

We compute the numerical error as given by the total energy flux radiated into and out of the black hole, 
\begin{equation}
    \dot{E}_{\text{total}} = \dot{E}^{\mathscr{I}^{+}} + \dot{E}^{\mathcal{H}}
    \label{totalenergyflux}
\end{equation}
against a reference value as provided by previous state-of-the-art frequency domain work \cite{thompson1811}, where a working precision of 32 digits with error estimates of about 16 digits was used.\footnote{For details please see Section IV.C of \cite{thompson1811} (or \cite{phdthesis-jonathan} for a more in-depth discussion).} The relative difference is defined as,
\begin{equation}
    \eta = \bigg| 1 - \frac{\dot{E}_{\text{total}}}{\dot{E}_{\text{total}}^{ref}} \bigg|.
    \label{numerical error}
\end{equation}

\begin{figure}
\includegraphics[width=96mm]{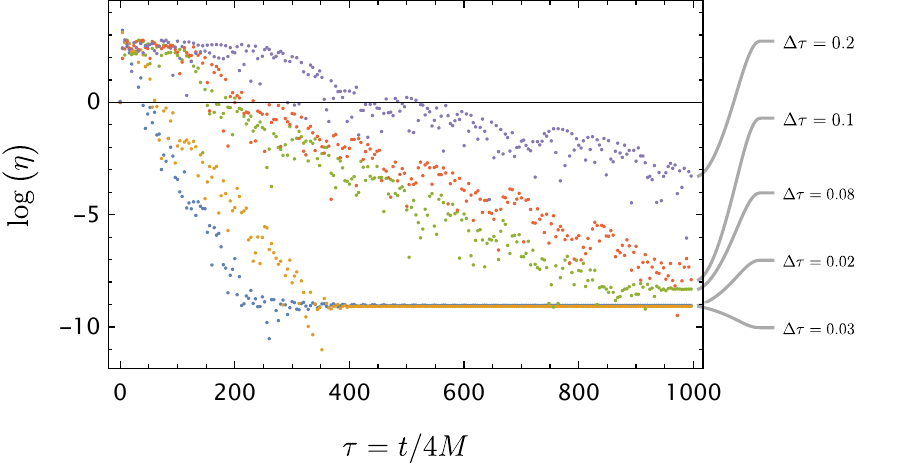}
\quad
\includegraphics[width=83mm]{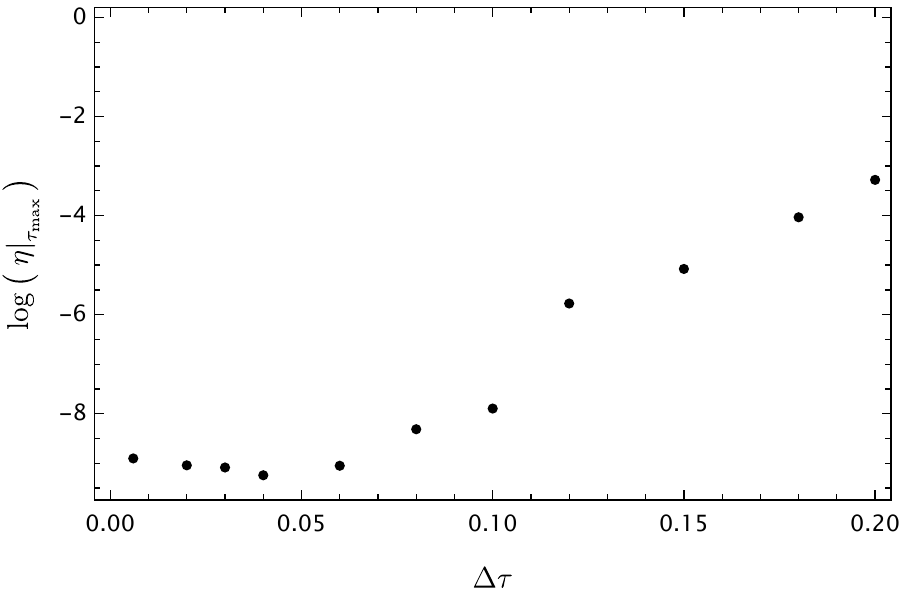}
\caption{\label{optimalConditions_FD_timestep} Convergence study to determine the optimal time discretisation step and the minimal time required for \textit{steady-state} evolution. Simulations were performed with $N=800$ equidistant nodes and $J=11$ jumps. \textbf{Top:} Convergence plot determining the numerical error as given in Eq.~\eqref{numerical error} for several discretisation time steps. Highest accuracy and saturation reached faster for a time step of $\Delta \tau = 0.02$. \textbf{Bottom:} Convergence rate showing the numerical error associated with the choice of time discretisation size at the final evolution time ${\tau_{\rm max}} = 400$.}
\end{figure}
In our case the numerical method has four main factors affecting its accuracy:
\ben[(i)]
\item \textit{number of nodes};
\item \textit{number of jumps};
\item \textit{minimal time for steady-state evolution}; 
\item \textit{time discretisation step size}.  
\een

\begin{figure}
\includegraphics[width=96mm]{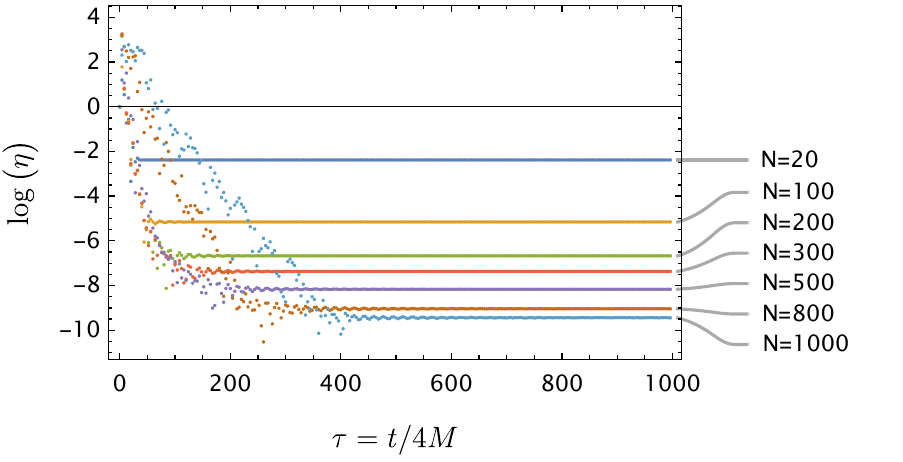}
\quad
\includegraphics[width=85mm]{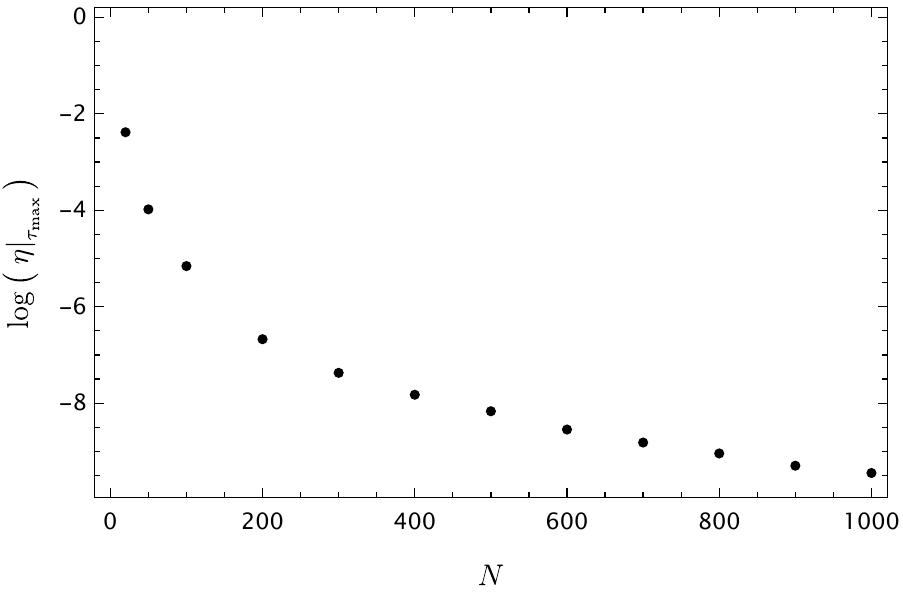}
\caption{\label{optimalConditions_FD_NODES} Convergence study to determine the optimal number of equidistant nodes, $N$,
for numerical evolution. Numerical error is calculated via Eq.~\eqref{numerical error} against the reference solution of \cite{thompson1811}. \textbf{Top:} Convergence study determining the optimal number of equidistant nodes required for accurate evolution. \textbf{Bottom:} Convergence rate showing the numerical error associated with the choice of nodes at the final evolution time ${\tau_{\rm max}}$.} \end{figure}

\begin{figure}
\includegraphics[width=95mm]{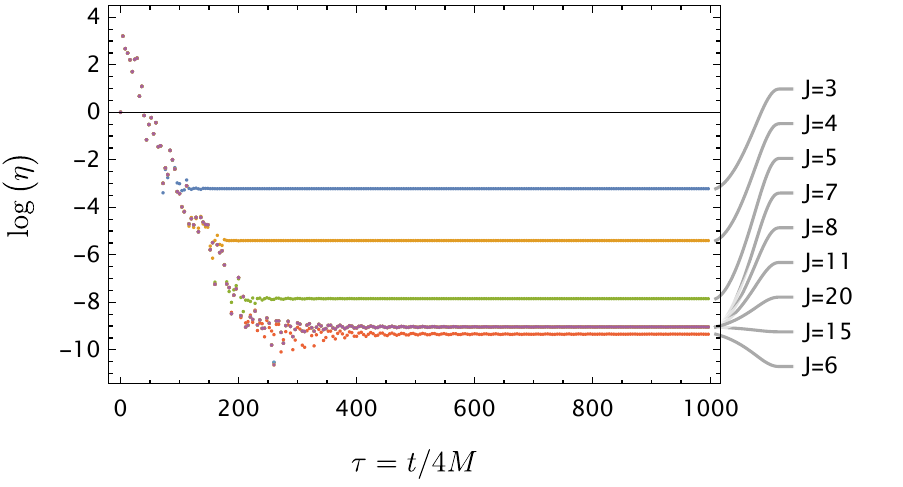}
\quad
\includegraphics[width=85mm]{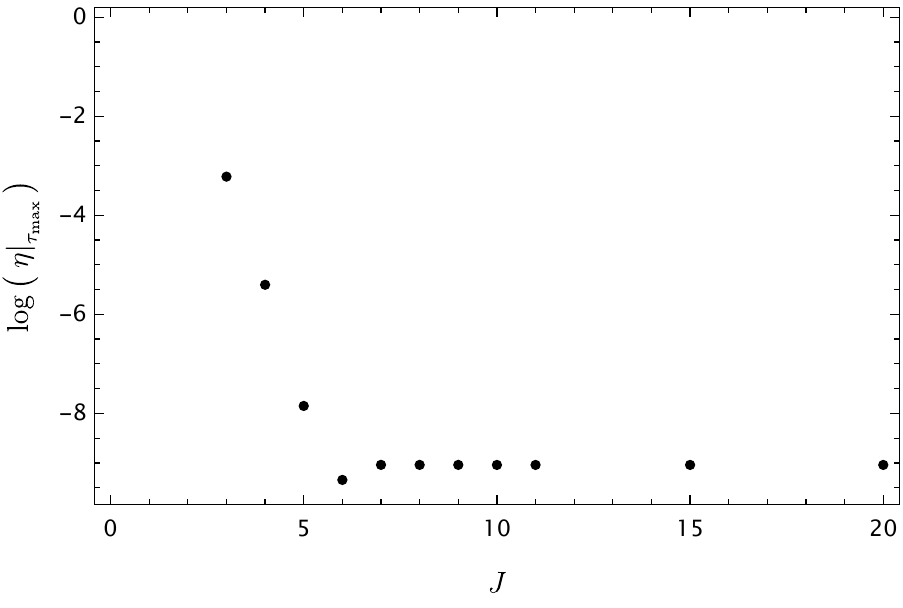}
\caption{\label{optimalConditions_FD_JUMPS} Convergence study to determine the optimal number of jumps, $J$,
for numerical evolution. Numerical error is calculated through Eq.~\eqref{numerical error} against the reference solution of \cite{thompson1811}. \textbf{Top:} Convergence study determining the optimal number of jumps required for accurate evolution. \textbf{Bottom:} Convergence rate showing the numerical error associated with the choice of jumps at the final evolution time ${\tau_{\rm max}}$.}
\end{figure}

To determine the optimal factors for this particular orbital set-up we performed several convergence tests as visualised in Figs.(\ref{optimalConditions_FD_timestep}-\ref{optimalConditions_FD_JUMPS}). We started our numerical studies by trying to assess the minimal time required for extraction of physical quantities and the time discretisation step that we should utilise. From previous Figs.~\ref{ch3_l2m2Fieldsolution},~\ref{ch3_l2m2solutionPhasePs}, we know at \textit{least} this should be $\tau > 100$. From Fig.~\ref{optimalConditions_FD_timestep} we observed the simulation to reach saturation for the same running time interval of, $\tau = [0, 400]$, with a discretisation time step of $\Delta \tau = 0.02$. Furthermore we can estimate the minimal time for \textit{steady-state} evolution and a reasonable extraction point to be $\tau_{\rm{max}} = 400$.

As observed in Fig.~\ref{optimalConditions_FD_NODES}, optimal results are attained with a number of nodes of $N=800$ with a finite-difference collocation scheme as given by Eq.~\eqref{ch3_sb2_EquiNodes}; past this number we did not observe a significant improvement that would justify the associated increase in simulation running time. From Fig.~\ref{optimalConditions_FD_JUMPS} it is furthermore clear the algorithm needs at least 6 jumps for optimal performance, and after this number accuracy seems to reach a plateau regardless of the number of jumps used for this particular physical quantity. Furthermore, given the required physical quantities as given in Eq.~\eqref{angflux_odds}, ~\eqref{angflux_evens} are the same as in the energy computation, it is reasonable to conclude the optimisation factors choices hold. 
\begin{table}
\begin{ruledtabular}
\begin{tabular}{l|| c ||c }
\textrm{$r_{p}$}& 
\textrm{$\dot{E}_{\text{total}}$}&
\textrm{\text{$\eta$}} \\ \colrule
$7.9456 M$ & $2.03294497 \times 10^{-4}$ & $1.3\times 10^{-9}$ \ \ \cite{thompson1811, durkan2022slowderivatives}  \\&
\textrm{$\dot{L}_{\text{total}}$}&
\textrm{\text{$\eta$}} \\
& $4.5531889267 \times 10^{-3}$ &  $2.8 \times 10^{-11}$ \ \ \cite{thompson1811, durkan2022slowderivatives}  \\ 
\end{tabular}
\caption{Comparison of the total energy $\dot{E}$, in units of \((M/\mu)^{2}\) , and angular momentum $\dot{L}$, in units of \((M/\mu^2)\), fluxes against the reference frequency domain values of \cite{thompson1811, durkan2022slowderivatives} for a total of $l_{\rm{max}} = 20$ modes.  }
\label{ch3g3_table_FD_FLUX_20MODES}
\end{ruledtabular} 
\end{table}
\begin{figure}[h!]
\includegraphics[width=85mm]{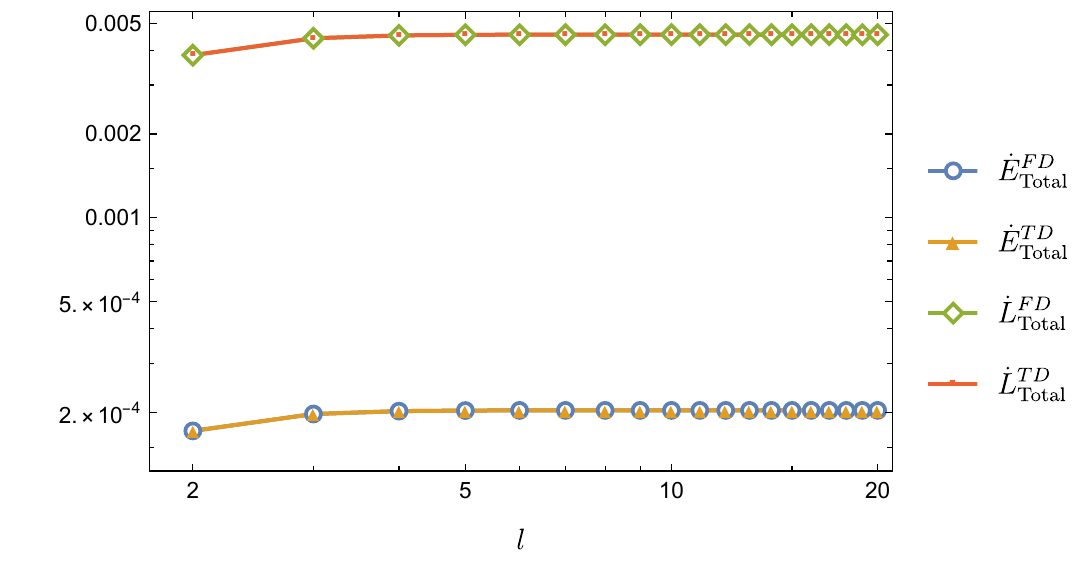}
\quad
\includegraphics[width=85mm]{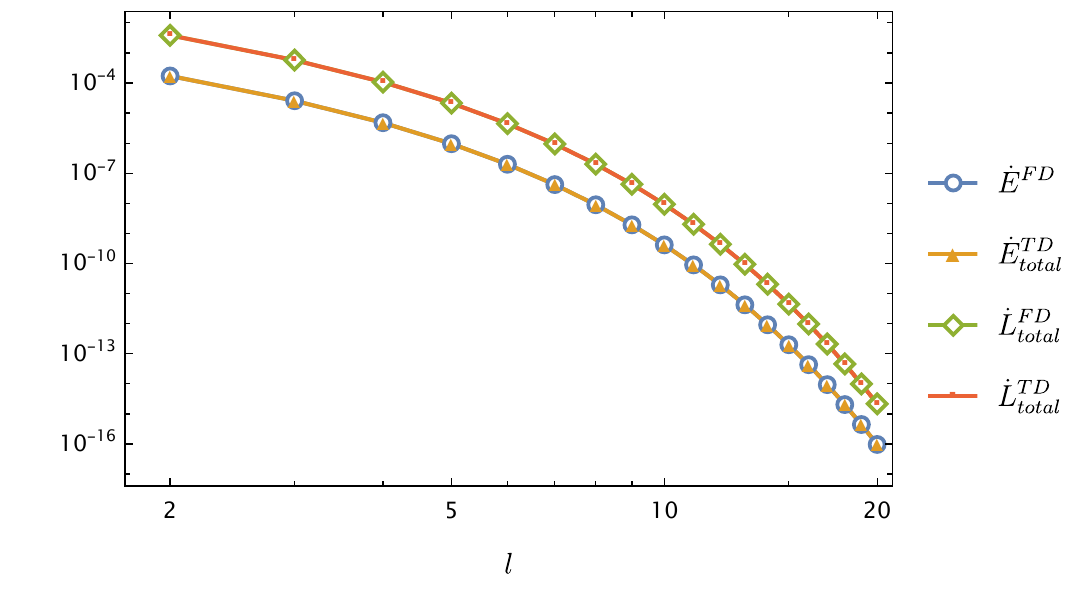}
\quad
\includegraphics[width=85mm]{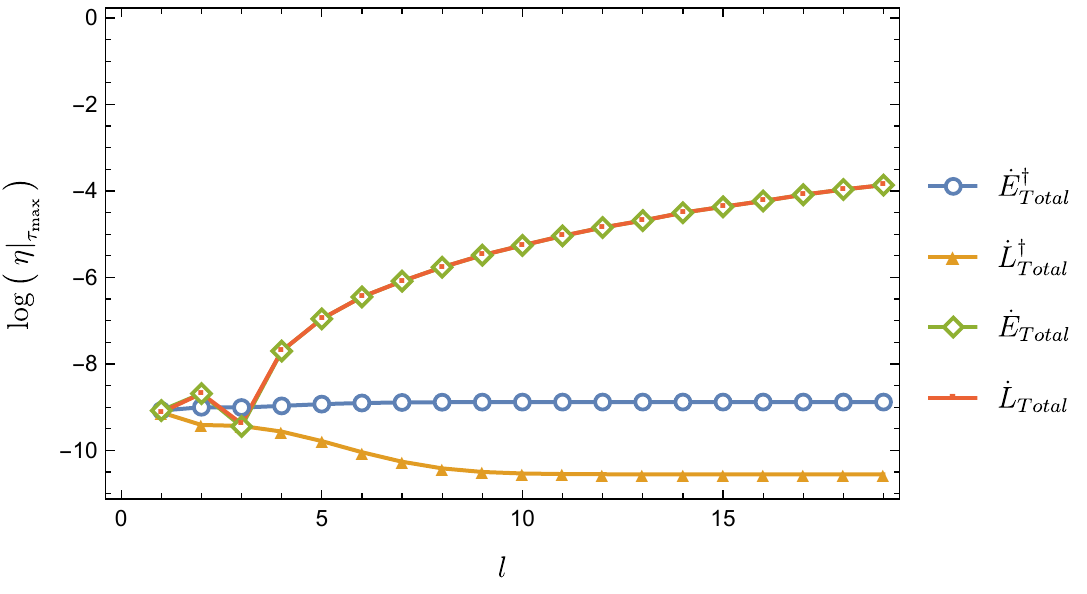}
\caption{\label{radiationfluxes} The $l$-mode contributions to the energy $\dot{E}_{\rm{Total}}$
and angular momentum $\dot{L}_{\rm{Total}}$ fluxes with units \((M/\mu)^{2}\), \((M/\mu^2)\), respectively, for a particle on a circular orbit $r_{p} = 7.9456 M$ for $l_{\rm{max}} =20$ modes from both our numerical time-domain algorithm and the reference frequency domain work of \cite{thompson1811}.  \textbf{Top:} $l$-mode contribution to the final result of the energy $\dot{E}_{\rm{Total}}$ and angular momentum $\dot{L}_{\rm{Total}}$ fluxes as displayed in Table \ref{ch3g3_table_FD_FLUX_20MODES}. \textbf{Middle:} $\dot{E}_{\rm{Total}}$ and $\dot{L}_{\rm{Total}}$ modes convergence. Around the $l_{\rm{max}} =20$ mode the individual contributions become increasingly less significant, where in the last few $l-$ modes is at around $10^{-15/16}$. \textbf{Bottom:}  Numerical error associated with the individual modal contributions $\dot{E}_{\rm{Total}}^{\dag}$/$\dot{L}_{\rm{Total}}^{\dag}$ and the error associated with the increase of modal contributions to the final value $\dot{E}_{\rm{Total}}$/$\dot{L}_{\rm{Total}}$.  }\end{figure}

From Tables \ref{ch3g3_table_energyInfinity_timedomain} and \ref{ch3g3_table_L_Infinity_timedomain} we observe a significant improvement with our numerical algorithm over the other aforementioned time-domain methods validating all our choices when addressing the three main difficulties concerning handling the discontinuous nature of the problem, the use of radiation boundary conditions and choosing a time integrator with conservative proprieties. To further assess our numerical algorithm's accuracy and how it compares to frequency domain methods, we extend our study to include $l_{\rm{max}}=20$ modes. From Table \ref{ch3g3_table_FD_FLUX_20MODES} we see our numerical algorithm retains good accuracy when summing over all 20 \(l\)-modes, showing a clear improvement from all previous time domain work \cite{barack2007gravitational, martel2004gravitational, sopuerta2006finite, field2009discontinuous, bernuzzi2011binary} compared in this draft. 

Furthermore from Fig.~\ref{radiationfluxes} we can get a clear understanding of how our algorithm performs with each increasing $l$-mode contribution. From the bottom sub-figure it is clear the error increases as the number of $l$-modes is increased; this is because overall each $(l,m)$ contribution decreases up to around $10^{-15/16}$ for the \(l=20\) (as demonstrated by the sub-figure in the middle). One could theoretically investigate further how our algorithm would perform for even higher modes using extended-precision computing, though at this stage, where our implementation is merely for proof-of-concept and to test the robustness of hyperboloidal algorithm we find this to be sufficient and motivate future work with more sophisticated computational resources \cite{lidiaHPC, phdthesis-lidia}.

\subsection{Numerical evaluation of the fields and its derivatives at the particle limit}\label{sectionivC}

\begin{figure}
\includegraphics[width=90mm]{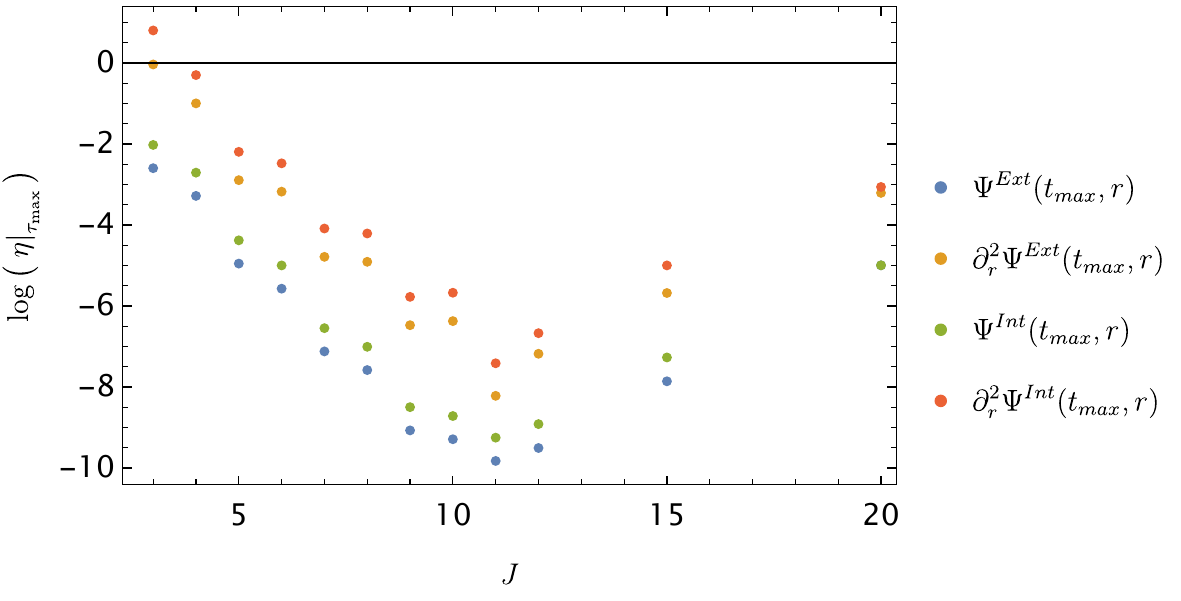}
\quad
\includegraphics[width=90mm]{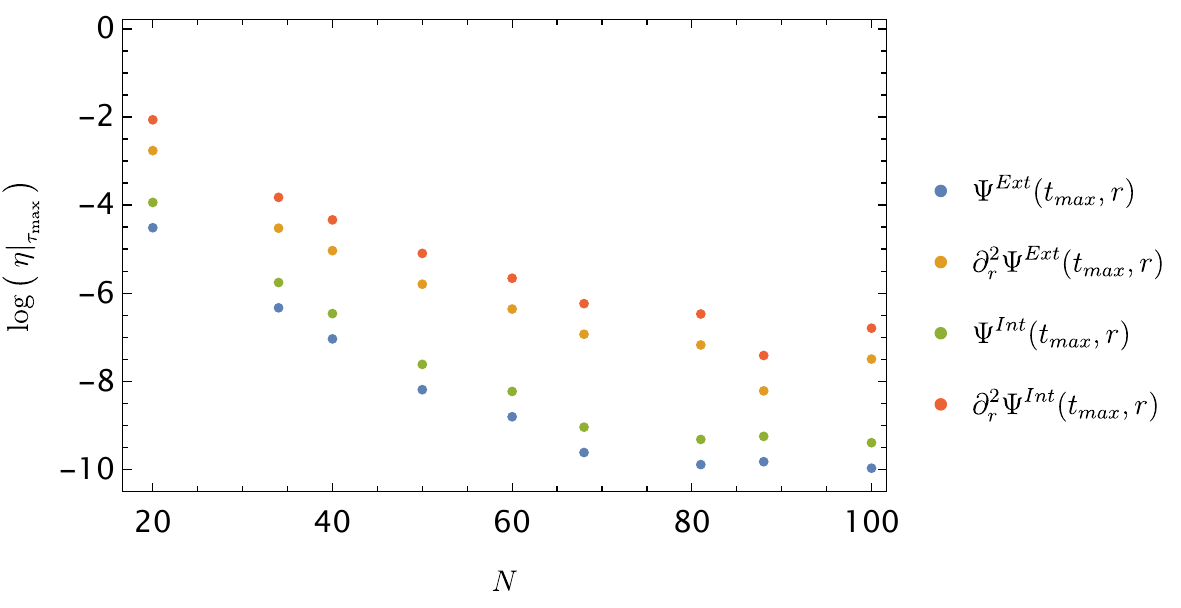}
\caption{\label{interpol_jumpsnodes} Numerical convergence study determining the optimal numerical evolution factors for interpolation of the master functions. \textbf{Top:} Convergence study to determine the optimal number of jumps, found to be $J= 11$. \textbf{Bottom:} Convergence study to determine the optimal number of Chebyshev collocation nodes, given by $N=89$.}%
\end{figure}

We now extend our studies to compare the fields and derivatives at the particle.  We do this with a \textit{two-fold purpose}: \textit{First}, it allows us to understand if our numerical algorithm has potential to provide competitive results to the frequency domain, and \textit{second}, it may serve as sensible benchmark for upcoming time domain codes. The numerical results that follow are a preview of our upcoming work \cite{paper2,paper3}, and the final optimal values can only be provided within the context of a full GSF computation as will become clearer in this section. Nevertheless, the quantities here evaluated should be a sensible metric for the applicability of any upcoming time-domain schemes. 

We directly use Eq.~\eqref{ch3_sb2_WFsolution} to interpolate the results at the grid points onto the particle position of $\sigma_{p}$.

\begin{figure}
\includegraphics[width=90mm]{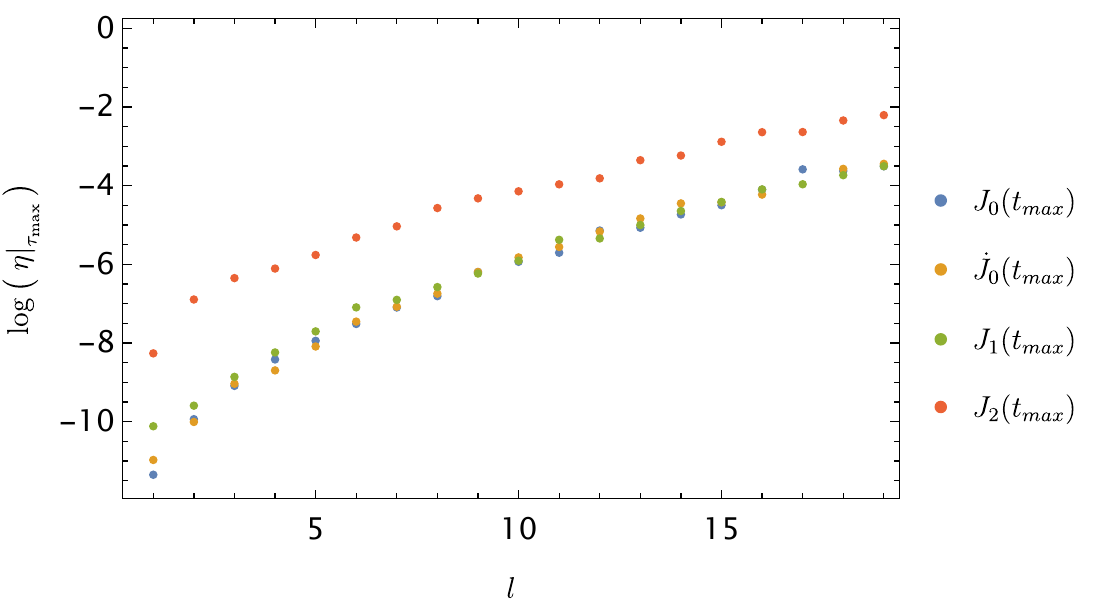}
\caption{\label{interpol_RECrelation} Individual $l$-mode contribution to the numerical error associated with the difference of our TD numerical approach versus the reference FD approach \cite{thompson1811}. We note here $J_{0,1,2}(t_{\rm{max}})$ correspond to the RWZ fields and its first two radial derivatives directly evaluated at the particle position of $r_{p} = 7.9456 M$. We decided to also include the first temporal jump, $\dot{J}_{1}(t_{\rm{max}})$, \textit{i.e.}, the first time derivative of the field evaluated at the particle position $\Pi(t_{\rm{max}}, r_{p})$.  }%
\end{figure}
Interpolation for the field derivatives, necessary for any metric perturbations and gravitational self-force computations (see for example Eqs.~(63-68) of \cite{thompson1811}), follows trivially from this formula. 
We note, though, that the chain rules \eqref{eq:hyp_chainrules} must be used when transforming back to the physical quantities from the hyperboloidal ones. 
At this stage, evaluating both the field $\Psi^{a/p}_{lm}(t,r)$ and its time derivative $\Pi^{a/p}_{lm}(t,r)$ along with the first and second order fields $\partial_{r}\Psi^{a/p}_{lm}(t,r), \ \partial_{r}^{2}\Psi^{a/p}_{lm}(t,r)$ is sufficient to assess any algorithm's suitability for a gravitational self-force computation through a mode-sum approach. 

For this part of our numerical work, we've chosen to work with pseudospectral collocation nodes as given by Eq.~\eqref{chebyshev_lobatto_nodes} due to significant accuracy improvements and shorter simulation running time for evaluations at the particle. As we have done in the previous section, we start by studying the systematic error induced by the four numerical parameters involved in the algorithm. We initially restrain our numerical evaluations to the $(l,m) = (2,2)$ mode of the RWZ master functions. It is important to note that our reference values and any numerical error calculations were done against the frequency domain data of \cite{thompson1811}, where we have to Fourier transform it to the hyperboloidal time-domain coordinates.

Interestingly, albeit somewhat expected, we found all the factors $i)-iv)$ to be best determined by the highest-order derivative, in this case, $\partial_{r}^{2}\Psi^{a/p}_{lm}(t,r)$. This is not only due to the fact that \textit{round-off} error tends to increase with the increase of the differentiation order but also due to the fact we transform our quantities back from hyperboloidal to physical coordinates in $(t,r)$. These transformations increase in complexity as the order of differentiation increases requiring more combinations of the physical fields arising from the chain rule. We found the second-order spatial derivative to require a longer evolution time to reach sufficient accuracy, and we will thereby for now leave the determination of factor $iv)$ for future work and determine factor $iii)$ based on the result for second-order radial field derivative. Nevertheless, we complement our numerical work with Fig.~\ref{ChebyTimeOFs} of Appendix \ref{AppD_complementSecIVC} for these four particular physical quantities.  We found that a reasonable extraction time to be around $\tau_{\rm max} = 10,000$. Even though this is $10 \times$ longer than the simulation extraction time of Section \ref{sectionivB} it is substantially faster due to the use of Chebyshev collocation method. The determination of factor $i)$ and factor $ii)$ is fully independent of whatever the field or its derivatives we are studying. From Fig.~\ref{interpol_jumpsnodes} bottom figure it is clear that there is a significant improvement as we increase resolution. For now we pick $N=89$ nodes as a compromise with simulation run-time. 

Perhaps one of the most relevant results in this paper is the striking contrast between the top plot of Fig.~\ref{interpol_jumpsnodes} and that of the previous section at the bottom of Fig.~\ref{optimalConditions_FD_JUMPS} for the determination of the optimal number of jumps, $J$. The latter very clearly reaches somewhat of a plateau, whereas the former decreases in accuracy after an optimal jump number is reached. Unlike in the previous section, where the fluxes are evaluated at the boundaries of the numerical domain, here we want to evaluate the fields and their derivatives at the particle limit (both the interior and exterior solutions) and thus, here, interpolation is \textit{unavoidable}.  
As one should be able to infer from Section \ref{sectioniii}, increase in \textit{round-off} error is certainly expected from the evaluation of the second part of Eq.~\eqref{cha2_spatial_disc_generic}, i.e $\Delta_{\Psi}$ (or whatever the equivalent for the field $t$- or $r$-derivatives). This quantity inherits error from our higher order jumps as given by Eq.~\ref{cha3_rec_relation_njumps_RWZ}, which by themselves increase in \textit{round-off} error with the increase of complexity of the symbolically-derived $m$ jump. We found $J=11$ to be the optimal number of jumps for these particular cases. 

To further corroborate our hyperboloidal jump recurrence relation Eq.~\eqref{cha3_rec_relation_njumps_RWZ}, we also studied the \textit{internal} error associated with the numerical method in obtaining the first 3 radial jumps i.e $J_{0,1,2}(t_{\rm{max}})$ which corresponds to the fields and their first order derivatives directly interpolated at the particle position of $r_{p} = 7.9456 M$. We note we compare it to \cite{thompson1811} but their values \textit{exactly at the particle} are the same as our \textit{input} $(t,r)$ jumps due to their internal high order accuracy and use of extended machine precision, therefore making this a sensible internal error measurement for our algorithm. In Fig.~\ref{interpol_RECrelation}, like in the previous section, we observe that error contribution increases with the individual $l$-mode increment. 

\begin{figure}
\includegraphics[width=85mm]{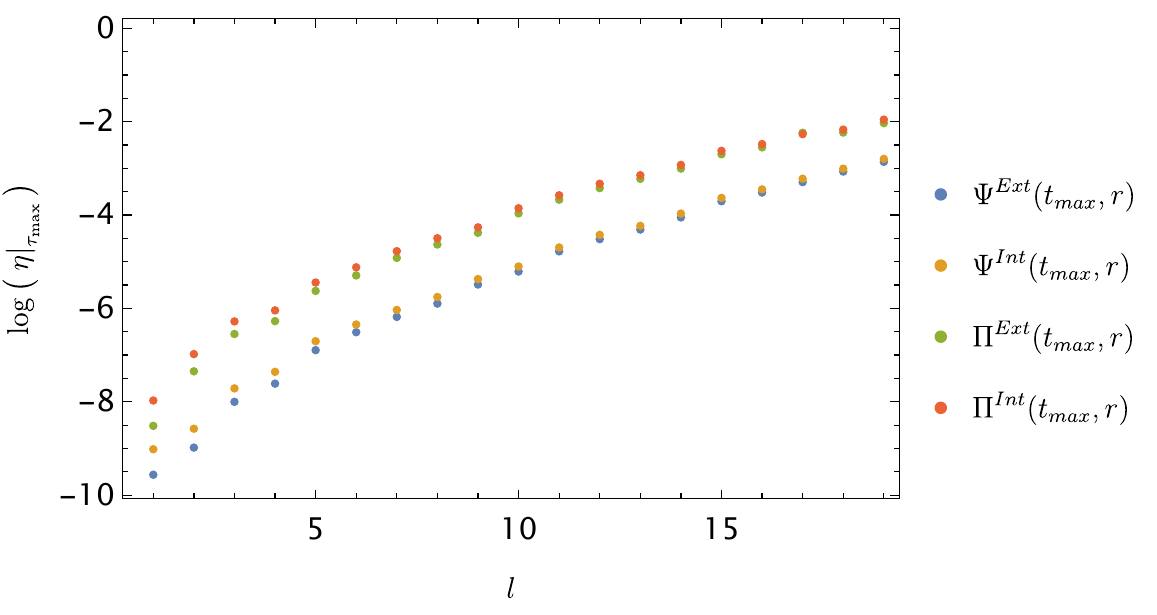}
\quad
\includegraphics[width=85mm]{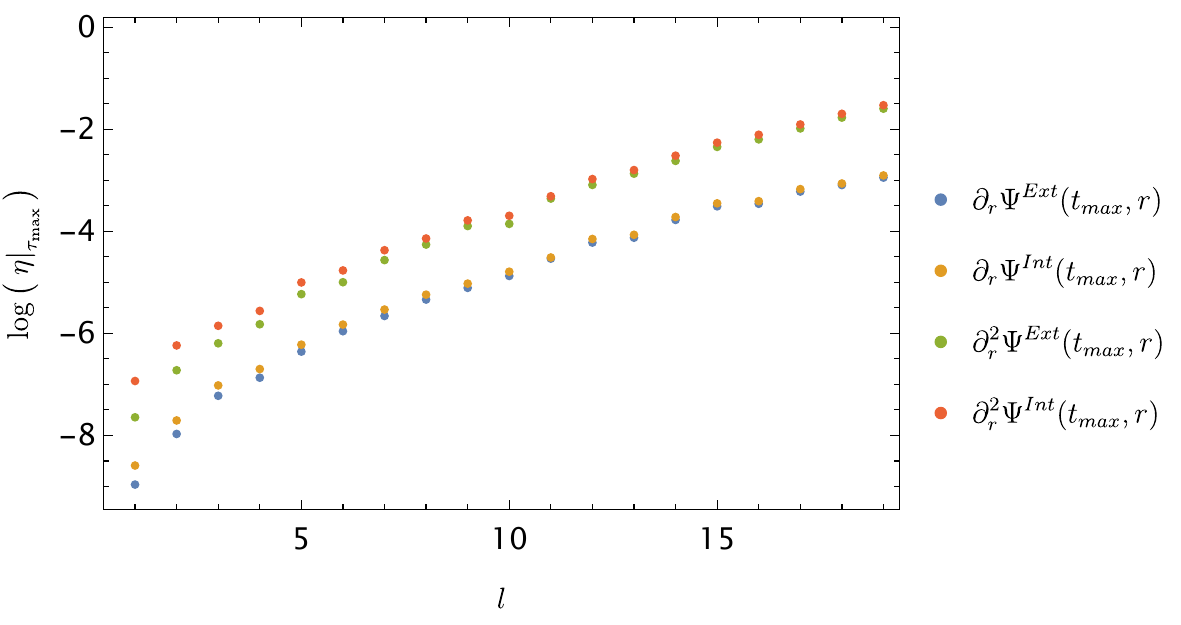}
\caption{\label{interpol_IntExt_IndError} Individual $l$-mode contributions to the numerical error between our time-domain results against the reference values of \cite{thompson1811}. \textbf{Top:} Numerical error individual $l$-mode contributions for the RWZ master functions and its first-order time derivative for both the interior and exterior solutions. \textbf{Bottom: } Numerical error individual $l$-mode contributions for both the interior and exterior solutions of the first- and second-order radial derivatives of the RWZ master functions. }
\end{figure}
Finally, in Fig.~\ref{interpol_IntExt_IndError}, we give the error projection expected with the increase of $l$-mode contributions for both the interior and exterior solutions of the master functions and their field derivatives, extracted at a final time of $\tau_{\rm{max}} = 10,000$. As we observed in all our numerical experiments there is a clear increase in error with the $(l,m)$ mode increment.  Furthermore, in Table \ref{ch3g3_table_evalIntExtAllFields} and Table \ref{ch3g3_table_evalIntExt_dr2psi}, we give their values for the first $l=5$ modes with the individual error associated with each $(l,m)$ mode compared to the FD reference values \cite{thompson1811} for the field and its second-order radial derivative. We note, for future reference we also include the first order time and radial derivative values in Table \ref{ch3g3_table_evalIntExt_Pi} and Table \ref{ch3g3_table_evalIntExt_drpsi}. 
\begin{table*}
\begin{ruledtabular}
\begin{tabular}{l|| c c ||c  c}
\textrm{$(l,m)$}& 
\textrm{$\Psi^{Int}(t_{max},r)$ \((M/\mu)\) }&
\textrm{$\eta$}&
\textrm{$\Psi^{Ext}(t_{max},r)$ \((M/\mu)\)}&
\textrm{$\eta$}\\
\colrule
(2,0) & $-0.765435516032$ & $[8.4\times  10^{-12}]$ & $-2.34650128369$ & $[3.1 \times 10^{-12}] $   \\
(2,1) & $0.1674885981 - 0.657754692 \ i $ &  $[4.0 \times 10^{-10}] $  & $-0.5517582433 + 2.1622479784 \ i $ & $[1.2\times 10^{-10}] $  \\ 
(2,2) & $-0.6095837026 - 0.3503516788 \ i $  & $[5.6 \times 10^{-10}] $  & $-2.3094419314- 1.2777865523 \ i $ & $[1.5\times 10^{-10}] $  \\ \hline
(3,0) & $0.6750964960$ & $[1.9\times  10^{-10}]$   & $-1.43360187478$ & $[6.9\times 10^{-11}] $   \\
(3,1) & $0.1333028327 - 0.5226482320 \ i $ &$[4.0 \times 10^{-10}] $   & $0.3103278798 - 1.2167230295 \ i $   & $[1.2 \times 10^{-10}] $ \\ 
(3,2) & $0.524482252 + 0.286083329 \ i $ & $[1.5 \times 10^{-9}] $   & $-1.165339811 - 0.6358758585 \ i $ &  $[6.9 \times 10^{-10}] $ \\ 
(3,3) & $0.3690237727 - 0.39461705537 \ i $ & $[8.4\times 10^{-10}] $  & $0.9988039774- 1.0717480472 \ i $ & $[2.7 \times 10^{-10}] $ \\ \hline 
(4,0) & $0.4391898012$ & $[9.0\times  10^{-10}]$  & $0.8425472696$ &  $[3.6\times 10^{-10}] $    \\
(4,1) & $-0.142965907 + 0.560536688\ i $ & $[5.2 \times 10^{-9}] $   & $0.253436339 - 0.993666026 \ i $ & $[3.0 \times 10^{-9}] $  \\ 
(4,2) & $0.382871421+ 0.208893175\ i $ & $[3.7\times 10^{-9}] $    & $0.756109534 + 0.412530182 \ i $ & $[2.0 \times 10^{-9}] $  \\ 
(4,3) & $-0.338922809+ 0.3644222016 \ i $ & $[2.6 \times 10^{-9}] $  & $0.624448668 - 0.671381527 \ i $ &  $[1.4\times 10^{-9}] $ \\ 
(4,4) & $0.251765092 + 0.391540614  \ i $  & $[6.9\times 10^{-9}] $    & $0.556176318 + 0.864498925 \ i $ & $[3.2 \times 10^{-9}] $  \\  \hline 
(5,0) & $-0.512066356$ & $[2.6\times  10^{-9}]$  & $0.809630295$ & $[4.9\times 10^{-10}] $   \\
(5,1) & $ -0.091895289 + 0.360300450\ i $ & $[9.1 \times 10^{-9}] $   & $-0.157513215 + 0.617573360 \ i $  & $[5.3\times 10^{-9}] $  \\ 
(5,2) & $-0.43186053 - 0.23562113\ i $ & $[1.0 \times 10^{-8}] $   & $0.689040619 + 0.375937407 \ i $ &$[6.6\times 10^{-9}] $  \\ 
(5,3) & $-0.253333769 + 0.272380868\ i $ & $[2.9 \times 10^{-9}] $   & $-0.4486443671 + 0.4823761295\ i $ & $[1.9 \times 10^{-10}] $  \\ 
(5,4) & $ -0.22626337 - 0.35153750\ i $ & $[1.3 \times 10^{-8}] $   & $0.3722220864 + 0.5783187181 \ i $ & $[7.1 \times 10^{-10}] $\\ 
(5,5) & $-0.39386599825+ 0.131432408 \ i $ & $[6.0\times 10^{-9}] $   & $ -0.7588309960+ 0.2532526950 \ i $ & $[3.1 \times 10^{-10}] $  \\ 
\end{tabular}
\caption{Results for the first $l=5$ modes for the RWZ master functions, i.e $\Psi(t_{\rm{max}},r)$ for both the interior and exterior solutions. We note the quantity in the square brackets is the numerical error computed as in Eq.~\eqref{numerical error} from converting the frequency domain data as given by \cite{thompson1811} to the time domain. }
\label{ch3g3_table_evalIntExtAllFields}
\end{ruledtabular} 
\end{table*}
\begin{table*}
\begin{ruledtabular}
\begin{tabular}{l|| c c ||c c }
\textrm{$(l,m)$}& 
\textrm{$\partial_{r}^{2} \Psi^{Int}(t_{max},r)$ \((M/\mu)^{2}\)}&
\textrm{$\eta$}&
\textrm{$\partial_{r}^{2} \Psi^{Ext}(t_{max},r)$\((M/\mu)^{2}\)}&
\textrm{$\eta$}\\
\colrule
(2,0) & $-0.06732365$ &  $[2.5\times  10^{-8}]$  & $-0.270077693$    & $[4.3\times 10^{-9}] $   \\
(2,1) & $0.01535267 - 0.06029044 \ i $  & $[5.3 \times 10^{-8}] $    & $-0.06577167 + 0.25777958 \ i $ & $[1.2 \times 10^{-8}] $  \\ 
(2,2) & $-0.04540006 - 0.02609435 \ i $ & $[3.8 \times 10^{-8}] $   & $-0.229499779 - 0.126538315\ i $   & $[6.0 \times 10^{-9}] $ \\  \hline 
(3,0) & $0.14535027$ &  $[ 6.7 \times  10^{-8}]$  & $-0.36781897$   & $[1.8\times 10^{-8}] $   \\
(3,1) & $0.0277920 - 0.1089657 \ i $  & $[1.4 \times 10^{-7}] $   & $0.07730855 - 0.30310879 \ i $  & $[2.7\times 10^{-8}] $  \\ 
(3,2) & $0.1057284 + 0.0576705 \ i $  & $[2.1 \times 10^{-7}] $   & $-0.281438950 - 0.15356627 \ i $   & $[5.0\times 10^{-8}] $ \\ 
(3,3) & $0.0668216 - 0.0714560 \ i $  & $[1.6\times 10^{-7}] $   & $0.21855359 - 0.23459612 \ i $   & $[9.4\times 10^{-8}] $  \\  \hline
(4,0) & $0.1652312$ &  $[1.8 \times  10^{-7}]$ & $0.36238101$   & $[2.6\times 10^{-8}] $    \\
(4,1) & $-0.0535780 + 0.2100671 \ i $   & $[3.0 \times 10^{-7}] $  & $0.1086065 - 0.4258212 \ i $    & $[1.3 \times 10^{-7}] $ \\ 
(4,2) & $0.1387490 + 0.0757010 \ i $ & $[2.5 \times 10^{-7}] $    & $0.3139990 + 0.17131660 \ i $   & $[1.6 \times 10^{-7}] $ \\ 
(4,3) & $-0.1176440 + 0.1264950 \ i $  & $[4.3 \times 10^{-7}] $  & $0.2490721- 0.2677930 \ i $  & $[2.2\times 10^{-7}] $  \\ 
(4,4) & $0.0808013+ 0.1256608 \ i $   & $[2.5\times 10^{-7}] $   & $0.2061707+ 0.3204449\ i $  & $[1.1\times 10^{-7}] $ \\  \hline 
(5,0) & $-0.2988317$ &  $[4.3 \times  10^{-7}]$ & $0.5264097$  &  $[1.1 \times 10^{-7}] $ \\
(5,1) & $ -0.0918953 + 0.3603005 \ i $ & $[3.5\times 10^{-7}] $   & $-0.1016144 + 0.3984065\ i $ & $[2.6 \times 10^{-7}] $   \\ 
(5,2) & $-0.2460227 - 0.1342289 \ i $  & $[3.2 \times 10^{-7}] $  & $0.4378832 + 0.2389070 \ i $   & $[2.9\times 10^{-7}] $  \\ 
(5,3) & $-0.1395838 + 0.15007829 \ i $  & $[2.9\times 10^{-7}] $   & $-0.2762416 + 0.2970110 \ i $ & $[2.2\times 10^{-7}] $  \\ 
(5,4) & $-0.1194673 - 0.1856122 \ i $    & $[5.8 \times 10^{-7}] $  & $0.2201230 + 0.34200318\ i $   & $[4.3 \times 10^{-7}] $ \\ 
(5,5) & $-0.1951357 + 0.0651164 \ i $ & $[3.5\times 10^{-7}] $  & $-0.4225279 + 0.1410168\ i $    & $[2.0\times 10^{-7}] $ \\  
\end{tabular}
\caption{Results for the first $l=5$ modes for the second-order radial derivative of the RWZ master functions, i.e $\partial_{r}^{2}\Psi(t_{\rm{max}},r)$ for both the interior and exterior solutions. We note the quantity in the square brackets is the numerical error computed as in Eq.\eqref{numerical error} from converting the frequency domain data as given by \cite{thompson1811} to the time domain.}
\label{ch3g3_table_evalIntExt_dr2psi}
\end{ruledtabular} 
\end{table*}

For completion we have decided to also compute the energy and angular momentum fluxes $\dot{E},\dot{L}$ at $\mathscr{I}^{+}$ with this particular choices of optimisation factors $i)-iv)$. From Table \ref{ch3g3_table_evalIntExt_dr2psi}, we have observed a drop in accuracy relative to the results in Section \ref{sectionivB}, which should not be interpreted as shortcoming of spectral methods, as demonstrated in previous work they are highly accurate \cite{jaramilloSopCan2011time, rpm_scalar_fd2202}, but it should be clear that computations at the \enquote{the infinities} and computations near the particle position require individual numerical calibrations. Numerical computational strategies that are optimal when directly evaluating observables near infinity and the horizon may not necessarily be optimal for evaluations at the particle's limit. These flux results still show a significant accuracy improvement to those in the literature, \cite{martel2004gravitational, sopuerta2006finite, field2009discontinuous, bernuzzi2011binary}, and we recommend further numerical studies specifically tackling measurement of these quantities at the outermost boundaries. 

\begin{table}
\begin{ruledtabular}
\begin{footnotesize}
\begin{tabular}{l||c c}
\textrm{$(l,m)$}& 
\textrm{$\dot{E}^{\infty}_{lm}$}&
\textrm{$\dot{L}^{\infty}_{lm}$}\\
\colrule
(2,2) & $1.70621955 \times 10^{-4}$  &$3.8214217 \times 10^{-3}$    \\
(2,1) &$8.163040 \times 10^{-7}$  &$1.8282769 \times 10^{-5} $ \\\hline
(3,3) &$2.547061 \times 10^{-5}$  & $5.704656 \times 10^{-4}$ \\
(3,2) &$ 2.519844\times 10^{-7}$  & $5.643699 \times 10^{-6}$ \\
(3,1) &$2.1730 \times 10^{-9}$   &$4.86693 \times 10^{-8}$ \\\hline
(4,4) &$4.72538 \times 10^{-6}$   &$1.058345 \times 10^{-4}$ \\
(4,3) & $5.7749 \times 10^{-8}$  &$1.29340\times 10^{-6}$ \\
(4,2) &$2.5090  \times 10^{-9}$ &$5.619\times 10^{-8}$  \\
(4,1) &$8.40 \times 10^{-13}$   &$1.880 \times 10^{-11}$ \\ \hline
(5,5) &$9.4560 \times 10^{-7}$  & $ 2.11785 \times 10^{-5}$  \\
(5,4) &$1.232\times 10^{-8}$   &$2.760 \times 10^{-7}$  \\
(5,3) &$ 1.093 \times 10^{-9}$  &$2.448\times 10^{-8}$   \\
(5,2) &$ 2.78 \times 10^{-12}$   &$6.24\times 10^{-11}$  \\
(5,1) &$1.0 \times 10^{-15}$  &$3.0\times 10^{-14}$  \\ \hline 
Total & $2.0290770 \times 10^{-4}$ & $4.5445256 \times 10^{-3} $ \\ \hline
$\eta$ & $3.8 \times 10^{-8}$ & $2.0 \times 10^{-8} $  
\end{tabular}
\caption{To complement Tables \ref{ch3g3_table_energyInfinity_timedomain}, \ref{ch3g3_table_L_Infinity_timedomain} and our numerical studies in this section we include our energy and angular momentum fluxes at infinity using $N=89$ Chebyshev collocation nodes as given in Eq.\eqref{chebyshev_lobatto_nodes} with the optimisation numerical factor choices $i)-iv)$ of Section \ref{sectionivC}. We further note units for the energy flux are given as $(M/\mu)^{2}$ and for the angular flux as $(M/\mu^{2})$. }
\label{ch3g3_table_energyInfinity_chebyshev}
\end{footnotesize}
\end{ruledtabular} 
\end{table}
Finally, we conclude our numerical algorithm is suitable for self-force computations regardless of the numerical domain regions of interest for the computation of physical quantities. 

\section{Summary and Conclusions}\label{conclusion}

In this work we present a new TD algorithm capable of solving a distributionally sourced PDE with competitive accuracy to FD methods. Previous implementations \cite{martel2004gravitational, sopuerta2006finite, field2009discontinuous, bernuzzi2011binary} in the TD for this problem have shown three main difficulties: $i)$ handling the Dirac-$\delta$ distributional source on the RHS of our problem; $ii)$ the outer radiation problem and $iii)$ developing a time-integration scheme that could handle an EMRI model with a long time-scale (months to years). In our work we show all these difficulties can be avoided by using a suitable novel numerical algorithm directly addressing them individually. 

There are two main results in this body of work: an accurate computation of TD energy and angular momentum fluxes, and an accurate computation of the master functions and their derivatives evaluated in the point-particle limit. The first of our numerical tests showed that highly accurate flux measurements are possible in the TD with competitive performance to FD codes relative to previous computations performed for the same numerical PDE problem in the Regge-Wheeler gauge \cite{martel2004gravitational, sopuerta2006finite, field2009discontinuous, bernuzzi2011binary} with both finite difference and pseudospectral collocation methods. Not only does our work show significant improvement for the first $l=5$ modes as has been previously tabulated in the literature, we also show it can retain accuracy for at least $l_{max} =20$ modes. 

Despite being of high physical relevance, flux measurements are not sufficient for a full self-force computation. As has been discussed there are four main computational strategies allowing the computation of both the conservative and dissipative components of the GSF. Flux balance laws are an incomplete strategy as they only allow for the computation of the dissipative component \cite{detweiler2008consequence, barack2007gravitational}. Other strategies like mode-sum require the use of the field and its derivatives evaluated at the particle limit. We thus included a second numerical test precisely showing the results for the field and its first higher order derivatives and include results that allow for benchmark against future codes for the first $l=5$ modes. 

It remains to discuss \textit{difficulty 4} concerning initial data choices. Following the discussion in \textit{Jaramillo~et~al} \cite{jaramilloSopCan2011time}, we too choose to handle it, by implementing \textit{zero} initial data values, which, given the nature of our algorithm, are passed on to the jump conditions in the evolution system correctly, and thus we did not expect to produce \textit{Jost} solutions. For radiation measurements at the outer boundaries, Section \ref{sectionivB}, we found we can confidently address \textit{concern I}, by identifying the presence of \textit{junk} radiation through aid of phase portraits describing the evolution of the system, as discussed in Section \ref{sectionivA} and the convergence studies in Section \ref{sectionivB}. From the latter, through the aid of Figs.\ref{optimalConditions_FD_timestep}-\ref{optimalConditions_FD_JUMPS} we found no reason that substantiates \textit{concern II}. 
Preliminary tests, to be included in future work, pertaining to Section \ref{sectionivC} suggest a more realistic approach is to perform convergence studies based on the reconstructed metric perturbations evaluated at the particle through comparison of the relative error against the FD results of \cite{thompson1811}.
This choice also allows us to establish a fairer comparison with another TD code computing the GSF in the Lorenz gauge for a point-particle on a circular orbit \cite{barack2007gravitational}, who also used the reconstructed metric perturbation values evaluated at the particle at different times to gauge whether there was any indication for the presence of $Jost$ solutions. So far, through the evaluation of the odd RWZ metric perturbations to be in \cite{paper3}, with the data showed in Tables \ref{ch3g3_table_evalIntExtAllFields}-\ref{ch3g3_table_evalIntExt_drpsi} we found it to be well-behaved and no further evidence to substantiate \textit{concern II}. Nevertheless, even though, from this paper and upcoming work \cite{paper2, paper3} it seems to be sufficiently accurate to use trivial ID, we further note it may be beneficial to re-evaluate the IVP at the light of our new numerical algorithms results. We find the results of \cite{o2021characteristic, dolan2023metric} promising and even if the accuracy remains unchanged, it may reduce extraction times, potentially, significantly reducing computational cost which will be of particular use when attempting a self-consistent evolution for example. 

This paper is the first of a series of upcoming papers studying circular geodesic motion in a Schwarzschild background. In upcoming work \cite{paper2} we will discuss an alternative hyperboloidal, time-domain algorithm for self force calculations based on the fully spectral scheme from ref.~\cite{PanossoMacedo:2014dnr}. Finally, in \cite{paper3} we will aim to corroborate the frequency domain results of \cite{thompson1811} and to show accurate \textit{gravitational self-force} computations are possible in the time-domain in the Regge-Wheeler gauge. 

For the gravitational work, we will use some of the results shown here, in particular we will compute the dissipative components from flux balance laws which have been tabulated in Table \ref{ch3g3_table_FD_FLUX_20MODES} and compare to the results we obtain though a local mode-sum approach. Even though this calculation is possible here, we prefer to include it once we obtained the full mode-sum results as done by \cite{barack2007gravitational}. We will compare all work to the implementation of \cite{thompson1811} but we will also include comparisons with \cite{durkan2022slowderivatives, barack2007gravitational}. A preliminary flux measurement as done in Table \ref{ch3g3_table_FD_FLUX_20MODES} shows an additional comparison to \cite{durkan2022slowderivatives} and comparable accuracy to what we obtained when comparing with \cite{thompson1811}. Even though our results show a significant improvement from Table IV of \cite{barack2007gravitational}, who computed their GSF results for an $l_{max} = 9$ modes, we can only provide a fair comparison when both flux and mode-sum strategy has been showed. Furthermore, it is likely that the numerical optimisation factors may vary to the choices we made here. 

Another important result of this work is that calibration should be performed on the highest order derivative of the field quantities under study, which will be clear when we lay out the full hyperboloidal black hole perturbation machinery needed \cite{paper3}. 

\begin{acknowledgments}
This work makes use of the Black Hole Perturbation Toolkit \cite{BHPToolkit}. LGdS would like to thank Andrew Spiers, Adam Pound and Leor Barack for invitation to give a seminar at Southampton and the invaluable discussion which helped identify some of the underlying implementation of this work pertaining to section \ref{sectioniiiC} and \ref{sectioniiiD}. We thank Josh Mathews, Niels Warburton, the Capra $26$'s and \enquote{Infinity on a Gridshell} participants for their valuable comments and suggestions. Finally, we thank the anonymous referee for their invaluable comments and suggestions, particularly the inclusion of difficulty 4. RPM acknowledges  the financial support provided partially by STFC via grant number ST/V000551/1, and partially by the VILLUM Foundation (grant no. VIL37766), the DNRF Chair program (grant no. DNRF162) by the Danish National Research Foundation, and the European Union’s H2020 ERC Advanced Grant “Black holes: gravitational engines of discovery” grant agreement no. Gravitas–101052587. Views and opinions expressed are however those of the author only and do not necessarily reflect those of the European Union or the European Research Council. Neither the European Union nor the granting authority can be held responsible for them. This work was supported by the NASA LISA Preparatory Science grant 20-LPS20-0005.
\end{acknowledgments}

\appendix

\section{Weak Form Solution of the RWZ master functions}\label{appA}
In this section we will for the first time introduce the full derivation of higher jumps for an equation of the type Eq.~\eqref{cha2_generalwave} in $(t,r)$ coordinates. Even though we will be solving Eq.~\eqref{ch3_rwz_fieldvariables_hyper} in the mapped hyperboloidal coordinates $(\tau,\sigma)$, it is important for corroboration with previous work as given in \cite{sopuerta2006finite, field2009discontinuous, hopper2010gravitational} to check that our generic algorithm returns the same jumps. Furthermore, one could also follow our numerical strategy and instead of working in hyperboloidal coordinates stay in the physical coordinates and use radiation boundary conditions. \\

For our derivation we transform the d'Alembert operator $\square \Psi$ back to its $(t,r)$ form, reverting Eq.~\eqref{ch3_general_pdes} into a PDE of the form 
\begin{equation}
    \bigg[-\frac{\partial^{2} }{\partial t^{2}} +  \eta(r) \frac{\partial^{2} }{\partial r^{2}} +  \varrho(r) \frac{\partial}{\partial r} - V_{l}(r)\bigg] \Psi_{lm}(t,r) = S_{lm}(t,r) \ \ \ \ 
    \label{aa_ugly_pde}
\end{equation}
where $\eta(r) = f^{2}$, $\rho(r) = ff'$.
At the end of deriving the recurrence relation for $J_{m+2}(t)$, we will then revert back to tortoise coordinate form. Our strategy is slightly different due to the nature of our algorithm but the final product is trivially the same as has been previously showed in the literature \cite{sopuerta2006finite, field2009discontinuous, hopper2010gravitational}. 

\subsection{Jump condition derivation through the Frobenius Method}\label{appA_frobs}

Our numerical algorithm can be seen as an application of the Frobenius method \cite{weisstein2002frobenius, frobenius1873} for the case where we have a function in \textit{weak form}. We will therefore revisit it briefly here. The Frobenius method states if $x_{0}$ is an ordinary point of an ODE describing a smooth function $\Phi(x)$ we can expand it in a Taylor series about $x_{0}$. We obtain the Maclaurin series by taking the expansion point as $x_{0} = 0$, resulting in
\begin{equation}
    \Phi =\sum_{n=0}^{\infty} a_{n}x^{n}.
    \label{maclaurinseries}
\end{equation}
We plug $\Phi$ back in the ODE and group the coefficients by power. We then obtain a recurrence relation for the $nth$ term and write the series expansion in terms of the $a_{n}$ coefficients. 
It's relevant to the calculations that follow to highlight the first two derivatives, for convenience we then list: 
\begin{eqnarray}
 \Phi &&=\sum_{n=0}^{\infty} a_{n}x^{n}. \label{frobenius_0}\\
 \Phi' &&= \partial_{x}\bigg( \sum_{n=0}^{\infty} a_{n}x^{n}\bigg),\\
  &&=\sum_{n=1}^{\infty} n a_{n}x^{n-1},\\
 &&=\sum_{n=0}^{\infty} (n+1) a_{n+1}x^{n}.\\
  \Phi'' &&= \partial_{x}^{2} \bigg(\sum_{n=0}^{\infty} a_{n}x^{n}\bigg),\\
   &&=\sum_{n=2}^{\infty} n(n-1) a_{n}x^{n-2},\\
 &&=\sum_{n=0}^{\infty}(n+2)(n+1) a_{n+2}x^{n}. \label{frobenius_der2} 
\end{eqnarray}

\subsubsection{Unit Jump relations}\label{appA1a}
As given in Section \ref{sectioniic}, the RWZ master functions assume a \textit{weak-form} as described by Eq.~\eqref{ch3_weakformsolutionRWZ}. We can also write the ansatz: 
\begin{equation}
    \Psi_{lm}(t,r) = \Psi^{S}_{lm}(t,r) + \Psi^{J}_{lm}(t,r),
    \label{AppA_WFansatz}
\end{equation}
where $\Psi^{S}_{lm}(t,r)$ like $\Psi^{+/-}(t,r)$ functions in Eq.~\eqref{ch3_weakformsolutionRWZ} are smooth functions satisfying everywhere the homogeneous PDE, 
\begin{equation}
     -\frac{\partial^{2} \Psi^{S}}{\partial t^{2}} +  \eta(r) \frac{\partial^{2} \Psi^{S}}{\partial r^{2}} +  \varrho(r) \frac{\partial \Psi^{S}}{\partial r} - V_{l}(r) \Psi^{S} = 0 ,
    \label{AppA_SmoothPart}
\end{equation}
and $\Psi^{J}_{lm}(t,r)$ describe the inhomogeneous part of the master function in an infinite series as with the Frobenius method. We then write, 
\begin{equation}
    \label{wave_eq_jumps1}
    \Psi^{J}(r,t;r_{p}(t)) = \sum^{\infty}_{n=0} J_{n}(t) \Psi_{n}(r;r_{p}(t)).
\end{equation}
 where \(\Psi_n\) are piecewise monomials centred at the discontinuity $r = r_{p}(t)$:
\begin{small}
\begin{eqnarray}
    \Psi_{n}\big(r,t;r_{p}(t)\big) &=&  \Psi_{n}\big(r - r_{p}(t)\big) = \Psi_{n},\nonumber \\
     &=& \frac{1}{2} \text{sgn} \big(r-r_{p}(t)\big)\frac{\big(r-r_{p}(t)\big)^{n}}{n!},
     \label{aa_sigmum_piecewisemonos}
\end{eqnarray}
\end{small}
here the $\text{sgn}$ is the signum function. The expansion in Eq.~\eqref{wave_eq_jumps1} allows one to compute the discontinuities
\begin{small}
\begin{eqnarray}
    \label{disco1}
\frac{\partial^{m} \Psi}{\partial{r}^{m}} \bigg\vert_{r =r_{p}^{+}} - \frac{\partial^{m} \Psi}{\partial{r}^{m}} \bigg\vert_{r = r_{p}^{-}}  &=& \frac{\partial^{m} \Psi^{J}}{\partial{r}^{m}} \bigg\vert_{r= r_{p}^{+}}  - \frac{\partial^{m} \Psi^{J}}{\partial{r}^{m}} \bigg\vert_{r= r_{p}^{-}}.\nonumber \\ 
&=& J_{m}(t)   
\end{eqnarray}
\end{small}

These functions have the property that all spatial derivatives have matching left and right limits at $r= r_{p}(t)$
\begin{eqnarray}
    \label{disco2}
\frac{\partial^{m} \Psi_{n}}{\partial{r}^{m}} \bigg\vert_{r = r_{p}^{+}} - \frac{\partial^{m} \Psi_{n}}{\partial{r}^{m}} \bigg\vert_{r = r_p^{-}} = \delta_{m,n},   
\end{eqnarray}
except at the $nth$ derivative, where one has a jump of one. 
\begin{equation}
    \label{disco3}
\frac{\partial^{n} \Psi_{n}}{\partial{r}^{n}} \bigg\vert_{r = r_{p}^{+}} - \frac{\partial^{n} \Psi_{n}}{\partial{r}^{n}} \bigg\vert_{r = r_{p}^{-}} = 1.
\end{equation}

By differentiating the monomials we obtain 

\begin{equation}\frac{\partial \Psi_n}{\partial r} = \left\{ 
\begin{array}{ccc}
\Psi_{n-1} &\text{for}& n\geq 1 \\
\delta(w) &\text{for}& n=0 
\end{array}
\right. .
\label{mondelta}
\end{equation}

\begin{equation}\frac{\partial^{2} \Psi_n}{\partial r^{2}} = \left\{ 
\begin{array}{ccc}
\Psi_{n-2} &\text{for}& n\geq 2 \\
\delta(w) &\text{for}& n=1 \\
\delta'(w) &\text{for}& n=0 
\end{array}
\right. .
\label{mondelta2}
\end{equation}

We note that here for simplicity we introduced,  $w = (r - r_{p}(t))$. We also want to credit that the idea to use unit jump functions to compute higher order jumps was given by \cite{pablosNotes} who initially demonstrated it for the wave equation sourced by a Dirac $\delta$ distribution as showed in \cite{2014arXiv1406.4865M}. 

\subsubsection{Spatial Jumps}\label{appA1b}
The inhomogeneous part of the solution is given by Eq.~\eqref{wave_eq_jumps1}. To determine the jumps in Eq.~\eqref{aa_ugly_pde}, it is useful to find the spatial and temporal derivatives of the unit jump functions. These derivations are an application of the Frobenius method. 
More explicitly, we have for the first order derivative of Eq.~\eqref{wave_eq_jumps1}:
\begin{eqnarray}
 \partial_{r}\Psi^{J}(w) &=&  \partial_{r} \bigg[\sum^{\infty}_{n = 0}\bigg(J_{n}(t) \Psi_{n}(w)\bigg) \bigg],\\
    &=& \sum^{\infty}_{n=1}\bigg(J_{n}(t) \Psi_{n-1}(w)\bigg),\\
     &=& \sum^{\infty}_{n=0}\bigg(J_{n+1}(t) \Psi_{n}(w) \bigg) + J_{0}(t)\delta(w).  \ \ \ \ \label{spatial_o1_fb} \ \ 
\end{eqnarray}
where we've used the selection property of Dirac-$\delta$ as given by Eq.~\eqref{appendix_a_dirac_selectionI} to ensure our problem is fully with respect to the the particle's worldline, $r_{p}(t)$. For second order we then have,  
\begin{eqnarray}
    \partial_{r}^{2} \Psi(w) &=& \partial_{r}^{2} \bigg[\sum^{\infty}_{n = 0}\bigg(J_{n}(t) \Psi_{n}(w)\bigg) \bigg], \nonumber\\
    &=& \sum^{\infty}_{n=0} \bigg[J_{n+2}(t) \Psi_{n}(w)  \bigg]
    + J_{1}(t)\delta(z) + J_{0}(t)\delta'(w).  \nonumber \\
    \label{spatial_o2_fb}
\end{eqnarray}
Given the coefficients present in the differential operators in Eq.~\eqref{aa_ugly_pde} and the potential term it is important to revisit Eq.~\eqref{aa_sigmum_piecewisemonos}. One of it's key properties is 
\begin{eqnarray}
    \text{sgn} = \frac{x}{|x|} = \frac{|x|}{x}.
    \label{sigmum_property}
\end{eqnarray}
Using this we can re-write Eq.~\eqref{aa_sigmum_piecewisemonos} as 
\begin{eqnarray}
\Psi_{n} &=& \frac{1}{2} \frac{|w|}{w} \frac{w^{n}}{n!},\label{wave_eq_jumps3} \\ 
  & = &\frac{1}{2n!} (w)^{n-1} |w|,\label{wave_eqs_jumps4} \\
 |w| \Psi_{n} & = &\frac{1}{2n!} |w| w^{n}. \label{wave_eqs_jumps5}
\end{eqnarray}
One can then generalise this for $\Psi_{n + m}$: 

\begin{eqnarray}
\Psi_{n+m}(w) &=& \frac{1}{(n+m)!} \big|w\big| \big(w\big)^{n + m - 1}, \label{wave_eqs_jumps6}\\
\Psi_{n+m}(w)(n+m)! &=&\frac{\big|w\big| \big(w\big)^{n+m-1}}{2}.   \ \ \ \label{wave_eqs_jumps7}
\end{eqnarray}

Substituting for the term $|w| \ w^{n}/2$ in the RHS of this equation by the LHS of Eq.~\eqref{wave_eqs_jumps5} multiplied by $n!$ we get, 
\begin{eqnarray}
    n! \ (w) \Psi_{n} &=& (w)^{1-m} (n+m)! \Psi_{n+m},  \nonumber \\ 
    (w)(w)^{m-1} \Psi_{n} &=& \frac{(m+n)!}{n!} \Psi_{n+m}, \nonumber \\ 
    (w)^{m} \Psi_{n} &=& \frac{(m+n)!}{n!} \Psi_{n+m}.
    \label{piecewise_mon_generic}
\end{eqnarray}
We want to be able to expand any coefficient in $r$ acting on our master functions $\Psi^{a/p}_{lm}(t,r)$ and its differential operators as given by Eq.~\eqref{aa_ugly_pde}. As an example we consider the potential term $V_{l}(r)$ acting on the master function, we can expand it around $r=r_{p}$ 
\begin{equation}
    V_{l}(r) = \sum^{\infty}_{k=0} V^{(k)}_{l}(r_{p}) \frac{(r-r_{p})^{k}}{k!} 
\end{equation}
where $k \leq n$ and we compute the product $V(r) \Psi^{J}(r;r_{p})$ following the ansatz of Eq.~\eqref{wave_eq_jumps1}, 

\begin{eqnarray}
   \nonumber V_{l}(r) \Psi^{J}(r;r_{p})&=&  \bigg[ \sum^{\infty}_{k=0} V^{(k)}_{l} (r_{p}) \frac{(r-r_{p})^{k}}{k!} \bigg] \\
    && \times\bigg[ \sum^{\infty}_{n=0}J_{n}(t) \Psi_{n}(r;r_{p}) \bigg]. \ \ 
    \label{proof_coefficients_1}
\end{eqnarray}
Furthermore, given $k \leq n $ we can define $n = m-k$ where at $n=0$, $m=k$, and the preceding equation simplifies to, 
\begin{eqnarray}
    V_{l}(r) \Psi^{J}(r;r_{p})= \nonumber \\
    \sum^{\infty}_{k=0} \sum^{\infty}_{m=k} V_{l}^{(k)}(r_{p}) \frac{(r-r_{p})^{k}}{k!} J_{m-k}(t) \Psi_{m-k}(r;r_{p}). \ \ \ 
    \label{proof_coefficients_2}
\end{eqnarray}
From Eq.~(\ref{piecewise_mon_generic}) with $m=k$ and $n=m-k$ we have, 
\begin{eqnarray}
    (r-r_{p})^{k} \Psi_{m-k} = \frac{(m-k+k)!}{(m-k)!} \Psi_{m-k+k}, \nonumber \\
    (r-r_{p})^{k} \Psi_{m-k} =  \frac{m!}{(m-k)!} \Psi_{m}.
    \label{monos_generilisationtok}
\end{eqnarray}
Eq.~\eqref{proof_coefficients_2} then simplifies to, 
\begin{small}
\begin{eqnarray}
   V_{l}(r) \Psi^{J}(r;r_{p}) = \sum^{\infty}_{m=0} \sum^{m}_{k=0} V_{l}^{(k)}(r_{p}) \frac{m!}{(m-k)!k!} J_{m-k}(t) \Psi_{m}(r;r_{p}), \nonumber  \\ 
     = \sum^{\infty}_{m=0} \bigg[ \sum^{m}_{k=0}  {m\choose k}  V_{l}^{(k)}(r_{p}) J_{m-k}(t) \bigg] \Psi_{m}(r;r_{p}). \ \ \ \nonumber \\
    \label{final_proof_coefficents_handling}
\end{eqnarray}
\end{small}
where to go from the first to the second line we adapt the factorial notation. 
Furthermore we can extend this to the coefficients of the first order derivative of the master functions as given by Eq.~\eqref{spatial_o1_fb}
\begin{eqnarray}
    \nonumber     \varrho(r) \partial_{r} \Psi(w) &=& 
        \sum^{\infty}_{m=0} \bigg[\sum^{m}_{k=0}  \varrho^{(k)}(r_{p}) {m\choose k} J_{n+1-k}(t)\Psi_{m}(w) \bigg]  \\  
        && + \varrho(r_{p}) J_{0}(t)\delta(w). 
        \label{corrected_firstorderdiff}
\end{eqnarray}
Furthermore for the second-order jump we have, Eq.~\eqref{spatial_o2_fb}, 
\begin{eqnarray}
   \eta(r) \partial_{r}^{2} \Psi(w) &=& \sum^{\infty}_{m=0} \bigg[\sum^{m}_{k=0} \eta^{(k)}(r_{p}) {m\choose k} J_{m+2-k}(t)\Psi_{m}(w) \bigg] \nonumber \\  
        &+& \eta(r) J_{1}(t)\delta(w) \nonumber \\
        &+& \eta(r) J_{0}(t)\delta'(w), \\
   &=& \sum^{\infty}_{m=0} \bigg[\sum^{m}_{k=0} \eta^{(k)}(r_{p}) {m\choose k} J_{m+2-k}(t)\Psi_{m}(w) \bigg] \nonumber \\  
        &+& \eta(r_{p}) J_{1}(t)\delta(w) + \eta(r_{p}) J_{0}(t)\delta'(w) \nonumber \\
        &-& 2 \varrho(r_{p})J_{0}(t)\delta(w),  \nonumber \\
        \label{corrected_secondorderdiff}
\end{eqnarray}
We note for both derivations we we applied selection properties as given by Eqs.~(\eqref{appendix_a_dirac_selectionI}, ~\eqref{appendix_a_dirac_selectionII}) on the coefficient of the Dirac delta distribution and its $r$-derivative respectively, to go from the second to the third line. 

\subsubsection{Temporal Jumps}\label{appA1b}
The first order temporal derivative and associated jumps of the function $\Psi^{J}(r,t;r_{p})$ are : 
\begin{eqnarray}
     \partial_{t}\Psi^{J}(w) &=& \partial_{t} \sum^{\infty}_{m = 0}\bigg[J_{m}(t) \Psi_{m}(w)\bigg], \nonumber \\
     &=& \sum^{\infty}_{m=0} \bigg[ \dot{J}_{m}(t) - \dot{r}_{p}  J_{n+1}(t) \bigg]\Psi_{n}(w) \nonumber \\
     &-& \dot{r_{p}} J_{0}(t)\delta(r-r_{p}).  \\
\end{eqnarray}

The second order derivative can be obtained by calculating: 
\begin{eqnarray}
    \label{temporal_o2_ansa}
    \partial_{t}^{2} \Psi^{J}(w) &=& \partial_{t} \bigg[\sum^{\infty}_{m=0} \dot{J}_{m}(t) \Psi_{m}(w) \bigg] \nonumber \\
    &-& \partial_{t} \bigg[ \sum^{\infty}_{m=0} \dot{r_{p}} J_{m+1}(t) \Psi(w)\bigg] \nonumber \\
    &-& \partial_{t}\bigg[\dot{r_{p}} J_{0}\delta(w)\bigg].
\end{eqnarray}

For simplicity and to clarify one will be showing the  explicit calculation this derivatives as $D1,D2,D3$ in the work that follows: 

\begin{small}
\begin{eqnarray}
   D1 &=& \partial_{t} \bigg[ \sum^{\infty}_{m=0} \dot{J}_{n}(t) \Psi_{m} \bigg], \nonumber  \\
      &=&  \sum^{\infty}_{m=0}\bigg[ \ddot{J}_{m}(t)  - \dot{r}_{p}\dot{J}_{m+1} \bigg] \Psi_{m}(w)  - \dot{r_{p}} \dot{J}_{0} \delta (w).  
     \label{temporal_D1}
\end{eqnarray}
\end{small}

\begin{eqnarray}
   D2 &=& - \partial_{t} \bigg[ \sum^{\infty}_{m=0} J_{n+1}(t)\dot{r}_{p} \Psi(w) \bigg], \nonumber \\
      &=& \sum^{\infty}_{m=0} \bigg( -\dot{r}_{p} \dot{J}_{m+1}(t) - \ddot{r}_{p} J_{m+1}(t) \nonumber \\
      &+& \dot{r}_{p}^{2} J_{m+2}(t) \bigg)\Psi_{m}(w) \nonumber \\
      &+& \dot{r}_{p}^{2} J_{1}(t) \delta(w). 
     \label{temporal_D2}
\end{eqnarray}

\begin{eqnarray}
   D3  &=& - \bigg[ \bigg(\dot{r}_{p} \dot{J}_{0} + \ddot{r}_{p} J_{0}(t)\bigg) \delta(w) - \dot{r}_{p}^{2} J_{0} \delta'(w) \bigg], \nonumber\\ 
    &=& r_{p}^{2} J_{0}(t)\delta'(w) - \dot{r}_{p}\dot{J}_{0}(t) \delta(w) \nonumber \\
    &-& \ddot{J}_{0}(t) \delta(w).
     \label{temporal_D3}
\end{eqnarray}

Plugging this results into Eq.~\eqref{temporal_o2_ansa} we obtain: 

\begin{eqnarray}
  \partial^{2}_{t}\Psi^{J}(w)  &=& \sum^{\infty}_{n=0} \bigg[ \ddot{J}_{n}(t) - 2\dot{r}_{p} \dot{J}_{n+1}(t) \nonumber \\
  &-& \ddot{r}_{p} J_{n+1}(t) + \dot{r}_{p}^{2} J_{n+2}(t) \bigg]\psi_{n}(w) \nonumber \\
    &+&  \dot{r}_{p}^{2} J_{1}(t) \delta(w)  - 2\dot{r}_{p}\dot{J}_{0}(t) \delta(w) \nonumber \\
    &-& \ddot{J}_{0}(t) \delta(w) +  r_{p}^{2} J_{0}(t)\delta'(w) \ \ \ \ \ 
   \label{temporal_fb_final}
\end{eqnarray}

Matching coefficients in $\delta'(w)$ with respect to the particle worldline gives our first initialising jump, 
\begin{equation}
    J_{0}(t) = \frac{F^{a/p}_{lm}(t,r_{p})}{\big(f^{2}_{p} - \dot{r}_{p}^{2}\big) }.
\end{equation}
Similarly for our second and final initialising jump we match the coefficients of the Dirac $\delta(z)$, 
\begin{eqnarray}
    J_{1}(t) &=&  - 2 \dot{r}_{p} \partial_{t}J_{0}(t) - \big(\ddot{r}_{p} - f_{p}f'_{p} \big) J_{0}(t) \nonumber \\
    &+& \big[ G^{a/p}_{lm}(t,r_{p}) - \partial_{r_{p}}F^{a/p}_{lm}(t,r_{p}) \big] / \big(f^{2}_{p} - \dot{r}_{p}^{2}\big).  \ \ \  \ \ 
\end{eqnarray}

These jumps match the documented jumps in the literature \cite{sopuerta2006finite, field2009discontinuous, hopper2010gravitational} without the $f$ factor and equations given in Eq.~(\ref{ch3_j0}, \ref{ch3_j1}) which match and follow the approach of \cite{hopper2010gravitational}. 

We could then to calculate the higher order jumps collect all remainder terms by solving for $J_{m+2}(t)$. However given we will be solving the RWZ master functions in the hyperboloidal coordinates we will revisit and give the recurrence relation in Appendix \ref{AppC_DiscSpace_Jumps}.

\begin{eqnarray}
    J_{m+2}(t) &=&  -\frac{1}{(f^{2}_{p}- \dot{r}^{2}_{p})} 
      \sum^{\infty}_{m=0}
    \bigg( \ddot{J}_{m}(t) \nonumber \\
    &+&  \sum^{m}_{k=0} {m \choose k} \bigg[ - \rho^{(k)}(r_{p}) J_{m+1-k}(t) \nonumber \\
    &+& V^{(k)}(r_{p}) J_{m-k}(t) \bigg]   \nonumber \\
&-& \sum^{m}_{k=1} {m \choose k} \eta^{(k)}(r_{p}) J_{m+2-k}(t)    \bigg).
\label{cha3_rec_relation_mjumps_PhysicalChart_RWZ}
\end{eqnarray}

\subsection{Dirac delta distribution proprieties}\label{appA2_DDDsProps}
In this work we have used the following selection proprieties 
\begin{eqnarray}
\label{appendix_a_dirac_selectionI}
f(a)\delta(a - b) =  f(b)\delta(b-a), \\
f(a)\delta'(a - b) = f(b)\delta'(a-b) - f(b)\delta(b-a). 
\label{appendix_a_dirac_selectionII}
\end{eqnarray}
Furthermore, the following composition rules were also needed: 
\begin{eqnarray}
\label{appendix_a_dirac_compI}
\delta(f(a)) = \frac{1}{|f'(b)|} \delta(a - b), \ \ \  \\
\delta'(f(a)) = \frac{f'(b)}{|f'(b)|^{3}} \delta'(a - b) + \frac{f''(b)}{|f'(b)|^{3}} \delta(a - b). \ \ \ 
\label{appendix_a_dirac_compositionII}
\end{eqnarray}
where the last distribution was brought to our attention from Appendix F of \cite{mathews2022self}.

\section{Black Hole Perturbation Theory in the Regge-Wheeler Gauge}\label{AppB_BHPT_RWG}
We summarise the source-term formalism presented in~\cite{thompson1611, thompson1811}. 

\subsection{Stress-energy projections}\label{AppB_projec}
Projecting Eq.~\eqref{cha2_sem_pointparticle} onto the tensor harmonic basis outlined in~\cite{thompson1611, thompson1811} on a circular orbit gives the source terms. We present here only those relevant to circular orbits, 

$l \geq 0 $, 
\begin{eqnarray}
  E^{lm}_{A} &=& -16 \pi \frac{\mu f\mathcal{E} }{r^{2}} \delta(r - r_{p}) Y^{*}_{lm} \bigg(\frac{\pi}{2}, \phi_{p}(t)\bigg), 
  \label{Ea}
\end{eqnarray}

$l \geq 1 $, 
\begin{eqnarray}
  E^{lm}_{C} &=& - \frac{16 \pi}{l(l+1)} \frac{\mu \mathcal{L} }{r^{3}}  \delta(r - r_{p}) \partial_{\theta}Y^{*}_{lm} \bigg(\frac{\pi}{2}, \phi_{p}(t)\bigg), \;\;\;\;
  \label{Ec}
\end{eqnarray}

$l \geq 2 $, 
\begin{eqnarray}
E^{lm}_{F} &=& -16 \pi \frac{(l-2)!}{(l+2)!}\frac{\mu f \mathcal{L}^{2}}{r^{4} \mathcal{E}} \delta(r - r_{p})  \nonumber  \\   &&\times [l(l+1) - 2m^{2}]Y^{*}_{lm}\bigg(\frac{\pi}{2}, \phi_{p}(t)\bigg).
  \label{Ef}
\end{eqnarray}

\subsection{Axial Parity Source Terms}\label{AppB_sourceterms_Axial}
The source term for the master axial function is, 
\begin{equation}
   S^{a}_{lm}(t,r)  = f (r^{2} \partial_{r} E_{C} + r E_{C} + r^{2} \partial_{t} E_{J}). 
\label{axial_s}
\end{equation}
To further relate to the form of Eq.~\eqref{cha2_sourceterm_wrt_partworldline} we re-formulate the source term as given by a combination of $\bar{G}^{a}_{lm}(t)$ and $\bar{F}^{a}_{lm}(t)$ functions, 
\begin{equation}
    S^{a}_{lm}(t,r) = \bar{F}^{a}_{lm}(t)\delta'(r-r_{p}) + \bar{G}^{a}_{lm}(t)\delta(r-r_{p}).
\end{equation}
For circular orbits Eq.~\eqref{axial_s} simplifies further as $E_{J} = 0$. We then must carefully handle the Dirac delta distribution present in $E_{C}$. 
We rewrite $E_{C}$ as combination of smooth functions in $r$ and $t$, $\mathfrak{C}_{lm}(r,t)$ and its Dirac delta distribution
\begin{equation}
    E_{C} = \mathfrak{C}_{lm}(r,t)\delta(r-r_{p}),
\end{equation}
where 
\begin{equation}
    \mathfrak{C}_{lm}(r,t) = c_{l}(r) \partial_{\theta}Y^{*}_{lm}\big(\frac{\pi}{2}, \phi_{p}(t)\big), 
\end{equation}
and from Eq.~\eqref{Ec} it follows, 
\begin{equation}
    c_{l}(r) = -\frac{16 \pi}{l(l+1)} \frac{f\mathcal{L} }{r^{3}}. 
\end{equation}

We then carefully take the partial derivative \textit{with respect to r}, 
\begin{eqnarray}
     \partial_{r}(\mathfrak{C}_{lm}(t,r) \delta(r - r_{p})) = \nonumber\\
     \partial_{r}(\mathfrak{C}_{lm}(t,r))\delta(r-r
     _{p}) + \mathfrak{C}_{lm}(t,r) \partial_{r}(\delta(r-r
     _{p})). \ \ 
    \label{axial_rule}
\end{eqnarray}
Finally, to get our source terms as a function of $t$ only and hereby ensuring our equation is fully with respect to the particle worldine, we use the selection property in Eq.~\eqref{appendix_a_dirac_selectionI}, ~\eqref{appendix_a_dirac_selectionII} to get, 
\begin{eqnarray}
   \bar{F}^{a}_{lm}(t)= f r^{2} \mathfrak{C}_{lm}(t,r)\bigg|_{r=r_{p}}, \ \ \  \\
   \label{axial_F}
   \bar{G}^{a}_{lm}(t) = \bigg[ f r^{2} \big(\partial_{r} \mathfrak{C}_{lm}(t,r)\big) \nonumber \\
   + f r \ \mathfrak{C}_{lm}(t,r) - \partial_{r} \big(f r^{2} \mathfrak{C}_{lm}(t,r)\big) \bigg]\bigg|_{r=r_{p}}.
\end{eqnarray}

\subsection{Polar Parity Source Terms}\label{AppB_sourceterms_Polar}
Similarly we have the simplified source term for the polar case, 
\begin{eqnarray}
  S^{p}_{lm}(t,r)  = \nonumber \\  - \frac{r f}{2 \kappa} \bigg[ - \frac{r[\lambda(\lambda - 2)r^{2}] + 2 M r (7 \lambda - 18) + 96 M^{2} }{2 r f \kappa} 
    \bigg]E_{A} \nonumber  \\ 
    + \  r^{2}\partial_{r} E_{A} 
  - \  (\lambda +2) \frac{\kappa}{2} E_{F},  \ \ \ \ \ \ \ \
\end{eqnarray}
which we then re-write as 
\begin{equation}
    S^{p}_{lm}(t,r) = \bar{F}^{p}_{lm}(t)\delta'(r-r_{p}) + \bar{G}^{p}_{lm}(t)\delta(r-r_{p}),
    \label{s_polar}
\end{equation}
as done for the axial case but now as a combination of $\mathfrak{A}_{lm}(t,r)$ and $\mathfrak{F}_{lm}(t,r)$ functions. 

\begin{eqnarray}
    E_{A} &=& \mathfrak{A}_{lm}(t,r) \delta(r-r_{p}),\\
    E_{F} &=& \mathfrak{F}_{lm}(t,r) \delta(r-r_{p})
\end{eqnarray}
where, 
\begin{eqnarray}
  \mathfrak{A}_{lm}(t,r) &=& a_{l}(r) Y^{*}_{lm}\bigg(\frac{\pi}{2}, \phi_{p}(t)\bigg), \ \ \ \\
  \mathfrak{F}_{lm}(t,r) &=& f_{l}(r) [l(l+1) - 2m^{2}] Y^{*}_{lm}\bigg(\frac{\pi}{2}, \phi_{p}(t)\bigg),  \ \   \ \ \ 
\end{eqnarray}

where $\lambda = (l-1)(l+2)$, $\kappa = \lambda r + 6 M$ and from Eqs.~(\eqref{Ea}, \eqref{Ef})

\begin{eqnarray}
    a_{l}(r) &=&  -16 \pi \frac{ \mu f \mathcal{E}}{r^{2}},\\
    f_{l}(r) &=& -16 \pi \frac{(l-2)!}{(l+2)!}\frac{\mu f \mathcal{L}^{2}}{r^{4} \mathcal{E}}. 
\end{eqnarray}

Like in the axial case we carefully handle the distributional source terms by correctly applying he selection proprieties and then evaluating at the particle worldine such that, 
\begin{eqnarray}
    \bar{F}^{p}_{lm}(t) &=& -  \frac{f r^{3}}{2 \kappa } \mathfrak{A}_{lm}(t,r)\bigg|_{r=r_{p}} \\
    \bar{G}^{p}_{lm}(t)&=& \bigg[ \mathfrak{A}_{1,lm}(t,r) + \mathfrak{A}_{2,lm}(t,r)  \nonumber \\
    &+& \frac{r f (\lambda + 2)}{4}  \mathfrak{F}_{lm}(t,r) \nonumber \\
    &-& \partial_{r}\bigg(-  \frac{f r^{3}}{2 \kappa } \mathfrak{A}_{lm}(t,r) \bigg) \bigg]  \bigg|_{r=r_{p}}, 
    \label{polar_f}
\end{eqnarray}

where, 
\begin{small}
\begin{eqnarray}
    \mathfrak{A}_{1,lm}(t,r) &=& \frac{r f}{2 \kappa}  \nonumber \\ 
    &\times&\bigg( \frac{r(\lambda (\lambda - 2)r^{2} + 2 r(7 \lambda -18) +96   ) }{2 r f \kappa}\bigg)\nonumber \\
    &\times& \mathfrak{A}_{lm}(t,r), \ \ \ \  \ \ \ \ \ \ \ \\
    \mathfrak{A}_{2,lm}(t,r) &=& -  \frac{r f}{2 \kappa} \bigg( r^{2} \partial_{r}(\mathfrak{A}_{lm}(t,r)) \bigg).
\end{eqnarray}
\end{small}

\section{Revisiting the weak field solution of the RWZ master functions in hyperboloidal coordinates}\label{AppC}
We complement our work in Section \ref{sectioniii} by including some of the explicit results necessary for the implementation of our numerical algorithm. 
\subsection{Discontinuous spatial discretisation}\label{AppC_DiscSpace}
\subsubsection{Discontinuous spatial discretisation associated source terms}\label{AppC_DiscSpace_SourceTerms}
Following Eqs.~(\eqref{ch3_generalhyperPDE}, \eqref{cha2_spatial_disc_generic}-\eqref{ch3_sb2_SpaceSpource}) we explicitly give the discretized form of all our differential operators corrected with our discontinuous collocation algorithm, 
\begin{small}
\begin{eqnarray}
  \tilde{\upchi}(\sigma) \partial_{\sigma}^{2}\Psi\bigg|_{\sigma=\sigma_{i}} = \nonumber \\ 
  \sum^{N}_{j=0} \bigg( \text{diag}(\tilde{\upchi}(\sigma_{i})) \cdot D^{(2)} \bigg)_{ij} \big[ \Psi_{j} + \Delta_{\Psi}(\sigma_{j}-\sigma_{p};\sigma_{i} - \sigma_{p}) \big], \nonumber   \\
  \label{chid2_discretization}
\end{eqnarray} 
\end{small}
\begin{small}
\begin{eqnarray}
  \tilde{\iota}(\sigma) \partial_{\sigma}\Psi\bigg|_{\sigma=\sigma_{i}} = \nonumber \\
  \sum^{N}_{j=0} \bigg( \text{diag}(\tilde{\iota}(\sigma_{i})) \cdot D^{(1)} \bigg)_{ij} \big[ \Psi_{j} + \Delta_{\Psi}(\sigma_{j}-\sigma_{p};\sigma_{i} - \sigma_{p}) \big],  \nonumber   \\
  \label{iotad1_discretization}
\end{eqnarray} 
\end{small}
\begin{small}
\begin{eqnarray}
  \tilde{ \upvarepsilon}(\sigma) \partial_{\sigma}\Pi\bigg|_{\sigma=\sigma_{i}} = \nonumber \\
  \sum^{N}_{j=0} \bigg( \text{diag}(\tilde{ \upvarepsilon}(\sigma_{i})) \cdot D^{(1)} \bigg)_{ij} \big[ \Pi_{j} + \Delta_{\Pi}(\sigma_{j}-\sigma_{p};\sigma_{i} - \sigma_{p}) \big].  \nonumber   \\
  \label{alpha_d1_discretization}
\end{eqnarray} 
\end{small}
The explicit form of $\tilde{\textbf{s}}(\tau)$ in Eq.~\eqref{s_spatial_disc_vector} containing all the necessary corrections to the differential operators is, 
\begin{small}
\begin{eqnarray}
    \tilde{s}_{\Psi}^{(1)} = \nonumber \\
    \sum^{N}_{j=0} \bigg( \text{diag}(\tilde{\iota}(\sigma_{i})) \cdot D^{(1)} \bigg)_{ij} \big[\Delta_{\Psi}(\sigma_{j}-\sigma_{p};\sigma_{i} - \sigma_{p}) \big], \ \ \ 
    \label{s1psi_iota}
\end{eqnarray}
\end{small}
\begin{small}
\begin{eqnarray}
    \tilde{s}_{\Psi}^{(2)} = \nonumber \\
    = \sum^{N}_{j=0} \bigg( \text{diag}(\tilde{\upchi}(\sigma_{i})) \cdot D^{(2)} \bigg)_{ij} \big[ \Delta_{\Psi}(\sigma_{j}-\sigma_{p};\sigma_{i} - \sigma_{p}) \big], \ \ \ 
    \label{s2psi_chi}
\end{eqnarray}
\end{small}
\begin{small}
\begin{eqnarray}
    \tilde{s}_{\Pi}^{(1)} = \nonumber \\
    =\sum^{N}_{j=0} \bigg( \text{diag}(\tilde{ \upvarepsilon}(\sigma_{i})) \cdot D^{(1)} \bigg)_{ij} \big[  \Delta_{\Pi}(\sigma_{j}-\sigma_{p};\sigma_{i} - \sigma_{p}) \big]. 
    \label{spi_epsilon}
\end{eqnarray}
\end{small}
To be precise Eqs.(\ref{s1psi_iota}, \ref{s2psi_chi}) contribute to the corrected $\textbf{L}_{1}$ operator associated with the master function $\Psi(\tau,\sigma)$ whereas Eq.~\eqref{spi_epsilon} contributes to the corrected $\textbf{L}_{2}$ operator. 

\subsubsection{Higher Order Jumps Recurrence Relation Derivation}\label{AppC_DiscSpace_REC}
For completion we include here the corrected hyperboloidal differential operators as described in detail Appendix \ref{appA}. The procedure follows directly and all the remainder terms give the $J_{m+2}(\tau)$ recurrence relation given in Eq.~\eqref{cha3_rec_relation_njumps_RWZ}. 
The function $\Psi^{J}(\sigma,\tau;\sigma_{p})$ first-order derivative has the form,  
\begin{eqnarray}
    \upvarrho(\sigma) \partial_{\tau}\Psi(\zeta) 
    &=& \sum^{\infty}_{m=0} \bigg[ \sum^{m}_{k=0} \upvarrho(\sigma_{p})^{(k)} {m\choose k} \bigg( \dot{J}_{m-k}(\tau) \nonumber \\ 
    &-& \dot{\sigma}_{p} J_{m+1-k}(\tau) \bigg)  \bigg] \Psi(\zeta) \nonumber \\ 
    &-&\upvarrho(\sigma_{p})\dot{\sigma}_{p} J_{0}(\tau)\delta(\zeta).   
    \label{hyper_partialtau_rho}
\end{eqnarray}
For the second-order temporal form we have, 
\begin{small}
\begin{eqnarray}
    \Gamma(\sigma) \partial_{\tau}^{2}\Psi(\zeta)
 &=& \sum^{\infty}_{m=0} \bigg[ \sum^{m}_{k=0} \Gamma(\sigma_{p})^{(k)} {m\choose k}  \bigg( \ddot{J}_{m-k}(\tau) \nonumber \\
 &-& \dot{\sigma}_{p} \dot{J}_{m+1-k}(\tau) - \ddot{\sigma}_{p} J_{m+1-k}(\tau) \bigg) \bigg]\Psi(\zeta) \nonumber \\
  &+& \Gamma(\sigma_{p}) \dot{\sigma}_{p}^{2}J_{m+1}(\tau)\delta(\zeta) - 2 \Gamma(\sigma_{p}) \dot{\sigma}_{p} \dot{J}_{0}(\tau) \delta(\zeta) \nonumber \\
    &-& \Gamma(\sigma_{p}) \ddot{\sigma}_{p} J_{0}(\tau) \delta(\zeta) + \Gamma(\sigma_{p}) \ddot{\sigma}_{p}J_{0}(\tau)\delta(\zeta). 
   \label{hyper_partiald2tau_Gamma}
\end{eqnarray}
\end{small}
The spatial form of $\Psi^{J}(\sigma,\tau;\sigma_{p})$ is given at first order as, 
\begin{eqnarray}
    \iota(\sigma)\partial_{\sigma}\Psi(\zeta) &=& \sum^{\infty}_{m=0} \bigg[ \sum^{m}_{k=0} \iota^{(k)}(\sigma_{p})  {m\choose k} \bigg( J_{n+1-k}(\tau) \bigg) \bigg]\Psi(\zeta), \nonumber \\
    &+& \iota(\sigma_{p})J_{0}(\tau)\delta(\zeta)
    \label{hyper_partialsigma_iota}
\end{eqnarray}
and second-order,
\begin{eqnarray}
     \upchi(\sigma)\partial_{\sigma}^{2}\Psi(\zeta) &=& \sum^{\infty}_{m=0} \bigg[ \sum^{m}_{k=0}  \upchi(\sigma_{p})^{(k)}  {m\choose k} \bigg( J_{m+2-k}(\tau))\bigg) \bigg]\Psi(\zeta) \nonumber \\
    &+&  \upchi(\sigma_{p}) J_{1}(\tau)\delta(\zeta) +  \upchi(\sigma_{p})J_{0}(\tau)\delta(\zeta). 
    \label{hyper_partiald2sigma_chi}
\end{eqnarray}
Finally the first order crossed derivative in $(\tau,\sigma)$ is given as, 
\begin{eqnarray}
   \upvarepsilon({\sigma})\partial_{\sigma}\partial_{\tau}\Psi(\zeta) 
    &=& \sum^{\infty}_{m=0} \bigg[ \sum^{m}_{k=0} \upvarepsilon(\sigma_{p})^{(k)}{m\choose k} \bigg( \dot{J}_{m+1-k}(\tau)  \nonumber \\ 
    &-&\sigma_{p}J_{m+2-k}(\tau) \bigg) \bigg] \Psi(\zeta) \nonumber \\
    &+& \upvarepsilon(\sigma_{p}) (\dot{J}_{0}(\tau) - \dot{\sigma}_{p}J_{1}(\tau))\delta(\zeta)  \nonumber \\
    &-& \upvarepsilon(\sigma_{p})\dot{\sigma}_{p} J_{0}(\tau)\delta'(\zeta).  
    \label{hyper_partialrhopartialtau_epsilon}
\end{eqnarray}
Collecting all higher order terms we get the recurrence relation given in Eq.~\eqref{cha3_rec_relation_njumps_RWZ}.  

\subsubsection{Initialising hyperboloidal jumps}\label{AppC_DiscSpace_Jumps}
In this work we choose to directly transform the jumps to the hyperboloidal chart, by using the selection rule as given in Eq.~(\eqref{appendix_a_dirac_compI}), 
\begin{equation}
    J_{0}(\tau) = \frac{1}{\chi(\sigma_{p})}f(\sigma_{p})\frac{g'(\sigma_{p})}{|g'(\sigma_{p})|^{3}} J_{0}(t(\tau,\sigma_{p})), 
    \label{hyper_j0}
\end{equation}
where here $g(\sigma) = r(\sigma) - r_{p}$ and $J_{0}(t)$ is given in Eq.~\eqref{ch3_j0} and further specified by Eqs.~(\eqref{axial_F}, \eqref{polar_f}) now in hyperboloidal time and hyperboloidal particle position $r \rightarrow r_{p} \rightarrow\ \sigma_{p}$. To obtain the second jump, we can directly use the first-order chain rule associated with the coordinate mapping of this work, $t \rightarrow t(\tau,\sigma), r\rightarrow r(\sigma)$, 
\begin{equation}
  \label{hyperx_j1}
    J_{1}(\tau) = - \lambda H'(\sigma_{p}) \dot{J}_{1}(t(\tau,\sigma_{p})) +  r'(\sigma_{p}) J_{1}(t(\tau,\sigma_{p}))
\end{equation}
where explicitly $ J_{1}(t),\dot{J}_{1}(t)$ are given in Section \ref{sectioniii} Eqs.(\eqref{ch3_j1}, \eqref{ch3_jt}) respectively. 

\subsection{Discontinuous time integration}\label{AppC_DiscTime}
Given in this work we only work with circular orbits, i.e there is no time dependence on the particle position, $\sigma_{p}$, the time integration scheme greatly simplifies as we do not need to account for any time jumps given as  $J(\Delta \tau, \Delta \tau_{i})$ in Eq.~\eqref{cha3_final_integral} as the contributions would only be residual, sufficing therefore to use the aforementioned formula. Nevertheless for completion and to simplify future work we include here the time jumps resulting from a discontinuous order 4 Hermite rule. Here we closely follow the methodology first demonstrated by \cite{2014arXiv1406.4865M} and later improved by \cite{phdthesis-lidia, 24thCapraTalk, 25thCapraTalk, reviews-lidia}. \\

Firstly, we consider the ODE, 
\begin{equation}
    \frac{dy}{dt} = f(y,t), 
    \label{disco_time_ode}
\end{equation}
approximating it's solution on a small interval $[t_{1},t_{2}]$ we have, 
\begin{equation}
    y(t_{2}) -y(t_{1}) = \int^{t_{2}}_{t_{1}} f(y,t) dt. 
    \label{disco_time_ode_int}
\end{equation}

They can be regarded together as a single piecewise polynomial as in Eq.\eqref{ch3_sb2_gen_p_poly}
\begin{equation}
    p(t) = \theta(t - \tau)p_{>}(t) + \theta(\tau -t)p_{<}(t), 
    \label{disco_time_singlepoly}
\end{equation}

To attain a fourth-order time integrator scheme we now need to construct a piecewise third order polynomial to the right and left, respectively,
\begin{eqnarray}
    p_{>}(x) = a_{0} + a_{1}x + a_{2}x^{2} + a_{3}x^{3}, \\
    \label{poly_disco4_right}
     p_{<}(x) = b_{0} + b_{1}x + b_{2}x^{2} + b_{3}x^{3}.
    \label{poly_disco4_left}
\end{eqnarray}
We solve this by using the algebraic method of undetermined coefficients with the following eight algebraic conditions for the aforementioned eight unknown coefficients, 
\begin{eqnarray}
    p_{4,<}(t_{1}) = p_{t_{1}}, \\ 
    \label{poly_disco4_co1}
    p_{4,>}(t_{2}) = p_{t_{2}}, \\ 
    \label{poly_disco4_co2} 
    p_{4,<}'(t_{1}) = dp_{t_{1}}, \\ 
    \label{poly_disco4_co1der}
    p_{4,>}'(t_{2}) = dp_{t_{2}}, \\ 
    \label{poly_disco4_co2der} 
    p_{4,>}(\tau) - p_{4,<}(\tau) = J_{0}, \\ 
    \label{poly_disco4_J0}
     p_{4,>}'(\tau) - p_{4,<}'(\tau) = J_{1}, \\ 
    \label{poly_disco4_J1}
     p_{4,>}''(\tau) - p_{4,<}''(\tau) = J_{2}, \\ 
    \label{poly_disco4_J2}
     p_{4,>}'''(\tau) - p_{4,<}'''(\tau) = J_{3}. 
    \label{poly_disco4_J3}
\end{eqnarray}
Solving the system, with the aid of Mathematica, and performing an Hermite $4^{rd}$ order time integration scheme in the interval where $t_{1} = \tau - \Delta \tau$ and $t_{2} = t_{1} + \Delta t$ we get, 

\begin{eqnarray}
     y(t_{2}) -y(t_{1}) &\approx& \frac{\Delta  t}{2} (f_{1} + f_{2})  + \frac{\Delta  t^{2}}{12} (f'_{1} + f'_{2}) \nonumber \\
     &+&   \textbf{J}(\Delta \tau, \Delta \tau_{i}), 
\end{eqnarray}
and we can thereby write $\textbf{J}(\Delta \tau, \Delta \tau_{i})$ to be given as, 

\begin{eqnarray}
     J(\Delta \tau, \Delta \tau_{i}) =  \textbf{J}_{0} + \textbf{J}_{1} + \textbf{J}_{2} + \textbf{J}_{3}, 
\end{eqnarray}
where the set of four time jumps is explicitly given as, 
\begin{eqnarray}
    \textbf{J}_{0} = \frac{1}{2} \big( \Delta t - 2 \Delta \tau \big) J_{0},\\
    \label{time_jump0}
     \textbf{J}_{1} = \frac{1}{12} \big( \Delta t^{2} -6 \Delta t \Delta \tau + 6 \Delta \tau^{2} \big) J_{1},\\
    \label{time_jump1}
    \textbf{J}_{2} = -\frac{1}{12}\Delta \tau \big( \Delta t^{2}  -3 \Delta t \Delta \tau + 2 \Delta \tau^{2} \big) J_{2},\\
    \label{time_jump2}
    \textbf{J}_{3} = \frac{1}{24}\Delta \tau^{2} \big( \Delta t - 2 \Delta\tau^{2} \big)^{2} J_{3}.
    \label{time_jump3}
\end{eqnarray}


\section{Complementary numerics and results to Section \ref{sectioniv}}\label{AppD_complement}
\subsection{Preliminary numerics - Price's Law test}\label{AppD_PriceLaw}
Given that we do not have an exact solution to our problem, we test our algorithm for the simpler case describing a scalar perturbation in Schwarzschild. The PDE problem given in Eqs.~(\eqref{ch3_rwz_fieldvariables_hyper}, \eqref{cha3_hyper_transformed_operator}) now simplifies to a homogeneous problem, i.e the RHS of Eq.~\eqref{ch3_rwz_fieldvariables_hyper} vanishes and has a scalar potential given as, 
\begin{equation}
    \tilde{V}_{l}^{S}(\sigma) = -\frac{1}{(1+\sigma)} \bigg( l(l+1) + \sigma \bigg).
    \label{scalar_potential}
\end{equation}
We can then validate our code by studying it's late time behaviour which should obey a Price's Law \cite{price1972nonspherical_i} , given as 
\begin{eqnarray}
\Psi_{l}|_{\sigma>0} \approx \tau^{-2 l - 3}, \\
\label{ch3g1_pricelawtoinf}
\Psi_{l}|_{\sigma=0} \approx \tau^{-l - 2}. \\
\label{ch3g1_pricelaw}
\end{eqnarray}

Following our previous work in \cite{o2022conservative}, we compute an effective power-law index for each $l$ mode given as, 
\begin{equation}
    \Gamma_{l} = |\tau \partial_{\tau} \ln \Psi_{l}| = \bigg|\frac{\tau \Pi_{l}}{\Psi_{l}} \bigg|,
    \label{ch3g1_pricetails_formula}
\end{equation}
where specifically we will evaluate $\Gamma_{l={0,1,2}}$ both at infinity, $\mathscr{I}^{+}$ (located at $\sigma=0$) and at the horizon, $\mathcal{H}$ (located at $\sigma=1$). 
As reported in \cite{price1972nonspherical_i}, we expect the code to converge at late times to 
\begin{eqnarray}
\Gamma_{l=0}|_{\sigma=0} = 2, \ \ \  \& \ \ \   \Gamma_{l=0}|_{\sigma=1} = 3, \nonumber \\
\Gamma_{l=1}|_{\sigma=0} = 3, \ \ \   \& \ \ \   \Gamma_{l=1}|_{\sigma=1} = 5, \nonumber \\
\Gamma_{l=2}|_{\sigma=0} = 4, \ \ \   \& \ \ \  \Gamma_{l=2}|_{\sigma=1} = 7. \nonumber \\
\label{ch3g1_gamma_expected}
\end{eqnarray}

\begin{figure}
\includegraphics[width=75mm]{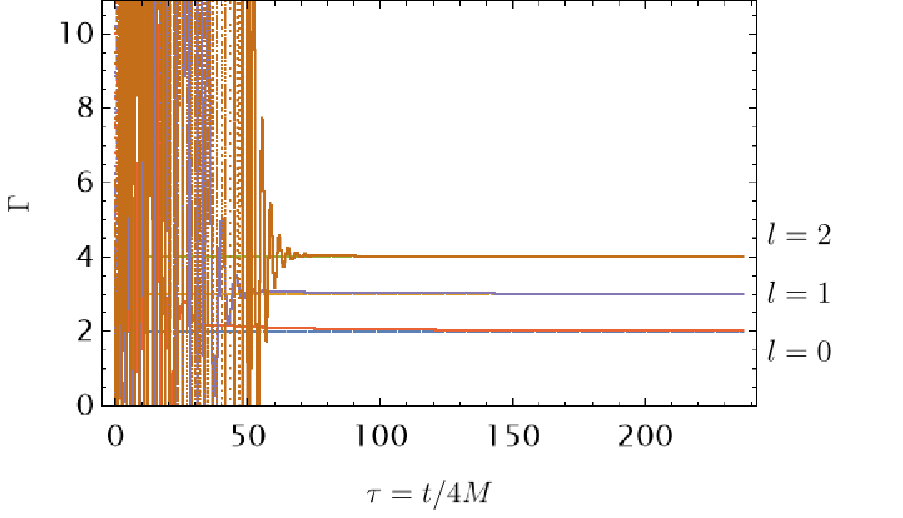}
\quad
\includegraphics[width=75mm]{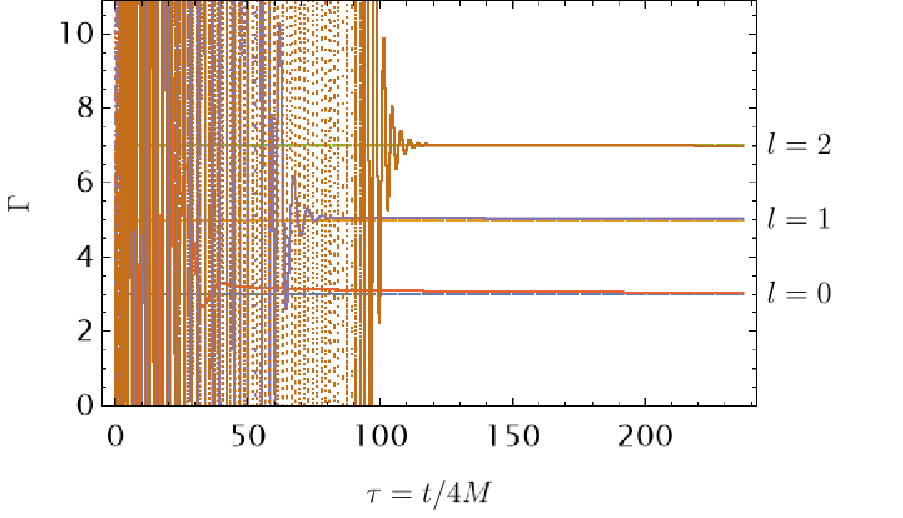}
\caption{\label{ch3_pricetails_fig}
Vacuum scalar field numerical evolution of the scalar perturbation problem with our numerical algorithm for the $l = {0,1,2}$ modes. We observe that the late time behaviour is as expected from the literature, highlighted in Eqs.~\eqref{ch3g1_gamma_expected} further validating our implementation in this regime.}
\end{figure}

To gauge the accuracy of our method we now implement the numerical recipe described in Section \ref{sectioniii} by Eq.~\eqref{reduced_form_evolution_homoge_integration}. 
For initial data we choose to inject a Gaussian pulse of the form, 
\begin{eqnarray}
\Psi_{\ell}(\tau=0,\sigma) = \exp{\bigg(-\frac{(\sigma-\sigma_{0})}{w^{2}} \bigg)},\nonumber \\
\partial_{\tau}\Psi_{\ell}(\tau=0,\sigma) = 0,
\label{ch3g1_ID_vaccuum}
\end{eqnarray}
where $w^{2} = 1/1000$, $\sigma_{0} = 0.6$ and we used $N=200$ pseudospectral collocation nodes as given in Eqs.~\eqref{chebyshev_lobatto_nodes}. 
As motivated in the preceding section, our numerical method is in a fully hyperboloidal framework and therefore we do not need to impose any BCs. Applying our fourth-order algorithm, we
verify in Fig.~\ref{ch3_pricetails_fig} that at late times the code for the modes $\ell={0,1,2}$ converges as expected for it's effective power law indices given in Eq.~\eqref{ch3g1_gamma_expected}.

\subsection{Complementary results to Section \ref{sectionivA}}\label{AppD_complementSecIVA}

\begin{figure*}
\includegraphics[width=80mm]{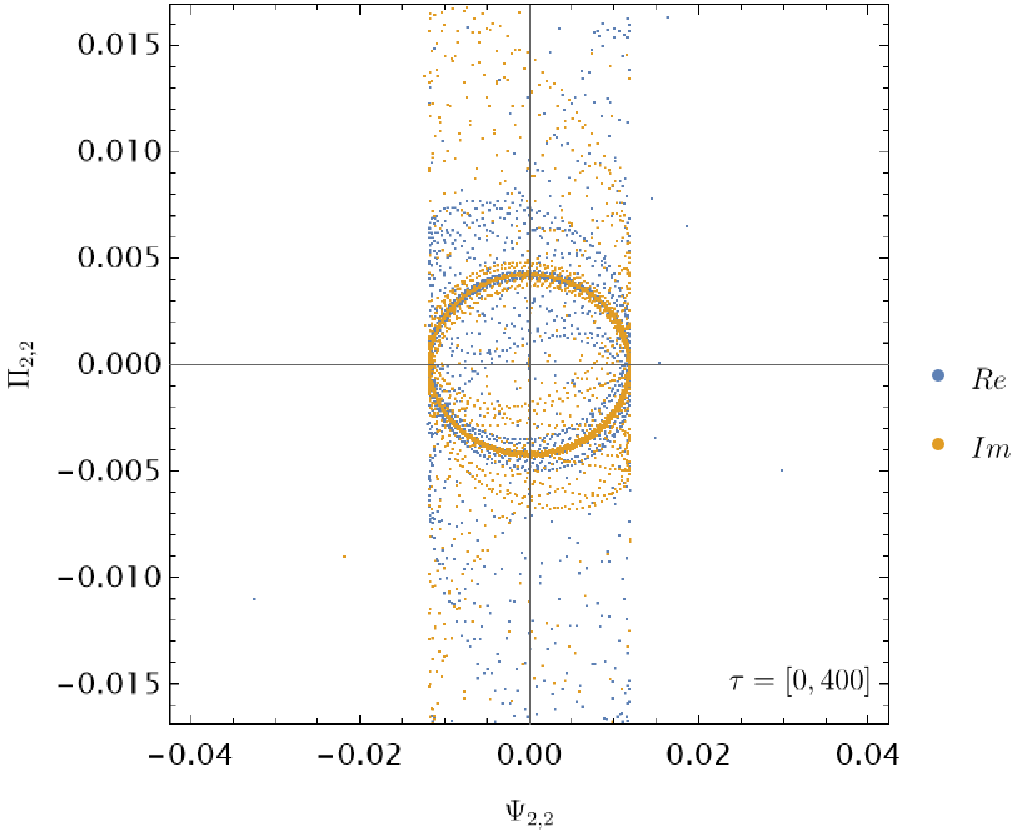}
\quad
\includegraphics[width=80mm]{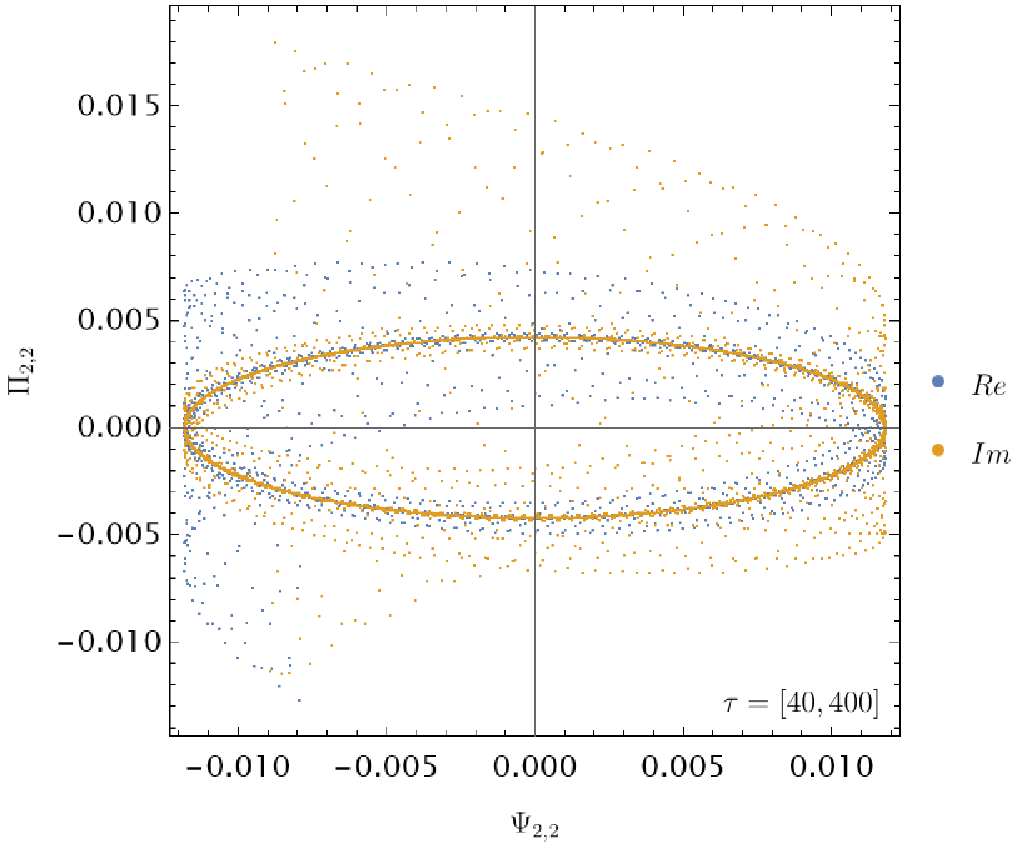}
\quad
\includegraphics[width=80mm]{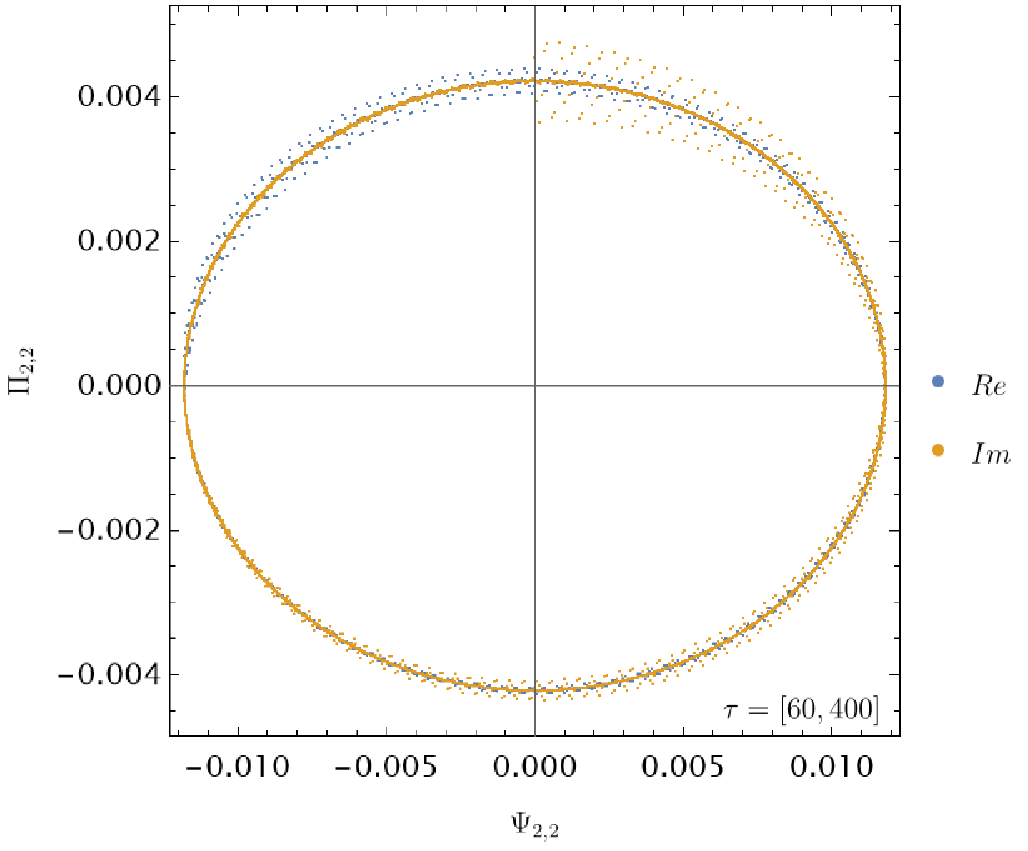}
\quad
\includegraphics[width=80mm]{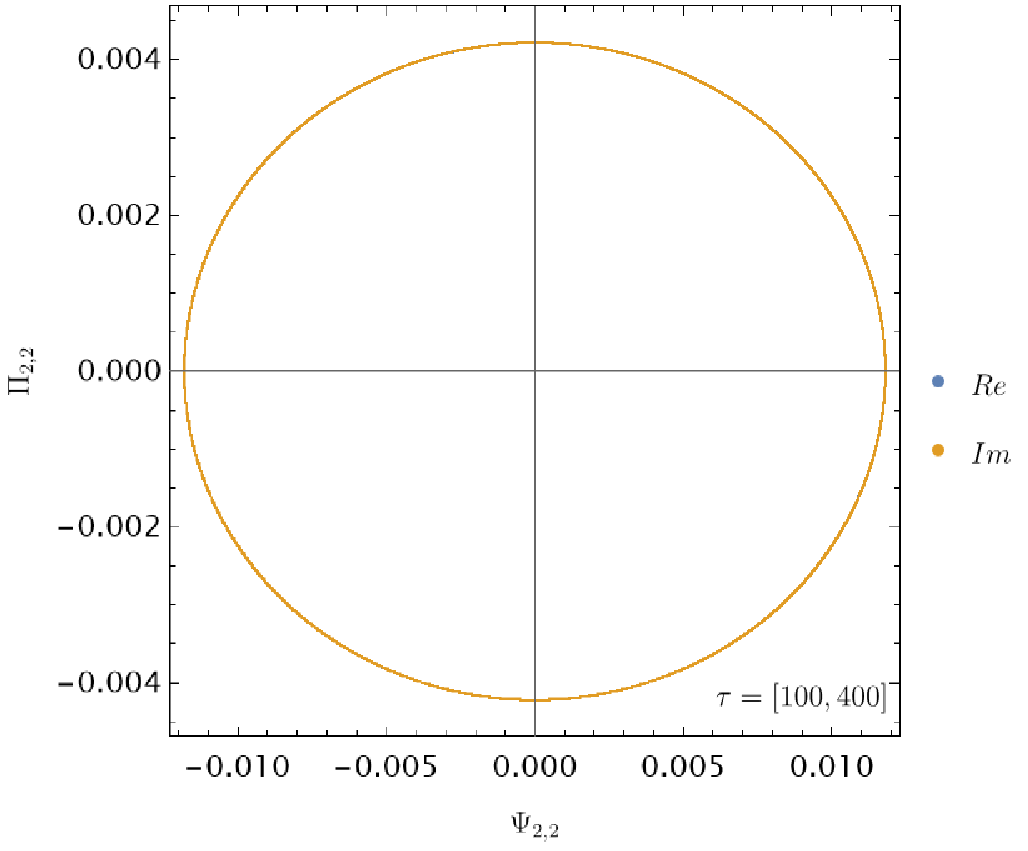}
\caption{\label{ch3_l2m2solutionPhasePsh}Here we show the phase portrait of our numerical evolution for the field $\Psi_{2,2}^{p}(\tau,\sigma)$ at $\mathcal{H}$. From left to right and then top to bottom, we include the evolution from an initial time of $\tau = \{0, 40, 60, 100\}$. We determine a reasonable time to assume all junk radiation to dissipate away and our simulation to be in \textit{steady-state} to be \textit{at least} $\tau > 100$, corresponding to ignoring the first 2.8 orbits as per Fig.~\ref{ch3_l2m2solutionPhasePsh}.}   %
\end{figure*}

To complement the discussion in Section \ref{sectionivA} and Fig.~\ref{ch3_l2m2solutionPhasePs} we include Fig.\ref{ch3_l2m2solutionPhasePsh}. The phase portraits show the radiation as it propagates towards $\mathcal{H}$. This corroborates further the choice for factor $iii)$ \textit{the minimal time for steady-state evolution} needs to be at least $\tau>100$. 

\subsection{Complementary results to Section \ref{sectionivC}}\label{AppD_complementSecIVC}

Even though the determination of numerical optimisation factors $iii) - iv)$ is left for future work we find pertinent to complement our numerical studies with the preliminary numerical tests. We observed, for an evolution time  interval of $\tau = [0, 10, 000]$ symplectic structure to be preserved and optimal extraction time for $\Delta \tau = 0.02$.\\
Furthermore, for the sake of completion and bench-marking we include the results for both the first order time and radial derivatives, given respectively by Table \ref{ch3g3_table_evalIntExt_Pi} and \ref{ch3g3_table_evalIntExt_drpsi} for the first $l = 5$ modes. 

\begin{table*}
\begin{ruledtabular}
\begin{tabular}{l|| c c||c c}
\textrm{$(l,m)$}& 
\textrm{$\Pi^{Int}(t_{max},r)$}&
\textrm{$\eta$}&
\textrm{$\Pi^{Ext}(t_{max},r)$}&
\textrm{$\eta$}\\
\colrule
(2,1) & $ -0.117471875 - 0.029912672 \ i $  & $[5.7 \times 10^{-9}] $  & $0.386167254 + 0.098541410 \ i $   & $[1.7 \times 10^{-9}] $ \\ 
(2,2) & $-0.125142303 + 0.217737529 \ i $ & $[4.9 \times 10^{-9}] $  & $-0.456413260 + 0.824910796 \ i $   & $[1.3 \times 10^{-9}] $ \\ \hline 
(3,1) & $-0.093342500 - 0.023807247 \ i $ & $[7.9\times 10^{-8}] $     & $-0.217300974 - 0.055423083\ i $  & $[3.4 \times 10^{-8}] $ \\ 
(3,2) & $0.10218625 - 0.18734009\ i $   & $[1.7\times 10^{-8}] $  & $-0.227128840 + 0.416248349\ i $    & $[7.5 \times 10^{-9}] $ \\ 
(3,3) & $-0.211430213 - 0.197717695 \ i $  & $[9.3\times 10^{-9}] $    & $-0.574227383 - 0.535144986 \ i $ & $[3.4 \times 10^{-9}] $  \\  \hline 
(4,1) & $0.1001092 + 0.0255330 \ i $ & $[2.7 \times 10^{-7}] $   & $-0.1774640 - 0.04526256 \ i $   & $[1.5 \times 10^{-7}] $  \\ 
(4,2) & $0.0746146- 0.1367581\ i $ & $[1.2 \times 10^{-7}] $   & $0.1473519 - 0.270075173\ i $   & $[6.1\times 10^{-8}] $  \\ 
(4,3) & $0.19525223 + 0.18159000  \ i $   & $[8.2 \times 10^{-8}] $  & $-0.35971670 - 0.33457075 \ i $    & $[4.4 \times 10^{-8}] $ \\ 
(4,4) & $0.27970922 - 0.17985623 \ i $ & $[5.1 \times 10^{-8}] $   & $0.61758168 - 0.39732184 \ i $  & $[2.4 \times 10^{-8}] $   \\  \hline 
(5,1) & $0.0643480 + 0.0164121 \ i $ & $[2.2\times 10^{-7}] $    & $0.1102957 + 0.0281311 \ i $ & $[1.3 \times 10^{-7}] $  \\ 
(5,2) & $ -0.0841617 + 0.1542565  \ i $   & $[2.7 \times 10^{-7}] $  & $0.1342813- 0.2461188 \ i $   & $[1.7 \times 10^{-7}] $  \\ 
(5,3) & $0.1459378 + 0.1357326\ i $  & $[2.0 \times 10^{-7}] $  & $0.2584503 + 0.2403772 \ i $  & $[1.2 \times 10^{-7}] $  \\ 
(5,4) & $-0.2511317 + 0.1616382 \ i $  & $[1.2\times 10^{-7}] $  & $0.413140 - 0.265908\ i $  & $[7.5 \times 10^{-8}] $  \\ 
(5,5) & $0.11736510 + 0.35171300 \ i $ & $[9.0 \times 10^{-8}] $   & $0.22614863 + 0.67761811 \ i $  & $[4.7 \times 10^{-8}] $  \\ 
\end{tabular}
\caption{Results for the first $l=5$ modes for the first-order time derivative of the RWZ master functions, i.e $\Pi(t_{\rm{max}},r)$ for both the interior and exterior solutions. We note the quantity in the square brackets is the numerical error computed as in Eq.~\eqref{numerical error} from converting the frequency domain data as given by \cite{thompson1811} to the time domain. }
\label{ch3g3_table_evalIntExt_Pi}
\end{ruledtabular} 
\end{table*}

\begin{table*}
\begin{ruledtabular}
\begin{tabular}{l|| c c||c c}
\textrm{$(l,m)$}& 
\textrm{$\partial_{r} \Psi^{Int}(t_{max},r)$}&
\textrm{$\eta$}&
\textrm{$\partial_{r} \Psi^{Ext}(t_{max},r)$}&
\textrm{$\eta$}\\
\colrule
(2,0) & $-0.2801982967$ &  $[3.8 \times  10^{-10}]$  & $0.6454758930$    & $[1.7 \times 10^{-10}] $  \\
(2,1) & $0.0625021010 - 0.2455016279 \ i $   & $[9.3 \times 10^{-10}] $   & $0.1530235017 - 0.6004153753 \ i $ & $[3.9\times 10^{-10}] $  \\ 
(2,2) &$ -0.212153203 - 0.1219054701 \ i $ & $[1.3 \times 10^{-9}] $   & $0.5544141559  + 0.2963301739 \ i $   & $[5.3\times 10^{-10}] $ \\ \hline  
(3,0) & $0.362412173$ &  $[1.4\times  10^{-9}]$ & $0.627804135$    & $[7.2\times 10^{-10}] $ \\
(3,1) & $0.07065076 - 0.27700468 \ i $ & $[1.0 \times 10^{-8}] $    & $-0.133359267 + 0.522872069 \ i $  & $[1.0\times 10^{-9}] $ \\ 
(3,2) & $0.275061905 + 0.150034786 \ i $  & $[4.4 \times 10^{-9}] $  & $0.487735845 + 0.266068715 \ i $  & $[2.6\times 10^{-9}] $  \\ 
(3,3) & $0.186245852 - 0.199162944 \ i $ & $[3.7\times 10^{-9}] $    & $-0.386249149 + 0.416375786 \ i $ & $[2.1 \times 10^{-9}] $  \\ \hline
(4,0) & $0.301720664 $ &  $[4.0\times  10^{-9}]$  & $-0.493539091 $   & $[2.6 \times 10^{-9}] $   \\
(4,1) & $-0.09810793 + 0.38465881 \ i $  & $[2.7\times 10^{-8}] $   & $-0.14799746 + 0.58026427 \ i $ & $[1.8 \times 10^{-8}] $  \\ 
(4,2) & $0.25928978 + 0.14146750 \ i $  & $[2.3\times 10^{-8}] $   & $-0.14799746 + 0.58026427 \ i $  & $[1.4 \times 10^{-8}] $ \\ 
(4,3) & $0.25928978 + 0.14146750\ i $ & $[2.3 \times 10^{-8}] $    & $-0.43258161 - 0.23601436 \ i $ & $[1.5 \times 10^{-8}] $  \\ 
(4,4) & $-0.22591315 + 0.24291010  \ i $ & $[1.8\times 10^{-8}] $   & $-0.34715906+ 0.37327203 \ i $  & $[3.2\times 10^{-8}] $  \\ \hline 
(5,0) & $-0.42904490$ &  $[1.0 \times  10^{-8}]$  & $-0.595388116$    & $[7.4\times 10^{-9}] $  \\
(5,1) & $ -0.076729723 + 0.300839729 \ i $ & $6.8 \times 10^{-9}] $   & $0.115287200- 0.452014792 \ i $  & $[4.7 \times 10^{-9}] $  \\ 
(5,2) & $-0.35835590 - 0.19551733 \ i $  & $[3.9\times 10^{-8}] $   & $-0.49942783 - 0.27248552 \ i $  & $[2.8 \times 10^{-8}] $ \\ 
(5,3) & $-0.20740485 + 0.22299877\ i $   & $[5.4 \times 10^{-8}] $  & $0.31873787 - 0.34270262 \ i $    & $[3.6 \times 10^{-8}] $ \\ 
(5,4) & $ -0.18217825 - 0.28304396\ i $  & $[4.5\times 10^{-8}] $  & $0.25750113 - 0.40007178 \ i $  & $[3.1\times 10^{-8}] $  \\ 
(5,5) & $-0.30930906 + 0.10321590 \ i $  & $[4.4\times 10^{-8}] $  & $ 0.50421824- 0.16832830\ i $    & $[2.8 \times 10^{-8}] $ \\ 
\end{tabular}
\caption{Results for the first $l=5$ modes for the first-order time derivative of the RWZ master functions, i.e $\partial_{r} \Psi(t_{\rm{max}},r)$ for both the interior and exterior solutions. We note the quantity in the square brackets is the numerical error computed as in Eq.~\eqref{numerical error} from converting the frequency domain data as given by \cite{thompson1811} to the time domain. }
\label{ch3g3_table_evalIntExt_drpsi}
\end{ruledtabular} 
\end{table*}

\begin{figure}
\includegraphics[width=80mm]{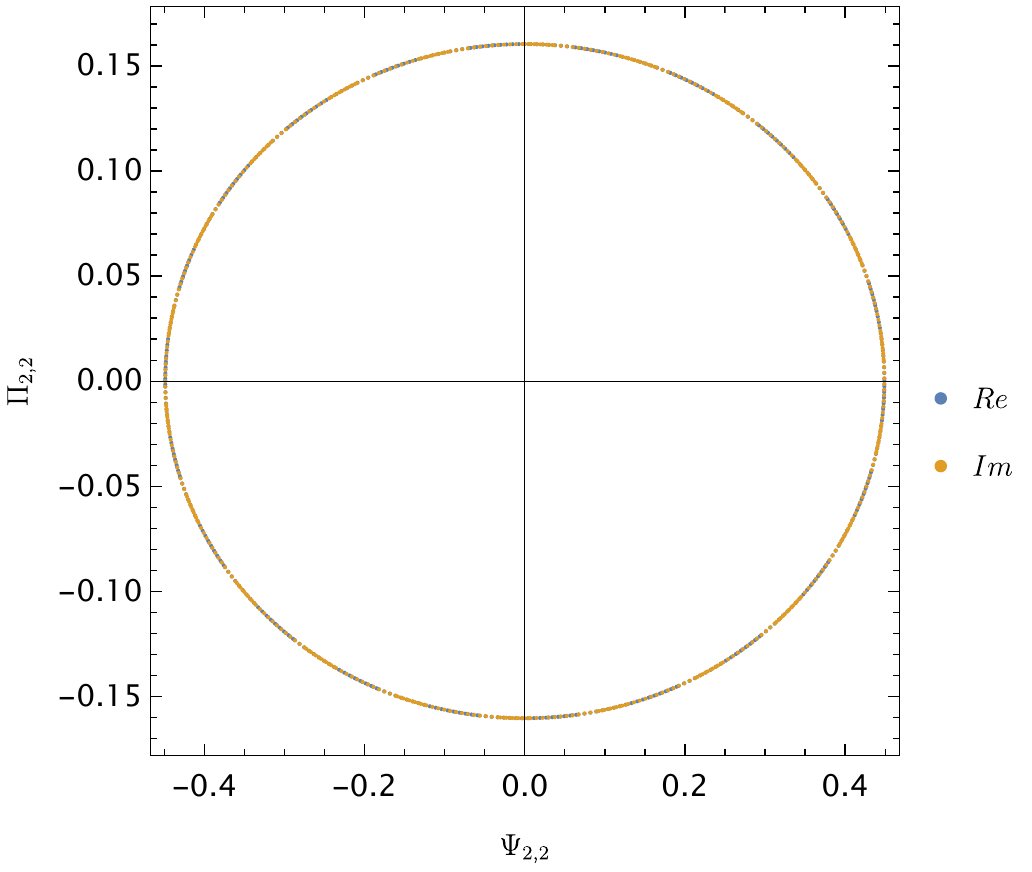}
\quad
\includegraphics[width=90mm]{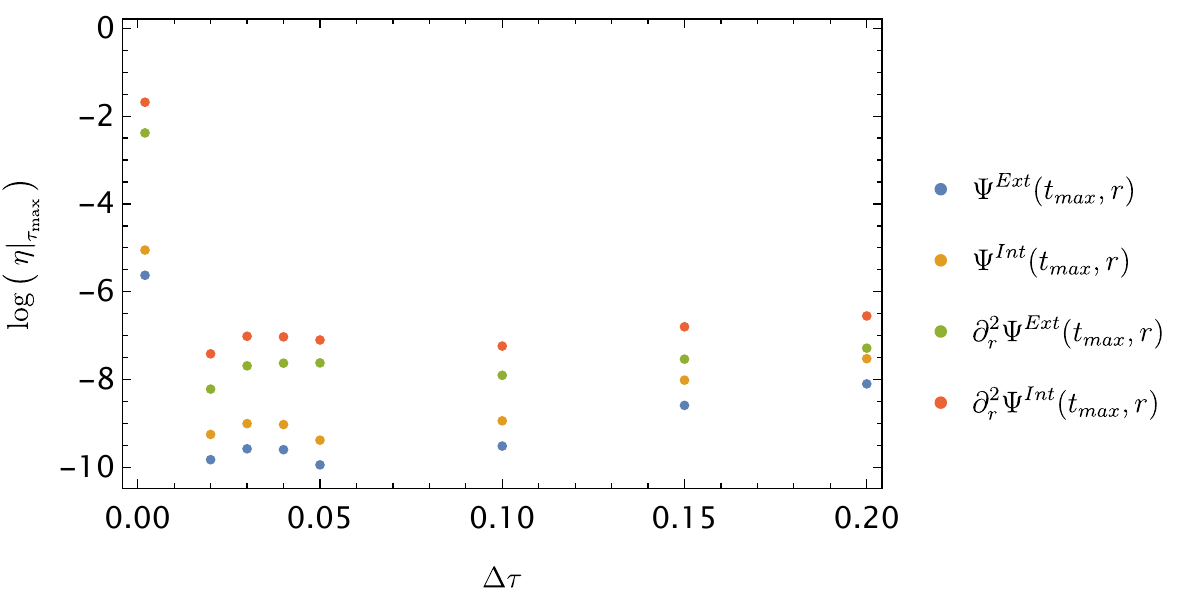}
\caption{\label{ChebyTimeOFs} \textbf{Top:} Phase portrait showing symplectic structure preservation for the $(l,m) = (2,2)$ mode at $\tau = 10,000$. \textbf{Bottom:} Numerical convergence \textit{preliminary} study for the optimal time discretisation step}%
\end{figure}

\clearpage

\providecommand{\noopsort}[1]{}\providecommand{\singleletter}[1]{#1}%

\end{document}